\documentclass[11pt]{article}
\pdfoutput=1
\usepackage{jheppub}

\usepackage{blindtext}

\usepackage{amsmath,amssymb,amsthm,amscd,graphicx}
\usepackage{psfrag}

\usepackage[utf8]{inputenc}
\usepackage{graphicx}
\usepackage{xcolor}
\usepackage{hyperref}
\hypersetup{colorlinks,allcolors=blue}

\definecolor{frangreen}{rgb}{0.040, 0.475, 0.435}

\usepackage[T1]{fontenc}

\usepackage[utf8]{inputenc}
\usepackage{graphicx}
\usepackage{amsmath,amsthm,amssymb,physics,mathtools}
\usepackage{tikz,tikz-cd, dynkin-diagrams}
\usepackage{epigraph}
\usepackage{hhline}
\usepackage{caption}
\usepackage{subcaption}
\usepackage{ytableau}
\usepackage{natbib}

\theoremstyle{definition}

\newcommand{\CZ}{{\cal Z}}

\newcommand{\bea}{\begin{eqnarray}}
\newcommand{\eea}{\end{eqnarray}}

\def\l{\ell}

\def\a{\boldsymbol{\alpha}}
\def\A{\boldsymbol{a}}
\def\b{\boldsymbol{\beta}}
\def\e{\boldsymbol{\eta}}
\def\ea{\epsilon_1}
\def\eb{\epsilon_2}
\def\s{\boldsymbol{\sigma}}

\def\l{\boldsymbol{\lambda}}

\def\n{\vb n}
\def\m{\vb m}

\def\v{\v}
\def\I{\sqrt{-1}}

\def\C{\mathbb{C}}
\def\N{\mathbb{N}}

\def\Z{\mathbb{Z}}
\def\12{\frac{1}{2}}
\newcommand{\be}{\begin{equation}}
\newcommand{\ee}{\end{equation}}
\newcommand{\ba}{\begin{aligned}}
	\newcommand{\ben}{\begin{eqnarray}\displaystyle}
	\newcommand{\een}{\end{eqnarray}}

\begin{document}

		\title{$tt^*$ Toda equations for surface defects in ${\mathcal N}=2$ SYM and instanton counting for classical Lie groups}
		
		\author{Giulio Bonelli, Fran Globlek, Alessandro Tanzini}
		\affiliation{International School of Advanced Studies (SISSA), via Bonomea 265, 34136 Trieste, Italy and INFN, Sezione di Trieste}
		\affiliation{Institute for Geometry and Physics, IGAP, via Beirut 2, 34136 Trieste, Italy}
    \emailAdd{bonelli@sissa.it}
    \emailAdd{fgloblek@sissa.it}
    \emailAdd{tanzini@sissa.it}
		
		\date{\today}
		
		\abstract{
			The partition function of $\mathcal{N}=2$ super Yang-Mills theories with arbitrary simple gauge group coupled to a self-dual $\Omega$-background is shown to be fully determined by studying the renormalization group equations relevant to the surface operators generating its one-form symmetries. The corresponding system of equations results in a {\it non-autonomous} Toda chain on the root system of the Langlands dual, the evolution parameter being the RG scale. A systematic algorithm computing the full multi-instanton corrections is derived in terms of recursion relations whose
			gauge theoretical solution is obtained just by fixing the
			perturbative part of the IR prepotential as its asymptotic boundary condition for the RGE. 
			We analyse the explicit solutions of the $\tau$-system for all the classical groups
		at the diverse levels, extend our analysis to affine twisted Lie algebras and provide conjectural bilinear relations for the $\tau$-functions of linear quiver gauge theory.
		}

		\maketitle
		
		
		\section{Introduction}

		In this paper we study the partition function of $\mathcal{N}=2$ super Yang Mills theories with general simple gauge group $G$ in presence of a surface defect. The latter is described by a two-dimensional $\mathcal{N}=(2,2)$ gauged linear $\sigma$-model living on the defect and coupled to the bulk four-dimensional theory. 
		This implies that the defect partition function obeys 
	 $tt^*$ equations \cite{Cecotti:1991me}, which for the theories under consideration correspond to a de-autonomized Toda system. The defect partition function is vector-valued according to the set of admissible boundary conditions, labeled by the roots of the affine Langlands dual Lie algebra $(\hat{\tt g})^\vee$\cite{Gukov:2006jk}. The de-autonomization corresponds to studying the gauge theory in the so-called self-dual $\Omega$-background $(\epsilon,-\epsilon)$. The limit $\epsilon\to 0$ reproduces the classical Seiberg-Witten theory \cite{Seiberg:1994aj} which is known to be described by the autonomous Toda chain of type $(\hat G)^\vee$ \cite{Gorsky:1995zq,Martinec:1995by}.    
		
		The system of equations we study is the radial reduction
		of $tt^*$-equations which describes complex deformations of a
		$Z(G)$-singularity, $Z(G)$ being the center of the gauge group. These are the equations of  
		{\it non-autonomous}
		twisted affine Toda chain of type $(\hat G )^\vee$, where $(\hat G)^\vee$ is the Langlands dual of the untwisted affine Kac-Moody algebra $\hat G$. 
		In order to clarify the appeareance of the Langlands dual
		group, we start from the analysis of the surface operators in the $\mathcal{N}=2^*$ theory in terms of the de-autonomized Calogero system, whose limit to super Yang-Mills naturally produces the relevant root system.
	    Each node of the resulting affine Dynkin diagram 
		defines a surface operator, the associated $\tau$-function being its vacuum expectation value. 
		A special r\^ole is played by the surface operators
		associated to the affine nodes. These are simple surface operators whose monodromy is twisted by elements of the center of the gauge group $Z(G)$. As such, they are bounded by fractional 't Hooft lines and generate the corresponding one-form symmetry of the gauge theory. This is manifest as
		a $Z(G)$-symmetry of the $\tau$-system and will be used to simplify its solution.
		Our analysis will be based on 
		the observation that the surface operators associated to affine nodes are described in a perturbative regime of the bulk gauge theory and as such the partition function of the theory in their  presence admits
		the ansatz \eqref{Kiev}.
		
		Indeed, the time flow of the non-autonomous system corresponds in the gauge theory to the renormalization group flow, the time playing the r\^ole of the gauge coupling constant. The system of equations produces recurrence relations for the coefficients of expansion in the gauge coupling \eqref{Kiev} thus providing a new effective algorithm to calculate instanton contributions for all classical groups $G$. 
		Actually, general recursion formulae based on bilinear relations can be provided for the $A,B$ and $D$ groups,
		while for gauge group of type $C$, $E$, $F$ and $G$ a case by case analysis is needed.
		
		On the mathematical side,
		the $\tau$-functions we obtain provide the general solution at the canonical rays for the Jimbo-Miwa-Ueno isomonodromic deformation problem 
		\cite{Jimbo:1981zz,Jimbo:1982zz}
		on the sphere with two-irregular punctures
		for all classical groups, which to the best of our knowledge was not known in the previous literature.

		The recursion relations we obtain are indeed different from the blow-up equations of \cite{Nakajima:2003pg} and further elaborated in \cite{Kim:2019uqw}.
		The latter necessarily involve the knowledge of the partition function in different $\Omega$-backgrounds,
		which makes the recursion relations and the results from blow-up equations more involved and difficult to handle.
		Nonetheless, we expect a relation between the two approaches to follow from surface defects blow-up relations.
		The isomonodromic $\tau$-function for the sphere with four regular punctures was obtained in a similar way from $SU(2)$ gauge theory with $N_f=4$ in \cite{Jeong:2020uxz}. An analogous analysis for general classical groups is still missing in the literature.

	Some of the results we discuss were anticipated in the letter
	\cite{Bonelli:2021rrg}. Here we elaborate on them, by giving more detailed proofs and checks and by extending to other cases.
	In particular we study the cases of twisted affine Lie algebrae and linear quiver theories. We find that the $\tau$ function for the twisted affine Lie algebra $BC_1$ interestingly satisfies the radial reduction of Bullough-Dodd equations, and it is related to the v.e.v. of surface defect in $\mathcal{N}=2$ $SU(2)$ gauge theory with one massless hypermultiplet in the fundamental representation. 
	We also study the 
 $BC_2$ for which we do not have at present a gauge theory interpretation.
	
	We conjecture bilinear relations satisfied by the $\tau$-functions of $SU(2)$ linear quiver theories which can be obtained from M-theory compactification on a Riemann sphere with two irregular punctures and $n-2$ regular ones \cite{Gaiotto:2009we}.
	From the mathematical viewpoint these $\tau$ functions describe isomonodromic deformations of $SL(2,\mathbb{C})$ flat connections on the very same Riemann sphere, and can be obtained from a suitable confluence limit of the Garnier system on the Riemann sphere with $n+2$ regular singularities. While the bilinear equations we write govern just the isomonodromic flow in the moduli of the two irregular punctures, we observe that a general solution can be found by imposing consistency of the system in suitable asymptotic limits. It would be interesting to complete the $\tau$-system with the equations governing the dependence on the moduli of the regular punctures and study the relation of the results we find with bilinear systems for the $\tau$-functions of the Garnier system \cite{Tsuda03birationalsymmetries}. 
	
	Let us comment on other interesting directions to investigate further.

		From the bulk four dimensional gauge theory perspective the $\tau$-system we find and its possible generalizations are expected to describe chiral ring relations in presence of a surface operator.
		Schematically, we expect eq.\eqref{hirotatausystem} to
		derive from the following fusion rule among the chiral operator ${\cal O}=\tr\phi^2$ and the surface operator $W_\beta$
		$$
		<:{\cal O}^2:W_\beta>=-{\frac{\b^\vee\cdot\b^\vee}{2}}\,t^{1/h^\vee} \prod\limits_{\a\in\hat\Delta,\a\neq\b} <W_{\boldsymbol\alpha}>^{-\a\cdot\b^\vee}\, ,
		$$
		while higher chiral observables should generate the flows of the full {\it non-autonomous} Toda hierarchy.
		
		The $\tau$-functions we compute in this paper could be used to describe through their zeroes the spectrum of the quantum Toda chain integrable system along the lines of 
		\cite{Bonelli:2016idi,Bershtein:2021uts}.
		
		It should be possible to apply the approach proposed here to general class-$\mathcal{S}$ theories \cite{Gaiotto:2009we} by studying the related isomonodromic deformation problem, for example for circular quivers, generalising to other classical groups the results of \cite{Bonelli:2019boe,Bonelli:2019yjd}.
		It would be also interesting to extend the analysis to non-self-dual $\Omega$-background, which should amount to the quantization of the $\tau$-systems, and its lift to five dimensional gauge theories on $\mathbb{R}^4\times S^1$, which should correspond to quantum $q$-difference $\tau$-systems \cite{Bonelli:2017gdk,Bershtein:2018srt,Bonelli:2020dcp,Nawata:2021dlk,Brini:2021wrm,Bonelli:2022dse}.
		
		The expansion in the large couplings regime should also be considered by extending the analysis of
		\cite{Its:2014lga,Bonelli:2016qwg}.
		Actually,
		the RG evolution at strong coupling can be analysed through late time expansion of the $\tau$-functions.
		In particular, in \cite{Bonelli:2017ptp} the solution in this regime for the $A_n$ series has been given 
		in terms of a matrix model describing the theory around the massless monopoles point which generalizes the $O(2)$ matrix model of \cite{McCoy:1976cd}. As a related problem, it would be also interesting to priovide a Fredholm determinant/Pfaffian representation for the $\tau$-functions presented here, see for example 
		\cite{Bertola:2021ugs} for the case of orthogonal groups.
		It would also be interesting to study the extension to defects in supergroup gauge theories, see for example \cite{Kimura:2021ngu}.

		{\bf Outline of the paper: }
		The content of the paper is organised as follows.
		In Section 2 we describe the non autonomous Calogero Moser system corresponding to $\mathcal{N}=2^*$ gauge thoery with general simple gauge group $G$ as an isomonodromic deformation problem on the elliptic curve and its limit to the non autonomous Toda system which instead describes the isomonodromic deformation problem on the Riemann sphere with two irregular singularities.
		We show how the limiting procedure directly gives the 
		twisted affine Toda chain of type $(\hat G)^\vee$. 
		In Section 3 we derive the $\tau$-form of the above non autonomous Toda system
		and fix the asymptotic conditions which are relevant to describe the gauge theory surface operators.
		In Section 4 we analyse the explicit solution of the $\tau$-system for all the classical groups
		at the diverse levels of accuracy we could reach.
		In Section 5 we extend our analysis to affine twisted Lie algebras and in Section 6 we comment on conjectural bilinear relations for the $\tau$-functions of linear quiver gauge theory.
		
		In the appendices we collect some technical results needed in the main text.
		In particular in Appendix B.5
		we discuss the blow-up formula in four dimensions 
		for general groups \eqref{blowup}
		and in appendix C we describe the universal asymptotic behaviour of the 
		instanton partition functions for large Coulomb moduli.

		\paragraph{\bf Acknowledgements:}
		We would like to thank 
		P. Gavrylenko, O. Lisovyy, N. Kubo,
		M. Mariño and T. Nosaka 
		for useful discussions and clarifications.
		This research  is  partially supported by the INFN Research Projects GAST and ST$\&$FI, PRIN "Geometria delle variet\`a algebriche", PRIN
		"Non-perturbative Aspects Of Gauge Theories And Strings",
		PRIN "String Theory as a bridge between Gauge Theories and Quantum Gravity" and INdAM. This work was partially done while A.T. was participating in the virtual program of the Institute for Mathematical Sciences, National University of Singapore, in 2021.

		\section{Isomonodromic deformations}
		
		In this section we derive the relevant equations on the $\tau$-functions for the Toda system related to any simple classical group. These are derived starting from the elliptic case in which Langland duality is manifest.
		
		On the one-pointed torus, $\mathcal{C}_{1,1}\cong\mathbb T_\tau\cong \mathbb C/\mathbb Z\oplus\tau\mathbb Z$, where $\tau\in\mathcal M\cong (\mathbb H_+)^{\text{PSL}(2,\mathbb Z)}\cup\{\sqrt{-1}\infty\}$ denotes a complex structure and corresponds to the isomonodromic time $\tau$
		The isomonodromic system is given by a Fuchsian system together with an isomonodromic flow
		\begin{gather}\nonumber
		\partial_{z}\Phi(z) = L(z)\Phi(z), \quad (2\pi\I)\partial_{\tau}\Phi(z) = - M(z)\Phi(z)
		\end{gather}
		where $z\in\mathcal{C}_{1,1}$ \cite{Takasaki1999}. The
		related autonomous integrable system is the elliptic Calogero-Moser system \cite{Krichever1980} which in gauge theory corresponds to $\mathcal{N}=2^*$. The reason for starting with an extra adjoint hypermultiplet as opposed to the pure theory is that the limit to pure theory gives the context as to why the Langland dual extended root system plays a r\^ole, since these are the only roots whose contributions survives in the decoupling limit to the de-autonomized Toda system.
		
		The deautonomized elliptic Calogero-Moser system 
		can be formulated for any complex simple Lie algebra $\mathfrak g$ of finite rank $k$, whose root system we realize in a finite dimensional $\C$-vector space $V$ equipped with an explicit basis $\{e_i\}_{i=1,...,\text{dim}V}$, so the root system is $R\subseteq V$. We identify $V^\vee\cong V$ using the canonical product.  Let $\boldsymbol \varphi:\mathcal M\to V$ be a vector valued function satisfying the \textit{deautonomized} elliptic Calogero-Moser system
		\begin{equation}\nonumber
		(2\pi \sqrt{-1}) \partial_\tau^2 \boldsymbol \varphi = -\frac{M^2}{2} \sum\limits_{\a \in R} \wp'( \a\cdot \boldsymbol \varphi |\tau)\boldsymbol\alpha
		\end{equation}
		where $\wp'(z|\tau)$ denotes the $z$-derivative of the Weierstrass elliptic function, and $M$ is the mass of the adjoint hypermultiplet. There is a well-defined autonomization procedure which maps the isomonodromic to the integrable system 
		\cite{takasaki1999elliptic}. These deautonomized systems are quite non-trivial. Indeed, in \cite{manin1996sixth} the so-called elliptic sixth Painlev\'e transcendent was defined as the solution to the equation $\partial_{\tau}^2 z = -\wp'(z|\tau)/(8\pi)^2$, and this is the simplest such system, corresponding to the Lie algebra $\mathfrak g = A_1$. 
		Let us briefly recall how the autonomization procedure works. Essentially, here we need to pass from the full problem formulated on the moduli space $\mathcal{M}$ of the one-punctured torus with complex structure $\tau$,  $\mathbb{T}_\tau$, to its tangent space at some fixed complex structure $\tau_0$, $T\mathcal M|_{\mathbb{T}_{\tau_0}}\cong H^0(\mathbb T_{\tau_0},\Omega^1)\cong \mathbb C$. As described in \cite{Levin2000}, we take $\tau=\tau_0+\epsilon t$, $\partial_\tau\mapsto \epsilon \partial_{\tau_0}$, and take the $\epsilon\to0$ limit, perhaps ridding ourselves of some convenient $2\pi i$ factors as well. In the context of gauge theory, this limit corresponds to turning off the Omega-background. 
		
		Let $\boldsymbol \rho^\vee$ and $h^\vee$ denote the dual Weyl vector and dual Coxeter number, respectively. The decoupling of the hypermultiplet which brings to pure $\mathcal{N}=2$ Super Yang-Mills or non-conformal AGT \cite{Marshakov:2009gn} is the Inosemtsev limit, achieved by setting
		\begin{align*}
		\tau &= \frac{1}{2\pi \sqrt{-1}}\log\left(\frac{\Lambda}{M}\right)^{2 h^\vee } \\
		\boldsymbol \varphi &\mapsto \boldsymbol \varphi+\frac{1}{2\pi \sqrt{-1}}\frac{1}{h^\vee} \log \left(\frac{\Lambda}{M}\right)^{2 h^\vee}\cdot \boldsymbol \rho^\vee
		\end{align*}
		and then sending $M\to\infty$. $\Lambda\in \C$ plays the role of the time. 
		
		To perform the limit, we quote the $q$-series of the relevant elliptic function \cite{Silverman1994}, which can be proved using the Lipschitz summation formula \cite[\S~2.2]{Zagier2008},
		\begin{equation}\nonumber
		(2\pi \sqrt{-1})^{-3}\wp'(z|\tau)=\sum\limits_{n\geq 0}q^n w_z\frac{1+q^n w_z}{(1-q^n w_z)^3} - \sum\limits_{n\geq1}q^n w_{-z} \frac{1+q^n w_{-z}}{(1-q^n w_{-z})^3}
		\end{equation}
		where $q=e^{2\pi \sqrt{-1} \tau}$ is the so-called \textit{nome} and $w_z=e^{2\pi \sqrt{-1} z}$. To perform the limit, first note that we may restrict ourselves to positive roots, as $\wp'( -\a,\vb Q|\tau)(- \a)=\wp'(\a,\vb Q|\tau)\a$. Second, examine the powers of $(\Lambda/M)^2$, and use the properties of positive simple coroots and the longest coroot given by the level function, as follows. Following \cite{DHoker1999}, define the \textit{level} function $\ell:R\to\mathbb R$ by $\boldsymbol\alpha\mapsto \ell(\boldsymbol\alpha):=\langle \boldsymbol\rho^\vee,\boldsymbol\alpha\rangle$, where $\boldsymbol \rho^\vee:=\frac{1}{2}\sum_{j=1}^r\boldsymbol\alpha_j^\vee\in \mathfrak g^\vee$ is the dual Weyl vector. Then, $\ell(\a^\vee)=1$ if and only if $\boldsymbol\alpha \in \Delta_+$, and $\ell(\a^\vee)=h^\vee-1$ if and only if $\a=\boldsymbol\theta^\vee$, where $h^\vee$ is the dual Coxeter number. Examining the terms remaining after the limit, we see we have contributions either from positive simple coroots, or from $\boldsymbol\theta^\vee$.
		
		The elliptic system reduces to a trigonometric one, and only the roots corresponding to the dual extended root system survive, namely the ones whose affine Cartan matrix got transposed. The significance of the dual affine system to SW theory is well-known \cite{Martinec:1995by,DHoker:1999yni}. The resulting system is
		\begin{equation} \label{Qsystem}
		\partial_{\log t}^2\boldsymbol \varphi = -t^{1/h^\vee}\sum\limits_{\a\in\hat\Delta_+} \a^\vee e^{\a^\vee\cdot\boldsymbol \varphi}
		\end{equation}
		where $t:=\Lambda^{2 h^\vee}$, $\hat\Delta_+=\{\boldsymbol\theta\}\cup\Delta_+$ are the extended positive roots, and we redefined $(2\pi \I)\boldsymbol \varphi$ $\mapsto\boldsymbol \varphi$ for simplicity. Once the asymptotic form of the solution is specified, the solution can be found by series expansion with a non zero (possibly infinite) convergence radius. The natural choice is to start with the homogenous solution and let $\boldsymbol \varphi = \vb a + \log t \cdot \vb b +  \boldsymbol \xi$ for constant $\vb a$ and $\vb b$. The prefactor $t^{1/h^\vee}$ can be eliminated by setting $\vb b = \boldsymbol \sigma - \frac{1}{h^\vee}\boldsymbol\rho^\vee$. After this, a solution in terms of a power series in $t$ and $\{t^{\sigma_i}\}_{i=1}^k$ can be found recursively from 
		\begin{equation}\label{Xsystem}
		\partial_{\log t}^2 \boldsymbol \xi = \boldsymbol\theta^\vee e^{-\boldsymbol\theta^\vee\cdot\vb a}t^{1-\boldsymbol\theta^\vee\cdot \s}e^{-\boldsymbol\theta\cdot \boldsymbol \xi} -
		\sum\limits_{\a\in\Delta_+}\a^\vee e^{\a^\vee\cdot \vb a}t^{\a^\vee\cdot\s}e^{\a^\vee\cdot \boldsymbol \xi}
		\end{equation} 
		from which we see that to ensure convergence, $\s\in\mathcal{W}_{\text{fund.}}^\vee$, the fundamental Weyl alcove of the dual root system, as
		\begin{equation}\nonumber
		\left.\begin{aligned}
		1-\boldsymbol\theta^\vee\cdot \s&>0\\
		\a_i^\vee\cdot \s&>0,\,i=1,...,r
		\end{aligned}\right\} \Rightarrow \s \in \mathcal{W}_{\text{fund.}}^\vee
		\end{equation}
		Therefore, solutions are in bijection with points of $\mathcal{W}_{\text{fund.}}^\vee$.
		
		The choice of the affine root is not unique if outer automorphisms of the affine Dynkin diagram exist. For the simplest case $A_1$, there is one root which we realize as $\boldsymbol\alpha=(1,-1)$ and the automorphism is the reflection around the origin. Then we have $\boldsymbol\rho=1/2\cdot \alpha = (1/2,-1/2)$ and $h^\vee=2$, so
		\begin{equation}\nonumber
		\vb b = \begin{pmatrix} 
		b_1 \\ b_2 
		\end{pmatrix} = \s - \frac{1}{h^\vee}\boldsymbol\rho^\vee =
		\begin{pmatrix} 
		\sigma_1 - 1/4 \\
		\sigma_2 + 1/4
		\end{pmatrix}
		\mapsto 
		\begin{pmatrix} 
		b_2 \\ b_1 
		\end{pmatrix} = 
		\begin{pmatrix} 
		\sigma_2 + 1/4 \\
		\sigma_1 - 1/4
		\end{pmatrix}
		\end{equation} 
		The effect of the reflection is $\sigma_1 \mapsto \sigma_2 + 1/2$, $\sigma_2\mapsto \sigma_1-1/2$. We should really be specializing to the $\mathfrak{sl}_2$ slice $\sigma_1+\sigma_2=0$, which we often neglect to make expressions simpler; setting $\sigma=\sigma_1=-\sigma_2$, however, we see that this is really the B\"acklund transformation $\tau(\sigma|t)\mapsto\tau(1/2-\sigma|t)$ of Painlev\'e III$_3$, analyzed in detail in \cite{Bershtein:2016uov}. In $A_n$, cyclic transformations may be seen to shift $\s$ by fundamental weights. We use this redundant to solve the system, 
		since as we will see it reduces the order of the equations drastically.

		\section{The Hirota relations}
		
		For any $\boldsymbol\alpha\in\hat\Delta_+$ we define the formal power series $\tau_{\boldsymbol\alpha}\in \mathbb{C}\left[ \left[ t,t^{\sigma_1},...,t^{\sigma_k} \right] \right]$ associated to $\boldsymbol \varphi$ as a solution to the following equation
		\begin{equation}
		\label{taudef}
		\partial_{\log t}^2 \log\tau_{\boldsymbol\alpha}(\boldsymbol \varphi,t) = t^{\frac{1}{h_{\mathfrak g}}} e^{\a^\vee\cdot\boldsymbol \varphi }
		\end{equation}
		up to constant and logarithmic terms. 
		
		We claim that the $\tau$ functions generate the Hamiltonian, in the sense that they satisfy
		\begin{equation}\nonumber
		\sum\limits_{\a\in\hat\Delta_+}\partial_{\log t} \log\tau_{\boldsymbol\alpha} = h^\vee t\, \mathcal{H}
		\end{equation}
		up to a constant, where
		\begin{equation}\nonumber
		t\,\mathcal{H}(\boldsymbol \varphi,\boldsymbol \pi,t)= \frac{1}{2}\boldsymbol\pi^2 + t^{\frac{1}{h^\vee}}\sum\limits_{\a\in\hat\Delta_+}e^{\a^\vee\cdot \boldsymbol \varphi}
		\end{equation}
		is the Hamiltonian of the deautonomised system.
		Indeed, we check that
		\begin{align*}
		(t\partial_t)\left(t \mathcal{H}\right)&=
		(t \partial_t)\left( \frac{\boldsymbol\pi^2}{2} + t^{\frac{1}{h^\vee}}\sum\limits_{\a\in\hat\Delta_+}e ^{\a^\vee\cdot \boldsymbol \varphi}\right)\\
		&=\boldsymbol \pi \cdot \partial_{\log t}\boldsymbol \pi + \left(t^{\frac{1}{h^\vee}}\sum\limits_{\a\in\hat\Delta_+}\a^\vee e ^{\a^\vee\cdot \boldsymbol \varphi}\right)\cdot\partial_{\log t}\boldsymbol \varphi +\frac{1}{h^\vee} t^{\frac{1}{h^\vee}}\sum\limits_{\a\in\hat\Delta_+} e ^{\a^\vee\cdot \boldsymbol \varphi}\nonumber
		\end{align*}
		and since the first two terms vanish on-shell, so the claim follows. Equipped with these $\tau$ functions, we note that \eqref{Qsystem} can be rewritten as 
		\begin{equation}\nonumber
		\partial_{\log t}^2\left(\boldsymbol \varphi + \sum_j \a_j^\vee \log\tau_{\a_j}\right)=0
		\end{equation} 
		We can integrate this and then reconstruct the solution $\boldsymbol \varphi$ in terms of $\tau$ functions from the components in the expansion
		 $\boldsymbol \varphi = \sum_i \varphi_i e_i$, namely
		\begin{equation}\label{QfromTau}
		\varphi_i = c_{1,i}+c_{2,i}\log{t}- \log\prod\limits_{\boldsymbol\alpha\in\hat\Delta_+}[\tau_{\boldsymbol\alpha}(\boldsymbol \varphi)]^{\a\cdot e_i}
		\end{equation}
		where the integration constants $c_{1,2}$ follow from the ambiguity in the definition of the $\tau$ functions. Feeding back into \eqref{taudef}, the isomonodromic system may be reformulated purely in terms of the $\tau$ functions as
		\begin{equation}\label{tausystem}
		\partial_{\log t}^2\log\tau_{\boldsymbol\alpha}=-t^{\frac{1}{h^\vee}} \prod\limits_{\boldsymbol\beta\in\hat\Delta_+}\left[ \tau_{\boldsymbol\alpha} \right]^{-\b^\vee\cdot\a^\vee}
		\end{equation}
		were the minus sign in the R.H.S. of \eqref{tausystem} is obtained by a rescaling the time variable as $t\to e^{\sqrt{-1}\pi h^\vee} t$. 
		We will find it useful to rewrite this expression in terms of a logarithmic Hirota derivative defined as \begin{equation}\nonumber
		D^2(f)=f^2\partial_{\log{t}}^2\log{f}=f\partial_{\log t}^2f - (\partial_{\log{t}}f)^2\end{equation}
		and satisfying
		\begin{align*}
		D^2(f^n)&=2n f^{2(n-1)}D^2(f^n)\\
		D^2(f \cdot g)&=f^2D^2(g)+g^2D^2(f) \, .
		\end{align*}
		We can then rewrite the system as
		\begin{equation}\label{hirotatausystem2}
		{\tau_{\a}}^{\a^\vee\cdot\a^\vee-2} D^2(\tau_{\a})=-t^{\frac{1}{h^\vee}} \prod\limits_{\b\neq\a}\left[ \tau_{\b} \right]^{-\b^\vee\cdot\a^\vee}
		\end{equation}
		where the factor of ${\tau_{\b}}^{\b^\vee\cdot\b^\vee}$ has been extracted from the product and carried to the other side and the definition of the Hirota derivative was used. Moreover, using the first Hirota derivative identity, we get
		\begin{equation}\nonumber
		    {\tau_{\a}}^{\a^\vee\cdot\a^\vee-2} D^2(\tau_{\a})=\frac{2}{\a^\vee\cdot\a^\vee}D^2([\tau_{\a}]^{\frac{\a^\vee\cdot\a^\vee}{2}})=-t^{\frac{1}{h^\vee}} \prod\limits_{\b\neq\a}\left[ \tau_{\b} \right]^{\frac{\b^\vee\cdot \b^\vee}{2}(-\a^\vee\cdot\frac{2 \b^\vee}{\b^\vee\cdot \b^\vee}) }
		\end{equation}
		and redefining $[\tau_{\a}]^{\frac{\a^\vee\cdot\a^\vee}{2}}\mapsto\tau_{\a}$ for every root $\a$ we finally get the equation
		\begin{equation}\label{hirotatausystem} D^2(\tau_{\a})=-{\a^\vee\cdot\a^\vee\over 2}\,t^{1/h^\vee} \prod\limits_{\b\in\hat\Delta,\b\neq\a}\left[ \tau_{\b} \right]^{-\a^\vee\cdot \b},
		\end{equation}
		The above redefinition leaves unchanged the $\tau$-functions corresponding to miniscule coweights.
		By the ambiguity in the constants of integration, both \eqref{tausystem} and \eqref{hirotatausystem2} may be modified by a constant or a power of $t$. Further, we will be writing $D^4:=D^2\circ D^2$, $D^{2n}:=D^2\circ D^{2n-2}$.	
		Finally, the $\tau$ function associated to the constant solution $\boldsymbol \varphi_0=\vb a$ is immediate from \eqref{taudef}, 
		\begin{equation}\nonumber
		\tau_{\boldsymbol\alpha}(\boldsymbol \varphi_0,t) = \exp{(h^\vee)^2 t^{\frac{1}{h^\vee}}e^{\a\cdot\A}}  
		\end{equation}
		
		Eq. \eqref{hirotatausystem} is the de-autonomization of the $\tau$-form of the standard Toda integrable system.
		From \cite{Gorsky:1995zq, Martinec:1995by} it is known that this  governs the classical Seiberg-Witten (SW) theory \cite{Seiberg:1994rs}.
		The de-autonomization is induced by coupling the theory to a self-dual $\Omega$-background $(\epsilon_1,\epsilon_2)=(\epsilon,-\epsilon)$ \cite{Bonelli:2016qwg}.
		In the autonomous limit $\epsilon\to 0$, the relevant $\tau$-functions boil down to Riemann $\theta$-functions on the classical SW curve \cite{Bonelli:2019boe}. These were used to provide recursion relations on the coefficients of the expansion of the SW prepotential in \cite{Edelstein:1998sp}. 
		
		The actual form of equations \eqref{hirotatausystem} depends on the Dynkin diagram. In particular, these reduce to bilinear 
		equations for the classical groups $A$, $B$ and $D$,   which we solve via general recursion relations. Instead,
		for $C$, $E$, $F$ and $G$ groups the equations of the $\tau$-system are of higher order and must be studied by a case by case analysis.
		The $\tau$-system displays a finite symmetry generated by the center of the group $G$, namely
		\begin{center}
			\begin{tabular}{c|c|c|c|c|c|c|c|c} 
				$\mathfrak g$ & $A_n$ & $B_n$ & $C_n$ & $D_{2n}$ & $D_{2n+1}$ & $E_{n}$ & $F_4$ & $G_2$ \\ 
				\hline 
				$Z(G)$ & $\Z_{n+1}$ & $\Z_2$ & $\Z_2$ & $\Z_2\times \Z_2$ & $\Z_4$ & $\Z_{9-n}$ & $1$ & $1$ \\  
			\end{tabular} 
		\end{center}
		
		The center is isomorphic to the coset of the affine coweight lattice by the
		affine coroot lattice, and coincides with the automorphism group of the affine Dynkin diagram. As in \footnote{ Bourbaki [Lie gps Ch. VIII \S 7]}, the coweights, and by extension the lattice cosets,  
		corresponding to these nodes are the minuscule coweights. We recall that a representation of $\mathfrak g$ is minuscule if all its weights form a single Weyl-orbit.
		This remark will be crucial to solve the $\tau$-system.
		
		The $\tau$-functions corresponding to the affine nodes,  namely the ones
		which can be removed from the Dynkin diagram while leaving behind that of an irreducible simple Lie algebra, 
		play a special r\^ole. In the gauge theory interpretation of the Introduction, these are related to simple surface operators associated to elements of the center $Z(G)$, and are bounded by 
		fractional 't Hooft lines. As such, they are the generators of the one-form symmetry of the 
		corresponding gauge theory, \cite{Gaiotto:2014kfa}. Since their magnetic charge is defined modulo the magnetic root lattice, a natural Ansatz for their expectation value is 
		\begin{equation}
		\label{Kiev}
		\tau_{\a_{\text{aff}}}\left(\s,\e|\kappa_{\mathfrak g}t\right) = \sum\limits_{\n\in Q_{\text{aff}}^\vee}
		e^{2\pi\I\e\cdot\n}
		t^{\12(\s+\n)^2}B(\s+\n|t)
		\end{equation}
		where $B(\s|t)=
		B_0(\s) \sum_{i\ge 0}t^i Z_i(\s)$ 
		with $Z_0(\s)\equiv 1$ 
		and  
		$Q_{\text{aff}}^\vee=\l_{\text{aff}}^\vee + Q^\vee$, $Q^\vee$ being the coroot lattice equipped with the canonical inner product normalized such that the norm of the short coroots is 2, and
		$(\l_{\text{aff}}^\vee,\a)=\delta_{\a{\text{aff}},\a}$ for any non-extended simple root $\a$.
		The constant 
		$
		\kappa_{\mathfrak g}= (-n_{\mathfrak g})^{r_{\mathfrak g,s}}
		$,
		where $n_{\mathfrak g}$ is the ratio of the squares of  long vs. short roots and $r_{\mathfrak g,s}$ is the number of short simple roots. For simply laced, all roots are long and $\kappa_{\mathfrak g}=1$.
		
		In the $A_n$ case, \eqref{Kiev} is known as the Kiev Ansatz. In particular, in the $A_1$ case, it was used to give the general solution of Painlev\'e III$_3$ equation in \cite{Its:2014lga} and further analysed in \cite{Mironov:2017lgl}. It was crucial for these results to identify the expansion coefficients of \eqref{Kiev} with the full Nekrasov partition function in the self-dual $\Omega$-background. 
		We will now show that this still holds for general classical groups. More precisely, this follows upon 
		the identification $\s={\bf a}/\epsilon$, where ${\bf a}$ is the Cartan parameter of the gauge theory. Let us remark that
	the variables $\e,\s\in Q^\vee$ are the integration constants of the second order differential equations \eqref{hirotatausystem} and correspond to the initial position and velocity of the 
		de-autonomized Toda particle.
		
		Let us set now the boundary conditions which we impose to the solutions
		of equations \eqref{hirotatausystem}.
		We consider the asymptotic behaviour of the solutions at $t\to0$ and $\s\to \infty$ as 
		\be \label{asy}
		\log (B_0)\sim  
		-\frac{1}{4} 
		\sum_{\vb r \in R}  (\vb r \cdot \s)^2  \log\left(\vb r \cdot\s\right)^2 
		\ee
		up to quadratic and $\log$-terms. 
		
		Notice that the $\tau$-system knows itself the one-loop exactness of the ${\mathcal N}=2$ gauge theory! Indeed, if one chooses a more general ansatz for the Wilsonian effective action as
			\unexpanded{$\log (B_0)\sim\sum_{\vb r \in R} c_{n,m}(\vb r \cdot \s)^{2n}\log\left((\vb r \cdot\s)^2\right)^m $}, then the consistency of the equation itself implies that $(n,m)=(1,1)$ and $(n,m)=(2,0)$ are the only allowed terms.
			
		We will show that the solution of \eqref{hirotatausystem} which satisfies the above asymptotic condition is
		\be\label{1loop}
		B_{0}(\s)=\CZ_{1-loop}(\s)\equiv \prod_{\vb r \in R}\frac{1}{G(1+\vb r \cdot\s)}
		\ee
		where $G(z)$ is the Barnes' G-function and $R$ is the adjoint representation of the group $G$. 
		The expansion of the above function matches the one-loop
		gauge theory result upon the appropriate 
		identification of the log-branch. This reads, in the gauge theory variables, as
		${\rm ln}\left[\sqrt{-1} \vb r \cdot\vb a /\Lambda\right]\in {\mathbb R}$ and in the $A_n$ case matches the canonical Stokes rays 
		obtained in \cite{Guest:2012yg}.
		Eq.\eqref{1loop} corresponds to the 1-loop term in the self-dual $\Omega$-background. To see this more clearly, recall the perturbative Coleman-Weinberg 1-loop term for a massless hypermultiplet in $4$ dimensions with IR regulator $\mu$,
		\begin{equation}\label{Coleman}
		\mathcal F_{\text{1 loop}}(\sigma)=\frac{3}{4}\tr\sigma^2-\frac{1}{4}\tr\sigma^2\log{\left(\frac{\sigma}{\mu}\right)^2}=\int_\mu^\infty \frac{\mathrm d s}{s^3}\tr e^{-s\sigma} + \mathcal O(\frac{1}{\mu^2})
		\end{equation}
		where the trace is taken in the relevant representation. In the self-dual $\Omega$-background with parameter $\epsilon$, this gets deformed to 
		\begin{equation}\nonumber
		\int_\mu^\infty \frac{\mathrm d s}{s}\frac{-\epsilon^2\cdot\tr e^{-s\sigma}}{(1-e^{s\epsilon})(1-e^{-s\epsilon})}
		\end{equation}
		which we can write in terms of the Barnes' G function by using its L\'evy-Khintchine type representation valid for $|z|<1$
		\begin{equation}\nonumber
		\frac{1}{G(1+z)}=\exp\left\{-\frac{\log{2\pi}-1}{2}z+\frac{1+\gamma}{2}z^2 - \int_{0}^\infty\frac{\mathrm d s}{s}\frac{e^{-zs}-1+zs-\frac{1}{2}s^2z^2}{(1-e^s)(1-e^{-s})}\right\}
		\end{equation}
		where the substractions in the integrand serve as an infrared regulator. For a general gauge group in the Coulomb phase, tracing over the Cartan yields
		\begin{equation}\nonumber
		e^{\mathcal F_{\text{1-loop}}}=\exp{\int_\mu^\infty \frac{\mathrm d s}{s}\frac{-\epsilon^2\cdot\tr_R e^{-s\s}}{(1-e^{s\epsilon})(1-e^{-s\epsilon})}}\rightsquigarrow \prod\limits_{\a\in R} \frac{1}{G(1+\s\cdot\a)}=:B_0(\s)
		\end{equation}
		The most important property of this expression is that,
		given some $\b\in R^\vee$, 
		\begin{equation}\label{1loopProp}
		B_0(\s+\b)=B_0(\s)\prod\limits_{\substack{\a\in R, n\geq 1\\ \a\cdot\b=n}}\frac{(-1)^{\left \lfloor{n/2}\right \rfloor}\Gamma(-\a\cdot\s)^n}{\Gamma(\a\cdot\s)^n(\a\cdot\s)^n\prod\limits_{k=1}^{n-1}(\a\cdot\s+k)^{2n-2k}}
		\end{equation}
		where we can pick only positive $n$'s since 
		the product runs over the whole root system.
		
		\section{Lie algebras}
		\subsection{\texorpdfstring{$A_n$}{An}}
		
		\color{black}
		
		\begin{center}
			\begin{dynkinDiagram}[extended,edge length=1.5cm, indefinite edge/.style={ultra thick,densely dashed}, edge/.style={ultra thick},o/.style={ultra thick,fill=white,draw=black}, root radius=.1cm,root/.style={ultra thick,fill=white,draw=black}]A{o.ooo.o}
				\node[below=.2cm] at (root 0) {$\tau_0$};
				\node[below=.2cm] at (root 1) {$\tau_1$};
				\node[below=.2cm] at (root 2) {$\tau_{j-1}$};
				\node[below=.2cm] at (root 3) {$\tau_{j}$};
				\node[below=.2cm] at (root 4) {$\tau_{j+1}$};
				\node[below=.2cm] at (root 5) {$\tau_{n}$};
			\end{dynkinDiagram}
		\end{center}
		The $A_n$ case is the simplest but already 
		illustrates most of the ideas of our analysis. The simplification in this case comes from the fact that every node of the extended Dynkin diagram corresponds to a miniscule (co)weight and that the resulting equations are strictly bilinear, none of which are true in general for different algebras. 
		
		We realise the roots using an orthonormal basis $\{e_i\}$ of $\mathbb{R}^{n+1}$ as $\{\pm(e_i-e_j)\}$ for $i\neq j$. The algebra is simply laced so the coroot lattice is the root lattice and is $Q^\vee=Q=\{\sum\limits_{i=1}^{n+1}c_i e_i |\sum\limits_{i=1}^{n+1}c_i=0\}$, while the fundamental weights
		\begin{equation}\nonumber
		\l_i = (1^{i},0^{n+1-i})-\frac{i}{n+1}(1^{n+1}) \, ,
		\end{equation}
		are all minuscule. Here $(1^p,0^{n+1-p})$ stands for a vector whose first $p$ entries are $1$ and the remaining entries vanish, while in $(1^{n+1})$ all entries are 1. Moreover
		we label the $\tau$-functions as $\tau_{\a_j}\equiv\tau_j$ and identify $\tau_j=\tau_{n+1+j}$ periodically. Then the 
		$\tau$-system can be written succinctly as 
		\begin{equation}\label{an}
		D^2(\tau_j)=-t^{\frac{1}{n+1}}\tau_{j-1}\tau_{j+1}\, .
		\end{equation}
		 Due to the $\mathbb{Z}_{n+1}$ outer automorphism group of the Dynkin diagram, each of the nodes of $A_n$ can be taken as the affine one so that the corresponding $\tau$-functions can be expressed through the Kiev Ansatz \eqref{Kiev}. Therefore, all the $\tau$-functions are determined by a single one, say $\tau_0$, as $\tau_j=\tau_0\vert_{Q\mapsto Q_j}$. 
		Owing to the $\mathbb{Z}_{n+1}$ symmetry, it is enough to solve \eqref{an} corresponding to $j=0$.
		Henceforth we adopt the shorthand $f(y\pm x)\equiv f(y+x)f(y-x)$. The Ansatz \eqref{Kiev} for $\tau_0$ reads
		\begin{equation}\label{kiev}
		\tau_0(\s,\e|t) = \sum\limits_{\n\in Q,\,i\geq 0}e^{2\pi\sqrt{-1}\n\cdot\e}
		t^{\12(\s+\n)^2+i}B_0(\s+\n)Z_i(\s+\n)
		\end{equation}
        Inserting the Kiev Ansatz \eqref{kiev} into \eqref{an} gives us
		\begin{align*}
		&\sum_{\substack{\n_1,\n_2\in Q\\i_1,i_2\geq 0}}e^{2\pi\sqrt{-1}(\n_1+\n_2)\cdot\e}
		t^{\12 \n_1^2+\12 \n_2^2+i_1+i_2+\s\cdot(\n_1+\n_2)}\notag \left(\12 \n_1^2-\12 \n_2^2+i_1-i_2+\s\cdot(\n_1-\n_2)\right)^2 \notag \\
		&\times B_0(\s+\n_1)B_0(\s+\n_2)Z_{i_1}(\s+\n_1)Z_{i_2}(\s+\n_2) \notag \\
		&=-\sum_{\substack{\m_1,\m_2\in Q\\j_1,j_2\geq 0}}e^{2\pi\sqrt{-1}(\m_1+\m_2)\cdot\e}t^{1+\12 \m_1^2+\12\m_2^2+e_1\cdot(\m_1-\m_2)+j_1+j_2+\s\cdot(\m_1+\m_2)}\times \notag \\
		&
		B_0(\s+\m_1+e_1)B_0(\s+\m_2-e_1)Z_{j_1}(\s+\m_1+e_1)Z_{j_2}(\s+\m_2-e_1)\label{strafu}
		\end{align*}
		This is solved as a power series in $t,t^{\sigma_1},...,t^{\sigma_n}$. To fix $B_0(\s)$, we look at the lowest order. The
		lowest order is linear in $t$ and produces the quadratic constraint
		\begin{equation}\label{quadratic}
		\12\n_1^2+\12\n_2^2+i_1+i_2=1+\12\m_1^2+\12\m_2^2+e_1\cdot(\m_1-\m_2)+j_1+j_2=1
		\end{equation}
		as well as $n+1$ linear constraints on the root lattice variables $(\n_1,\n_2)$ and $(\m_1,\m_2)$.
		Let us fix $p,q\in\{0,...n+1\}$, $p\neq q$ and look for terms with $t^{\sigma_p-\sigma_q}$. The linear constrains are $\n_1+\n_2=\m_1+\m_2=e_p-e_q$. Up to Weyl reflections, the only solution to the above mentioned constraints is given by
		$\n_1=e_p-e_q$, $\n_2=0$ and $\m_1=e_p-e_1$, $\m_2=-e_q+e_1$, with $i$s and $j$s in \eqref{quadratic} vanishing, leading to the functional equation
		\begin{equation}\label{An1loop}
		\left(1+(e_p-e_q)\cdot\s\right)^2 B_0(\s+e_p-e_q)B_0(\s) =-B_0(\s+e_p)B_0(\s-e_q)  \, .
		\end{equation}
		Now we suppose that $B_0(\s)=f(\s)
		\prod_{\vb r \in R}\frac{1}{G(1+\vb r \cdot\s)}$. 
		First of all we show that of ratios of $\Gamma$-functions which arise from manipulating the Barnes' G-functions cancels. Namely, consider, for $\b\in Q^\vee+\l^\vee$ in a general Lie algebra
		\begin{equation}\nonumber
		\hat\Gamma(\b):=\prod\limits_{\substack{\a\in R, n\geq 1\\ \a\cdot\b=n}}\left({\Gamma[-\a\cdot\s]\over \Gamma[\a\cdot\s]}\right)^n
		\end{equation}
		which is the product of $\Gamma$-functions in \eqref{1loopProp}. Noting that
		\begin{align*}
		\hat\Gamma(\b)&=\prod\limits_{\substack{\a\in R\\ \a\cdot\b\geq 0}}\left({\Gamma[-\a\cdot\s]\over \Gamma[\a\cdot\s]}\right)^{\a\cdot\b}\\
		&= \prod\limits_{\substack{\a\in R\\ \a\cdot\b\geq 0}}\left({1\over \Gamma[\a\cdot\s]}\right)^{\a\cdot\b}\left({1\over \Gamma[-\a\cdot\s]}\right)^{-\a\cdot\b}= \prod\limits_{\a\in R}\left({1\over \Gamma[\a\cdot\s]}\right)^{\a\cdot\b}
		\end{align*}
		we get
		\begin{equation}\nonumber
		\hat\Gamma(\b_1)\hat\Gamma(\b_2)=\hat\Gamma(\b_1+\b_2)\,.
		\end{equation}
		In particular if $\sum_k \b_k = \sum_k  \boldsymbol{\gamma}_k$, which corresponds to the linear constrains,
		\begin{equation}\nonumber
		\prod_k\hat\Gamma(\b_k)=\hat\Gamma(\sum_k \b_k )=\hat\Gamma(\sum_k \boldsymbol{\gamma}_k)=\prod_k\hat\Gamma(\boldsymbol{\gamma}_k)
		\end{equation}
		Therefore, these products of $\Gamma$-functions cancels from all formulas, as we obtain all of them by matching equal powers of $t^{\sigma_1},...,t^{\sigma_n}$. This discussion is valid for all Lie algebras. For the $A_n$ case, the LHS of \eqref{An1loop}, after discarding products of $\Gamma$-functions, becomes
		\begin{equation}\nonumber
		 \frac{\left(1+\sigma_p-\sigma_q\right)^2f(\s+e_p-e_q)f(\s)}{-(\sigma_p-\sigma_q)^2(1+\sigma_p-\sigma_q)^2\prod\limits_{\b\cdot(e_p-e_q)=1}\b\cdot\s} = -\frac{f(\s+e_p-e_q)f(\s)}{(\sigma_p-\sigma_q)^2\prod\limits_{p\neq k \neq q}(\sigma_p^2-\sigma_k^2)(\sigma_k^2-\sigma_q^2)}\,.
		\end{equation}
		This has to equal the RHS 
		\begin{equation}\nonumber
		-\frac{f(\s+e_p)}{\prod\limits_{k \neq p}(\sigma_p^2-\sigma_k^2)}\frac{f(\s-e_q)}{\prod\limits_{k \neq q}(\sigma_k^2-\sigma_q^2)}\,.
		\end{equation}
        Simple arithmetics converts this to $f(\s+e_p-e_q)f(\s)=f(\s+e_p)f(\s-e_q)$ which 
        implies that $f$ is periodic on the lattice. 
        The asymptotic condition \eqref{asy} reads as $f\sim 1$ when $\s\to\infty$, so that $f=1$.
		
		The higher order terms in $t,t^{\sigma_1},...,t^{\sigma_n}$ provide the recursion relations
		\begin{gather*}
		k^2 Z_k(\s)=-\sum\limits_{\substack{\n^2+j_1+j_2=k\notag\\ \n\in e_1+ Q,\,j_{1,2}<k}}\frac{B_0(\s\pm\n)}{B_0(\s)^2}Z_{j_2}(\s-\n)Z_{j_1}(\s+\n)
		\\+\sum\limits_{\substack{\n^2+i_1+i_2=k\notag\\ \n\in Q,\, i_{1,2}<k}}\left(i_1-i_2+2\n\cdot\s\right)^2 \frac{B_0(\s\pm\n)}{B_0(\s)^2}Z_{i_1}(\s+\n)Z_{i_2}(\s-\n) \, ,
		\end{gather*}
		where $B_0(\s)$ is given by \eqref{1loop}.
		In particular, $k=1$ gives the simple expression
		\ytableausetup{smalltableaux}
		\begin{equation}\nonumber
		Z_1(\s) = -\sum\limits_{i=1}^{n+1}\frac{B_0(\s\pm e_i)}{B_0(\s)^2}=(-1)^{n+1}\sum\limits_{i=1}^{n+1}\frac{1}{\prod_{j\neq i}(\sigma_i-\sigma_j)^2}\,.
		\end{equation}
		Upon abbreviating $\sigma_{ij}=\sigma_i-\sigma_j$, the $k=2$ term gives 
		\begin{equation}\nonumber
		Z_2(\s)= -\frac{1}{4}\sum\limits_{i=1}^{n+1}\frac{B_0(\s\pm e_i)}{B_0(\s)^2}
		[Z_1(\s+ e_i)+Z_1(\s- e_i)]+
		\sum\limits_{i<j}^{n+1}(\sigma_i-\sigma_j)^2\frac{B_0(\s\pm(e_i-e_j))}{B_0(\s)^2}
		\end{equation}
		which we can write as 
		\begin{gather*}
		Z_2(\boldsymbol{\sigma}) =
		\frac{1}{4}\sum_i \frac{1}{\prod_{j\neq i}\sigma_{ij}^2}\left(\sum_k {1 \over \prod_{l\neq k}(\sigma_{kl}+\delta_{kl}-\delta_{li})^2}
		+{1 \over \prod_{l\neq k}(\sigma_{kl}-\delta_{kl}+\delta_{li})^2 }\right)
		\\
		-\sum_{i<j}\frac{1}{(\sigma_{ij}+1)^2(\sigma_{ij}-1)^2\sigma_{ij}^2\prod_{i \neq k\neq j}\sigma_{ik}^2\sigma_{jk}^2}
		\\
		= \frac{1}{4}\sum_i \frac{1}{\prod_{k\neq i}\sigma_{ki}^2 \cdot \prod_{k\neq i}(\sigma_{ki}-1)^2}
		+\frac{1}{4}\sum_i \frac{1}{\prod_{k\neq i}\sigma_{ki}^2 \cdot \prod_{k\neq i}(\sigma_{ki}+1)^2}
		\\
		+\sum_{i<j}\frac{\sigma_{ij}^2}{(\sigma_{ij}^2-1)^2}\frac{1}{\prod_{k\neq i}\sigma_{ki}^2\cdot \prod_{k\neq j}\sigma_{kj}^2}
		\end{gather*}
		
	where in the second step we cancelled the off-diagonal terms in the double product, to simplify the comparison with  Nekrasov formulae for $k=2$ for $\epsilon_1=-\epsilon_2=1$. Indeed, the three sums above correspond exactly to $Z^{SU(n+1)}(\vec Y)$ of \eqref{AnCountDiagram} with $\vec Y$ having two boxes $\raisebox{1mm}{\tiny{\ydiagram{1,1}}}\normalsize$ or two boxes ${\tiny{\ydiagram{2}}}\normalsize$ in the $i$-th position and the last double sum is over $\vec Y$ such that one box is in the $i$-th and another in the $j$-th position. These are all the possible tuples $\vec Y$ such that $|\vec Y|=2$. 
	\ytableausetup{nosmalltableaux}

		\noindent To summarise, the above coincide with one and two instanton contributions to the $SU(n+1)$ Nekrasov partition function as computed from supersymmetric localization
		\cite{Nekrasov:2003af, Nekrasov:2003rj}. 
		Let us remark that the use of the $\tau$-system \eqref{an} provides a completely independent tool to compute all instanton corrections just starting from the asymptotic behaviour \eqref{asy}. This procedure
		extends to all classical groups.

		\subsection{\texorpdfstring{$B_n$}{Bn}, \texorpdfstring{$D_n$}{Dn}}
		Due to our strategy of solving the problem by attaching Kiev Ans\"atze to nodes corresponding to minuscule coweights, we treat the algebras $B_n$ and $D_n$ simultaneously. The coroot lattices are likewise the same, so the only difference between $B_n$ to $D_n$ is the asymptotic condition the extra roots of $B_n$ impose.
		
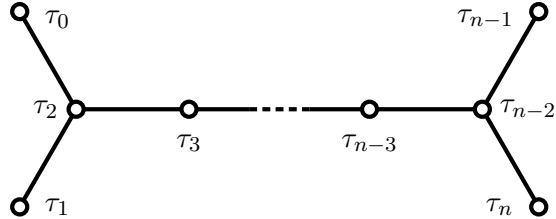
\begin{center}
			\begin{dynkinDiagram}[extended,reverse arrows, edge length=1.5cm, indefinite edge/.style={ultra thick,densely dashed}, edge/.style={ultra thick},o/.style={ultra thick,fill=white,draw=black}, root radius=.1cm,root/.style={ultra thick,fill=white,draw=black}, arrow width = 0.4cm, arrow style={length=5mm, width=5mm,line width = 1pt}]D{ooo.oooo}
				\node[below=.1cm,right=.2cm] at (root 0) {$\tau_0$};
				\node[right=.2cm] at (root 1) {$\tau_1$};
				\node[left=.1cm] at (root 2) {$\tau_2$};
				\node[below=.2cm] at (root 3) {$\tau_{3}$};
				\node[below=.2cm] at (root 4) {$\tau_{n-3}$};
				\node[right=.1cm] at (root 5) {$\tau_{n-2}$};
				\node[below=.1cm,left=.2cm] at (root 6) {$\tau_{n-1}$};
				\node[left=.2cm] at (root 7) {$\tau_{n}$};
			\end{dynkinDiagram}
		\end{center}
		$D_n$ is a simply laced algebra, whose coroot lattice is the checkerboard lattice $Q=Q^\vee =\{\sum_{i=1}^n c_i e_i | \sum_{i=1}^n c_i\in 2\Z \}$. In this section we consider only $n\geq 4$ and leave the special cases of $n=2,3$ to the appendix \ref{appendix:D23}. There are four minuscule weights, $\l_0=(0^n)$, $\l_1=(1,0^{n-1})$, $\l_{n-1}=((\12)^{n-1},-\12)$, $\l_{n}=((\12)^{n-1},+\12)$ and these correspond to the "legs" of the affine diagram. Whatever the rank we consider, we always have the consistency conditions
		\begin{gather}\label{Dtau}
		D^2(\tau_0)=D^2(\tau_1),\quad  D^2(\tau_{n-1})=D^2(\tau_n)
		\end{gather}
		which immediately follow from the equations $D^2(\tau_0)=-t^{1/2n}\tau_2$, $D^2(\tau_1)=-t^{1/2n}\tau_2$ and the analogue ones at the other end of the diagram. The second consistency condition is morally just the first one with $\sigma$ shifted by $((\12)^{n})$. In the special case $n=4$ we have a further equality, due to the enhanced symmetry of $D_4$,
		\begin{equation}\nonumber
		D^2(\tau_0)=D^2(\tau_1)= D^2(\tau_3)=D^2(\tau_4).
		\end{equation}
		Practically, however, the first condition is sufficient to solve the problem. 
		
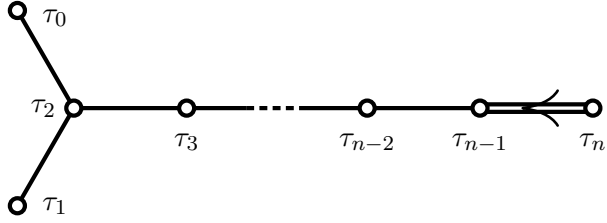
\begin{center}
			\begin{dynkinDiagram}[extended,reverse arrows, edge length=1.5cm, indefinite edge/.style={ultra thick,densely dashed}, edge/.style={ultra thick},o/.style={ultra thick,fill=white,draw=black}, root radius=.1cm,root/.style={ultra thick,fill=white,draw=black}, arrow width = 0.4cm, arrow style={length=5mm, width=5mm,line width = 1pt}]B{ooo.ooo}
				\node[below=.1cm,right=.2cm] at (root 0) {$\tau_0$};
				\node[right=.2cm] at (root 1) {$\tau_1$};
				\node[left=.1cm] at (root 2) {$\tau_2$};
				\node[below=.2cm] at (root 3) {$\tau_{3}$};
				\node[below=.2cm] at (root 4) {$\tau_{n-2}$};
				\node[below=.2cm] at (root 5) {$\tau_{n-1}$};
				\node[below=.2cm] at (root 6) {$\tau_{n}$};
			\end{dynkinDiagram}
		\end{center}
		
		$B_n$ is not simply laced. In addition to the roots of the corresponding $D_n$, $\{e_i\pm e_j\}_{i\neq j}$, it has shorter roots $\{e_i\}$. This is the first case, however, in which we have to worry about looking at the Langlands dual algebra, and send each root to the coroot via $R\ni \a \mapsto 2\a/(\a\cdot\a)\in R^\vee$. Therefore, the extended Dynkin diagram above has reversed arrows compared to the usual, since the roots $\{e_i\}$ get rescaled to $\{2 e_i\}$.
		The coroot lattice is still the checkerboard lattice $Q^\vee=\{\sum_{i=1}^n c_i e_i | \sum_{i=1}^n c_i\in 2\Z \}$ of $D_n$, and the two minuscule weights are $\l_0^\vee=(0^n)$ and $\l_1^\vee=(1,0^{n-1})$, corresponding to the "antennae" of the new diagram, provided $n>3$. The additional $\mathbb{Z}_2$ symmetry of $D_n$ is broken. The $\tau$-system coincides with that of $D_{n+1}$, with the modification that (i) there is no $\tau_{n+1}$ node and (ii) that
		\begin{equation}\nonumber
		D^2(\tau_{n-1})=-2t^{\frac{1}{2n-1}}\tau_{n-2}\tau_{n},\quad
		D^2(\tau_n)=-t^{\frac{1}{2n-1}}\tau_{n-1}^2.
		\end{equation}
		The case $n=3$ is discussed separately along with the algebra $C_2$ in section \ref{subsubsection:C2}.
        We limit the present discussion to $n>3$ so the analysis proceeds as for $D_n$, except we can only consider the first equation in \eqref{Dtau}.
		This unifies the approach to both $D_n$ and $B_n$. Explicitly, inserting \eqref{Kiev} and $\tau_1(\s|t)=\tau_0(\s+\l_1|t)$ into the first equation of \eqref{Dtau} we get 
		\begin{align*}
			\sum_{\substack{\n_1,\n_2\in Q^\vee\\i_1,i_2\geq 0}}&e^{2\pi\sqrt{-1}(\n_1+\n_2)\cdot\e}t^{\12 \n_1^2+\12 \n_2^2+i_1+i_2+\s\cdot(\n_1+\n_2)}
			\\
			&\left(\12 \n_1^2-\12 \n_2^2+i_1-i_2+\s\cdot(\n_1-\n_2)\right)^2
			\\
			&B_0(\s+\n_1)B_0(\s+\n_2)Z_{i_1}(\s+\n_1)Z_{i_2}(\s+\n_2)
			\\
			=\sum_{\substack{\m_1,\m_2\in Q^\vee\\j_1,j_2\geq 0}}&e^{2\pi\sqrt{-1}(\m_1+\m_2)\cdot\e}t^{1+\12 \m_1^2+\12 \m_2^2+\l_1\cdot(\m_1+\m_2)+j_1+j_2+\s\cdot(\m_1+\m_2+2\l_1)}
			\\
			&\left(\12 \m_1^2-\12 \m_2^2+j_1-j_2+(\s+\l_1)\cdot(\m_1-\m_2)\right)^2
			\\
			&B_0(\s+\m_1+\l_1)B_0(\s+\m_2+\l_1)Z_{j_1}(\s+\m_1+\l_1)Z_{j_2}(\s+\m_2+\l_1)
			\end{align*} 
		In the following, $p,q = {1,...,n}$, $p\neq q$, and following the discussion in the previous section, we look for the lowest terms in powers of $t$ and $\{t^{\sigma_i}\}$. Explicitly, the term to consider is $t^{1+\s\cdot(e_p+e_q)}$, which we got by putting $\n_1=e_p+e_q$ and $\n_2=0$ on the LHS, up to symmetry. To get this term we need to impose $\m_1=e_p-e_1$, $\m_2=e_q-e_1$ on the RHS, with all $i$'s and $j$'s vanishing. The functional equation we get, analogous to \eqref{An1loop}, is 
		\begin{equation}
		\label{1loopfix}
		(1+(e_p+e_q)\cdot\s)^2B_0(\s)B_0(\s+e_p+e_q)=\left((e_p-e_q)\cdot\s\right)^2B_0(\s+e_p)B_0(\s+e_q) \, .
		\end{equation}
		The two cases are distinguished by the different asymptotic conditions \eqref{asy} the root systems impose.  Indeed, we have
		\begin{gather*}
		 B_{0}^{[D_{n}]}(\s)=\prod\limits_{i<j}^n {1\over G(1\pm\sigma_i\pm\sigma_j)
		} \\
		 B_{0}^{[B_{n}]}(\s)
		=\left(\prod\limits_{k=1}^n{1\over G(1\pm \sigma_k)}\right)B_{0}^{[D_{n}]}(\s)
		\end{gather*}
	One can show that the large $\sigma$ asymptotics of these different solutions are consistent with the full $\tau$ system, not only the reduced consistency condition we are considering. Next, since the equation and the Ansatz are the same, the recursion relations are as well, and turn out to be
		\begin{gather}\nonumber
		k^2 Z_k(\s)=\sum\limits_{\substack{(\n-\l_1)^2+j_1+j_2=k\\ \n\in \l_1+ Q^\vee\, ,j_{1,2}<k}}Z_{j_1}(\s+\n)Z_{j_2}(\s-\n)\left(j_1-j_2+2\n\cdot\s\right)^2\frac{B_0(\s\pm\n)}{B_0(\s)^2}
		\\-\sum\limits_{\substack{\n^2+i_1+i_2=k\\\n\in Q^\vee,\, i_{1,2}<k}}Z_{j_1}(\s+\n)
		\times Z_{j_2}(\s-\n)
		\left(i_1-i_2+2\n\cdot\s\right)^2\frac{B_0(\s\pm\n)}{B_0(\s)^2}
		\label{Drec}
		\end{gather}
		
		\noindent This result is in line with the contour integral formulae for the relevant Nekrasov partition functions. Indeed the poles in the $D_n$ and $B_n$ cases are the same, but with different residues, as noticed in \cite{Marino:2004cn}.	
		%
		From the above recursion relation we can compute the 1-instanton terms 
		\bea \nonumber
		Z_1(\s)=\sum_{k=1}^n 4\sigma_k^2\frac{B_0(\s\pm e_k)}{B_0(\s)^2}=
		\begin{cases*}
			\sum_{k=1}^n \frac{-4}{\prod\limits_{j\neq k}(\sigma_k^2-\sigma_j^2)^2}, & $B_n$\\
			\sum_{k=1}^n \frac{4\sigma_k^2}{\prod\limits_{j\neq k}(\sigma_k^2-\sigma_j^2)^2}, & $D_n$
		\end{cases*}
		\notag
		\eea
		and the 2-instantons
		\bea
		Z_2(\s) = \sum\limits_{\a\in Q^\vee, \a^2=2}\frac{-1}{(\a\cdot\s)^2((\a\cdot\s)^2-1)^2\prod\limits_{\b\cdot\a=1}(\b\cdot\s)^2}\notag\\
		+\sum\limits_{k=1}^n \frac{Z_1(\s+e_k)(\sigma_k+\12)^2+Z_1(\s-e_k)(\sigma_k-\12)^2}{\prod\limits_{\b\cdot e_k=\pm 1}(\b\cdot\s)}\notag
		\eea
		and so on. These are easily compared to the instanton counting from appendix \ref{appendix:instanton_counting}, and the appendix of \cite{Marino:2004cn} where the results were first presented.

	    \subsection{\texorpdfstring{$C_n$}{Cn}}
		
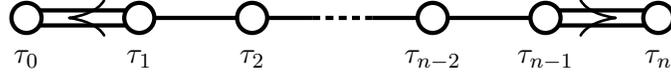
		\begin{center}
			\begin{dynkinDiagram}[extended,reverse arrows, edge length=1.5cm, indefinite edge/.style={ultra thick,densely dashed}, edge/.style={ultra thick},o/.style={ultra thick,fill=white,draw=black}, root radius=.2cm,root/.style={ultra thick,fill=white,draw=black}, arrow width = 0.4cm, arrow style={length=5mm, width=5mm,line width = 1pt}]C{oo.ooo}
				\node[below=.3cm] at (root 0) {$\tau_0$};
				\node[below=.3cm] at (root 1) {$\tau_1$};
				\node[below=.3cm] at (root 2) {$\tau_2$};
				\node[below=.3cm] at (root 3) {$\tau_{n-2}$};
				\node[below=.3cm] at (root 4) {$\tau_{n-1}$};
				\node[below=.3cm] at (root 5) {$\tau_{n}$};
			\end{dynkinDiagram}
		\end{center}
		Here there is a potential issue of normalizing the roots, so we must make note of our conventions. In writing \eqref{Kiev} we have stressed that the bilinear form is fixed by demanding $|\a|^2 = 2$ for all long roots $\a$. If we decide to choose roots of $C_n$ as $\{D_n\text{ roots}\}\cup\{\pm 2 e_i\}$, clearly $|2e_i|^2=4$. So we should normalize them as $\{\pm \frac{1}{\sqrt 2}e_i\pm \frac{1}{\sqrt 2}e_j\}\cup\{\pm \sqrt{2} e_i\}$.
		The dual lattice is then $Q^\vee=\sqrt{2}\Z^n$. In literature, the factors of $\sqrt 2$ are sometimes avoided, which can be accommodated in this approach by rescaling time and working with  
		\begin{equation}
		\tau_{i} = \sum\limits_{\m\in \Z^n+\lambda_i^\vee} e^{\m\cdot \e}t^{\frac{1}{2}\sum_{i=1}^n (\sigma_i+m_i)^2}B(\s+\m|\sqrt{t})
		\label{taui}
		\end{equation}
	    The minuscule weights are $\l_0=\vb{0}$ and $\l_n=((\frac{1}{\sqrt{2}})^n)$. Bilinear relations are only available for $n=1,2$, where accidental isomorphisms map the algebras to those already considered. We explore the lower ranks explicitly up to and including $C_4$.
	    
	    As for the analysis of the higher order algebras, these
		produce more complicated recurrence relations to be solved by a case by case analysis, unlike in the $A,B,D$ types which allow for a unified treatment.
		We performed explicit checks for $C_{5}$ and $C_{6}$ up to one-instanton, 
		again in agreement with \cite{Marino:2004cn}. 
		
		\subsubsection{\texorpdfstring{$C_1$}{C1}}
		
		This is the simplest case, in fact isomorphic to $A_1$. The coroot lattice is $Q^\vee=\sqrt{2}\Z$, $\lambda_1 = 1/\sqrt{2}$, and the equations are formally the same as $A_1$,
		\begin{gather*}
		D^2(\tau_0)=-t^{\frac{1}{2}}\tau_1^2,\quad
		D^2(\tau_1)=-t^{\frac{1}{2}}\tau_0^2
		\end{gather*}
			
		\subsubsection{\texorpdfstring{$C_2$}{C2}}
		\label{subsubsection:C2}
		For the subsequent rank, the lattice is $Q^\vee=\sqrt{2}\Z^2$, $\l=[(\frac{1}{\sqrt{2}})^2]$. The full system
		\begin{gather*}
		D^2(\tau_0)=-t^{\frac{1}{3}}\tau_1,\quad
		D^2(\tau_1)=-2 t^{\frac{1}{3}}\tau_0^2\tau_2^2,\quad
		D^2(\tau_2)=-t^{\frac{1}{3}}\tau_1
		\end{gather*}
		leads to the single equation
		\begin{equation}\nonumber
		D^2(\tau_0)=D^2(\tau_1)
		\end{equation}
		As with the rank 1 case, there is an accidental isomorphism at this level, namely, $C_2\cong B_2$, i.e. $\mathfrak{sp}_2\cong \mathfrak{so}_5$, leading to the same equation. The isomorphism is realised by 
		\begin{gather*}
		2\sigma_1^{[C_2]}=(\sigma_1+\sigma_2)^{[B_2]}\quad
		2\sigma_2^{[C_2]}=(\sigma_1-\sigma_2)^{[B_2]}\,.
		\end{gather*}
		As such, we find a recurrence relation for the equivariant volumes of the instanton moduli space  which resembles the other recurrence relations we have already found, but it does not generalize to higher rank and pertains only to $C_2$. 
		
		One easily finds that
		\begin{equation}\nonumber
		B_0(\s)={1\over G(1\pm\sqrt{2}\sigma_1) G(1\pm\sqrt{2}\sigma_1) G(1\pm\frac{1}{\sqrt{2}}(\sigma_1\pm\sigma_2)) 
	}
		\end{equation}
         as well as the simple recurrence relation which we can write as
		
		\begin{align*}
		2(\frac{k}{2})^2 Z_k(\s)&=\sum\limits_{\12 \m+\l\cdot\m+j_1+j_2=\frac{k-1}{2}}\left(j_1-j_2 +2(\l+\m)\cdot\s\right)^2\\
		&Z_{2j_1}(\s+\l+\m)Z_{2j_2}(\s-\l-\m)\frac{B_0(\s\pm(\l+\m))}{B_0(\s)^2}
		\\
		&-\sum\limits_{\substack{\n^2+i_1+i_2=\frac{k}{2}\\i_{1,2}<k/2}}\left(i_1-i_2+2\n\cdot\s\right)^2Z_{2i_1}(\s+\n)Z_{2i_2}(\s-\n)\frac{B_0(\s\pm\n)}{B_0(\s)^2}
		\end{align*} 
		
		\subsubsection{\texorpdfstring{$C_3$}{C3}}
		In higher ranks one gets higher order relations among the $\tau$-functons. In particular, while for $C_n$ with $n$ even the central node is seen to be invariant, $n$ being odd presents an interesting challenge. In the following, $Q^\vee=\sqrt{2}\Z^3$, and $\l_3=((\frac{1}{\sqrt{2}})^3)$, and the $\tau$-system is
		\begin{gather}
		D^2(\tau_0)=-t^{\frac{1}{4}}\tau_1\\
		D^2(\tau_1)=-2t^{\frac{1}{4}}\tau_0^2\tau_2\\
		D^2(\tau_2)=-2t^{\frac{1}{4}}\tau_1\tau_3^2\\\label{lastC3}
		D^2(\tau_3)=-t^{\frac{1}{4}}\tau_2
		\end{gather}
		By multiplying \eqref{lastC3} by $\tau_0^2$, we obtain
		\begin{equation}\nonumber
		\tau_0^2D^2(\tau_3) =- t^{\frac{1}{4}}\tau_0^2\tau_2=\frac{1}{2}D^2(\tau_1)=\frac{1}{2}t^{-\frac{1}{2}}D^4(\tau_0)
		\end{equation}
		Dividing by $\tau_0$ and using the $Y$ operators defined in appendix \ref{appendix:Y} we rewrite this as the cubic system
		\begin{equation}\nonumber
		Y^3(\tau_0)=2 t^{1/2}\tau_0 D^2(\tau_3)
		\end{equation} 
		Inserting \eqref{taui} we obtain
		\begin{align*}
		&\sum_{\substack{ \n_{1,2,3}\in \sqrt{2}\Z^3\\ i_{1,2,3}\in{\mathbb N_0}}} \prod\limits_{k=1}^3 e^{2\pi\I \e\cdot\n_k }t^{\12(\s+\n_k)^2+i_k} B_0(\s+\n_k) Z_{i_k}(\s+\n_k) \notag
		\\
		&\frac{1}{3!}\prod\limits_{k_1<k_2}\left(\12\n_{k_1}^2+i_{k_1}-\12\n_{k_2}^2-i_{k_2}+(\n_{k_1}-\n_{k_2})\cdot\s\right)^2\notag
		\\
		=2t^{1/2}&\sum_{\substack{ \m_1\in \sqrt{2}\Z^3 \\ \m_{2,3}\in \sqrt{2}\Z^3+\l_3\\ j_{1,2,3}\in{\mathbb N_0}}}\prod\limits_{k=1}^3 e^{2\pi\I \e\cdot\n_k }t^{\12(\s+\n_k)^2+i_k} B_0(\s+\n_k) Z_{i_k}(\s+\n_k) \notag
		\\
		&\left(\12\m_{2}^2+j_{2}-\12\m_{3}^2-j_{3}+(\m_{2}-\m_{3})\cdot(\s+\l_3)\right)^2
		\end{align*}
		Then, seeing that $2\times\12\l_3^2=\frac{3}{2}$ and rewriting $\m_{1}=\m_{1}^{(0)}$ $\m_{2,3}=\m_{2,3}^{(0)}+\l_3$ where $\m_{1,2,3}^{(0)}\in\sqrt{2}\Z^3$, we reduce to the constraints
		\begin{gather} 
		\sum_{k=1}^3 \12 \n_i^2+ i_k =2+\l_3\cdot(\m_2^{(0)}+\m_3^{(0)})+\sum_{k=1}^3 \12 (\m_k^{(0)})^2 + j_k \label{C3sqr} \\ \label{C3lin}
		\sum_{k=1}^3 \n_i =2\l_3 + \sum_{k=1}^3 \m_i^{(0)}
		\end{gather}
		Let $p_{1},p_2,p_3$ be a permutation of $\{1,2,3\}$. We consider factors of $t^{\sqrt{2}\s\cdot(e_{p_1}+e_{p_2})+2}$, in other words \eqref{C3sqr}=2 and \eqref{C3lin}$=\sqrt{2}(e_{p_1}+e_{p_2})$. For the LHS we find the solutions $\n_1=\sqrt{2}(e_{p_1}+e_{p_2})$, $\n_2=\n_3=\vb{0}$ and permutations thereof, for which the LHS vanishes due to degeneracy, and $\n_1=\sqrt{2}e_{p_1}$, $\n_2=\sqrt{2}e_{p_1}$, $\n_3=\vb{0}$. For the RHS, there are two solutions $\m_1^{(0)}=\m_2^{(0)}=\vb{0}$, $\m_3^{(0)}=-\sqrt{2}e_{p_3}$, and $\m_1^{(0)}=\m_3^{(0)}=\vb{0}$, $\m_2^{(0)}=-\sqrt{2}e_{p_3}$. We are led then to the equation 
		\begin{gather*}(1+\sqrt{2}\sigma_{p_1})^2(1+\sqrt{2}\sigma_{p_2})^2(\sigma_{p_1}-\sigma_{p_2})^2 B_0(\s+\sqrt{2}e_{p_1})B_0(\s+\sqrt{2}e_{p_2}) \\
		=4\sigma_{p_3}^2 B_0(\s+\l_3)B_0(\s+\l_3-\sqrt{2}e_{p_3})
		\end{gather*}
		Using \eqref{1loopProp} we find on the LHS
		\begin{equation}\nonumber
	    (\sigma_{p_1}-\sigma_{p_2})^2 { 2 \over \sigma_{p_1}^2(\sigma_{p_1}^2-\sigma_{p_2}^2)(\sigma_{p_1}^2-\sigma_{p_3}^2)}
	    { 2 \over \sigma_{p_2}^2(\sigma_{p_1}^2-\sigma_{p_2}^2)(\sigma_{p_2}^2-\sigma_{p_3}^2)}
		\end{equation}
		and on the RHS
		\begin{equation}\nonumber
		4\sigma_{p_3}^2 { 1 \over \sigma_{p_1}\sigma_{p_2}\sigma_{p_3}(\sigma_{p_1}+\sigma_{p_2})(\sigma_{p_1}+\sigma_{p_3})(\sigma_{p_2}+\sigma_{p_3})}
		{ 1 \over \sigma_{p_1}\sigma_{p_2}(-\sigma_{p_3})(\sigma_{p_1}+\sigma_{p_2})(\sigma_{p_1}-\sigma_{p_3})(\sigma_{p_2}-\sigma_{p_3})}
		\end{equation}
		as there are no roots $\beta^\vee$ such that $\beta^\vee\cdot\l_3=2$. Due to this an equality, we get \begin{equation}\nonumber
		B_0(\s)=\prod_{i=1}^3{1\over G(1\pm\sqrt 2\sigma_i)}\prod_{i<j=1}^3{1\over G(1\pm\frac{1}{\sqrt 2}(\sigma_i\pm\sigma_i))}
		\end{equation} 
		
		Keeping \eqref{C3sqr}=2, but letting \eqref{C3lin}=$\sqrt{2}e_{p_1}$, we find one nonvanishing solution for the LHS,  $\n_{1}=\sqrt{2}e_{p_1}$, $\n_{2}=\n_{3}=0$, and $i_{2}=1$ or $i_{3}=1$, with the rest zero. This leads to the term 
		\begin{equation}\nonumber
		-\frac{4}{(\sigma_{p_1}\pm\sigma_{p_2})(\sigma_{p_1}\pm\sigma_{p_3})}Z_1(\s)
		\end{equation}
		On the RHS we can describe the four solutions as
		the two couples $\m_{1}=0$, $\m_2=1/\sqrt{2} e_{p_1}\pm 1/\sqrt{2} (e_{p_2} + e_{p_3})$ and $\m_3=1/\sqrt{2} e_{p_1}\mp 1/\sqrt{2} (e_{p_2} + e_{p_3})$ and
		$\m_{1}=0$, $\m_2=1/\sqrt{2} e_{p_1}+ 1/\sqrt{2} (\pm e_{p_2} \mp e_{p_3})$ and $\m_3=1/\sqrt{2} e_{p_1}+ 1/\sqrt{2} (\mp e_{p_2} +\pm e_{p_3})$. This gives the RHS
   		\begin{equation}\nonumber
		-\frac{16}{\sigma_1^2\sigma_2^2\sigma_3^2(\sigma_{p_1}\pm\sigma_{p_2})(\sigma_{p_1}\pm\sigma_{p_3})}
		\end{equation}		
		so that $Z_1(\s)=\frac{4}{\sigma_1^2\sigma_2^2\sigma_3^2}$, which is indeed the 1 $Sp(6)$ instanton equivariant volume, with the v.e.v.'s rescaled by $\sqrt 2$ factors.

		Continuing to two instantons, we have to collect $t^{\sqrt{2}\s\cdot e_{p_1}+3}$ terms, as we find that $t^{\sqrt{2}\s\cdot(e_{p_1}+e_{p_2})+3}$ ones don't involve $Z_2$ and lead to an identity involving shifts of $Z_1$ and rational functions. The structure of solutions is more involved. By picking $(p_1,p_2,p_3)=(1,2,3)$ for readability, we find the relation $\left(\sqrt{2} \sigma _1-1\right){}^2 \sigma _2^2 \sigma _3^2Z_2(\s)=$
		
		\resizebox{.95\linewidth}{!}{
		\begin{minipage}{\linewidth}
		\begin{align*}
		&=\sum_{(w_1,w_2)\in\{(-1,-1),(-1,1),(1,-1)\}}\frac{2 \left(\sigma _1+\sigma _2 w_1+\sigma _3 w_2\right){}^2}{\sigma _1^2 \left(\sqrt{2} \sigma _1+1\right){}^2 \left(\sigma _1+\sigma _2 w_1\right){}^2 \left(\sigma _1+\sigma _3 w_2\right){}^2 \left(\sigma _2 w_1+\sigma _3 w_2\right){}^2}+2 Z_1\left(\s\right)\\
			&+\sum_{(w_1,w_2)\in\{(-1,-1),(-1,1),(1,-1),(1,1)\}} \frac{\left(\sqrt{2} \sigma _2 w_1+\sqrt{2} \sigma _3 w_2+1\right){}^2 Z_1\left(\sigma _1+\frac{1}{\sqrt{2}},\sigma _2+\frac{w_1}{\sqrt{2}},\sigma _3+\frac{w_2}{\sqrt{2}}\right)}{2 \left(\sigma _2 w_1+\sigma _3 w_2\right){}^2}\\
			&+\sum_{w=\pm 1}\frac{2 \sigma _2^2}{\sigma _3^2 \left(\sigma _2^2-\sigma _3^2\right){}^2 \left(1-\sqrt{2} \sigma _3 w\right){}^2 \left(\sigma _3 w+\sigma _1\right){}^2}+\frac{2 \sigma _3^2}{\sigma _2^2 \left(\sigma _2^2-\sigma _3^2\right){}^2 \left(1-\sqrt{2} \sigma _2 w\right){}^2 \left(\sigma _2 w+\sigma _1\right){}^2}\\
			&+\sum_{w=\pm 1}-\frac{2 \sigma _3^2 \left(\sqrt{2} \sigma _1+\sqrt{2} \sigma _2 w+1\right){}^2}{\left(\sqrt{2} \sigma _1+1\right){}^2 \sigma _2^2 \left(\sigma _2^2-\sigma _3^2\right){}^2 \left(\sigma _1-\sigma _2 w\right){}^2 \left(\sqrt{2} \sigma _2 w+1\right){}^2}-\frac{2 \sigma _2^2 \left(\sqrt{2} \sigma _1+\sqrt{2} \sigma _3 w+1\right){}^2}{\left(\sqrt{2} \sigma _1+1\right){}^2 \sigma _2^2 \left(\sigma _2^2-\sigma _3^2\right){}^2 \left(\sigma _1-\sigma _3 w\right){}^2 \left(\sqrt{2} \sigma _3 w+1\right){}^2}\\
			&-\frac{32 \sigma _2^2}{\left(\sqrt{2} \sigma _1+1\right){}^2 \left(1-2 \sigma _3^2\right){}^2 \left(\sigma _2^2-\sigma _3^2\right){}^2}-\frac{32 \sigma _3^2}{\left(\sqrt{2} \sigma _1+1\right){}^2 \left(1-2 \sigma _2^2\right){}^2 \left(\sigma _2^2-\sigma _3^2\right){}^2}-\frac{1}{4} \left(\sqrt{2} \sigma _1+2\right){}^2 \sigma _2^2 \sigma _3^2 Z_1\left(\s\right) Z_1\left(\sigma _1+\sqrt{2},\sigma _2,\sigma _3\right)
			\end{align*}
			\end{minipage}
			}
     \vspace*{0.5cm}
		\\
		\noindent
		which gives the correct 2-instanton equivariant volume compared to instanton counting, although in a vastly different presentation. 
		
		\color{black}
		
		\subsubsection{\texorpdfstring{$C_4$}{C4}}
		In this case $n$ is even, so under shifts, the middle node gets mapped to itself, up to some power of $t$ as required by asymptotics. The relevant lattice is $Q^\vee=\sqrt{2}\Z^4$, the shift $\l_4=((\frac{1}{\sqrt{2}})^4)$, and the full system is
		\begin{gather*}
		D^2(\tau_0)=-t^{\frac{1}{5}}\tau_1,\quad
	    D^2(\tau_1)=-2t^{\frac{1}{5}}\tau_0^2\tau_2,\quad
		D^2(\tau_2)=-2t^{\frac{1}{5}}\tau_1\tau_3\\
		D^2(\tau_3)=-2t^{\frac{1}{5}}\tau_2\tau_4^2,\quad
		D^2(\tau_4)=-t^{\frac{1}{5}}\tau_3
		\end{gather*}
		We can eliminate the middle node tau function $\tau_2$ from the following 
		\begin{gather}\nonumber
		D^4(\tau_0)=-2t^{-\frac{1}{5}}\tau_0^2\tau_2,\quad D^4(\tau_4)=-2t^{-\frac{1}{5}}\tau_2\tau_4^2
		\end{gather}
		to write   
		\begin{equation}\label{C4}
		\tau_4 Y^3(\tau_0)=\tau_0 Y^3(\tau_4)\,.
		\end{equation}
		We can repeat the calculation in the previous section, this time in short. Inserting \eqref{taui} we obtain
		\begin{align*}
		&\sum_{\substack{ \n_1\in \sqrt{2}\Z^4+\l_4 \\\n_{2,3,4}\in \sqrt{2}\Z^3\\ i_{1,2,3,4}\in{\mathbb N_0}}} \prod\limits_{k=1}^4 e^{2\pi\I \e\cdot\n_k }t^{\12(\s+\n_k)^2+i_k} B_0(\s+\n_k) Z_{i_k}(\s+\n_k) \notag
		\\
		&\prod\limits_{k_1<k_2=2}^4\left(\12\n_{k_1}^2+i_{k_1}-\12\n_{k_2}^2-i_{k_2}+(\n_{k_1}-\n_{k_2})\cdot\s\right)^2\notag
		\\
		=&\sum_{\substack{ \m_1\in \sqrt{2}\Z^4 \\ \m_{2,3,4}\in \sqrt{2}\Z^4+\l_4\\ j_{1,2,3,4}\in{\mathbb N_0}}}\prod\limits_{k=1}^4 e^{2\pi\I \e\cdot\n_k }t^{\12(\s+\n_k)^2+i_k} B_0(\s+\n_k) Z_{i_k}(\s+\n_k) \notag
		\\
		&\prod\limits_{k_1<k_2=2}^4\left(\12\m_{k_1}^2+j_{k_1}-\12\m_{k_2}^2-j_{k_2}+(\m_{k_1}-\m_{k_2})\cdot\s\right)^2\notag
		\\
		\end{align*}
		Then, as  $2\times\12\l_4^2=2$, we decompose the vectors in terms of the coroot lattice as $\n_{1}=\n_{1}^{(0)}+\l_4$, $\n_{2,3,4}=\n_{2,3,4}^{(0)}$, $\m_{1}=\m_{1}^{(0)}$ $\m_{2,3,4}=\m_{2,3}^{(0)}+\l_4$
		which implies the constraints
		\begin{gather} 
		\l_4\cdot\n_1^{(0)}+\sum_{k=1}^4 \12 (\n_i^{(0)})^2+ i_k =2+\l_4\cdot(\m_2^{(0)}+\m_3^{(0)}+\m_4^{(0)})+\sum_{k=1}^4 \12 (\m_k^{(0)})^2 + j_k \label{C4sqr} \\ \label{C4lin}
		\sum_{k=1}^4 \n_i^{(0)} =2\l_4 + \sum_{k=1}^4 \m_i^{(0)}
		\end{gather}
		Let $p_{1},p_2,p_3,p_4$ be a permutation of $\{1,2,3,4\}$. To obtain the functional equations for the one-loop term we consider factors of $t^{\sqrt{2}\s\cdot(e_{p_1}+e_{p_2})+2}$,  \eqref{C4sqr}=2 and \eqref{C4lin}$=\sqrt{2}(e_{p_1}+e_{p_2})$. For the LHS the only nonvanishing solutions are $\n_1^{(0)}=0$ and $\n_{2,3,4}$ permutations of $\{\sqrt2 e_{p_1},\sqrt2 e_{p_2},0\}$, while on the RHS the only nonvanishing ones are $\m_1^{(0)}=0$ and $\m_{2,3,4}^{(0)}$ permutations of $\{-\sqrt2 e_{p_3},-\sqrt2 e_{p_4},0\}$. Some factors cancel, leading to
    \begin{gather*}2(1+\sqrt{2}\sigma_{p_1})^2(1+\sqrt{2}\sigma_{p_2})^2(\sigma_{p_1}-\sigma_{p_2})^2 B_0(\s+\sqrt{2}e_{p_1})B_0(\s+\sqrt{2}e_{p_2}) \\
		=4\sigma_{p_3}^2(\sigma_{p_3}-\sigma_{p_4})^2\sigma_{p_4}^2 B_0(\s+\l_3-\sqrt{2}e_{p_3})B_0(\s+\l_3-\sqrt{2}e_{p_4})
		\end{gather*}
		We checked that \eqref{1loop} satisfies this relation also in this case.
		To find the one-instanton term, we need to collect factors of $t^{\sqrt{2}\s\cdot e_{p_1}+2}$. The solutions on the LHS are either with all $i$'s vanishing, that is with $\n_1^{(0)}=-\n_{k}^{(0)}=-\sqrt 2 e_{p}$ for any $k,p\in\{2,3,4\}$ and the remaining two vectors equal to $\sqrt 2 e_{p_1}$ and zero respectively, or with one out of $i_{2,3,4}$ being 1 with a single vector -- of index different from both $1$ and from the index of the $i$s -- being equal to $\sqrt 2 e_{p}$. On the RHS, $j$'s vanish, $\m_1^{(0)}=0$ and the rest are a permutation of $\{\sqrt 2 e_{\tilde p_2},\sqrt 2 e_{\tilde p_3}+\sqrt 2 e_{\tilde p_4},0\}$ with $\tilde p_{2,3,4}$ a permutation of $\{p_2,p_3,p_4\}$. 
		After some cancellation of rational functions,
		we find the correct one-instanton term  $Z_1(\s)=-8/\sigma_1^2\sigma_2^2\sigma_3^2\sigma_4^2$.
		
		One continues similarly up to higher order. We have checked agreement with instanton counting until four instantons.
		
		\subsection{\texorpdfstring{$E_6$}{E6}}
		Even though the equations presented up to this point were novel, the instanton volumes were able to be obtained by means of instanton counting as in appendix \ref{appendix:instanton_counting}. We now turn to non-classical Lie algebras and describe novel ways of obtaining instanton volumes where instanton counting is unavailable. We note that yet another way to obtain them is via blowup relations, likewise conjectural at time of writing, and these serve as a cross-check. They are also described in appendix \ref{appendix:instanton_counting}. Computationally, however, the equations we find are quicker, because they only involve instanton volumes at the same  $\Omega$-background. We begin with the simplest simply laced exceptional Lie algebra, $E_6$. 
		
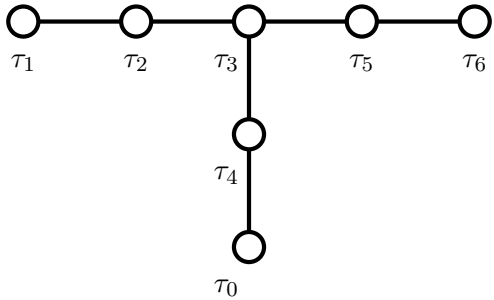
\begin{center}
			\begin{dynkinDiagram}[upside down, ordering=Carter, extended,edge length=1.5cm, indefinite edge/.style={ultra thick,densely dashed}, edge/.style={ultra thick},o/.style={ultra thick,fill=white,draw=black}, root radius=.2cm,root/.style={ultra thick,fill=white,draw=black}, arrow width = 0.4cm, arrow style={length=5mm, width=5mm,line width = 1pt}]E{oooooo}
				\node[left=.3cm,below=.3cm] at (root 0) {$\tau_0$};
				\node[below=.3cm] at (root 1) {$\tau_1$};
				\node[below=.3cm] at (root 2) {$\tau_2$};
				\node[left=.3cm,below=.3cm] at (root 3) {$\tau_3$};
				\node[left=.3cm,below=.3cm] at (root 4) {$\tau_4$};
				\node[below=.3cm] at (root 5) {$\tau_5$};
				\node[below=.3cm] at (root 6) {$\tau_6$};
			\end{dynkinDiagram}
		\end{center}
		Due to the similarities of the root systems of the $E$-type algebras, we will give a brief overview of $E_8$ at this point and describe the others as its reductions. The root system is the union of $D_8$ roots $\{e_i\pm e_j\}_{i\neq j}$ and $(x_1,...,x_8)\in \mathbb{R}^8$ of length $2$ such that all $x_i\in\mathbb Z+\12$ and $\sum_i x_i$ is even. The coroot lattice can be obtained from two cosets of the $D_8$ one as $Q^{[E_8]}=Q^{[D_8]}\cup \left(Q^{[E_8]}+((\12)^{n-1},-\12)\right)$. $E_6$ is then obtained by projecting all of the roots to have the last three coordinates equal, $(x_1,...,x_5,x_6,x_6,x_6)$. Clearly, this forces the $D_8$-type roots to an embedding of $D_5$, with the last three coordinates zero. 
		Unlike $E_8$, which is unimodular and has no minuscule coweights, $E_6$ has three: $\l_0=0$, $\l_1=(1,0^4,(-{1\over3})^3)$, and $\l_6=(0^5,(-{2\over3})^3)$. The Dynkin diagram exhibits an outer $\mathbb{Z}_3$ symmetry. For this exceptional algebra we obtain the $\tau$-system
		\begin{gather}
		 \label{E6first}
		\tau_4=-t^{-\frac{1}{12}}D^{2}(\tau_0),\quad\tau_2=-t^{-\frac{1}{12}} D^{2}(\tau_1),\quad\tau_5= -t^{-\frac{1}{12}} D^{2}(\tau_6)\\
		D^{2}(\tau_3)=-t^{\frac{1}{12}}\tau_2\tau_4\tau_5\\
		-t^{\frac{1}{12}}\tau_3=\tau_0^{-1}D^{2}(\tau_4)=\tau_1^{-1}D^{2}(\tau_2)=\tau_6^{-1}D^{2}(\tau_5) \label{E6last}
		\end{gather}
		Focusing on the legs with $\tau_0$ and $\tau_6$, inserting \eqref{E6first} in the last equation \eqref{E6last} and using the operators defined in \eqref{Yoperators} gives us
		\begin{equation}
		\label{E6}
		Y^{3}(\tau_0)=Y^{3}(\tau_6)
		\end{equation}
		The Kiev Ansatz we insert likewise has $\s,\e\in\mathbb{C}^8$, but with the last three components restricted to be same. The equation to be solved becomes
		\begin{align*}
    		&\sum_{\substack{ \n_{1,2,3}\in Q \\ i_{1,2,3}\in{\mathbb N}}} \prod\limits_{k=1}^3 e^{2\pi\I \e\cdot\n_k }t^{\12(\s+\n_k)^2+i_k} B_0(\s+\n_k) Z_{i_k}(\s+\n_k) \notag
    		\\
    		&\prod\limits_{k_1<k_2}\left(\12\n_{k_1}^2+i_{k_1}-\12\n_{k_2}^2-i_{k_2}+(\n_{k_1}-\n_{k_2})\cdot\s\right)^2 \notag
    		\\=
    		&\sum_{\substack{ \m_{1,2,3}\in Q \\ j_{1,2,3}\in{\mathbb N}}} t^2\cdot \prod\limits_{k=1}^3 e^{2\pi\I \e\cdot\m_k }t^{\12(\s+\m_k)^2+\l_6\cdot(\s+\m_k)+i_k} B_0(\s+\m_k+\l_6) Z_{j_k}(\s+\m_k+\l_6) \notag
    		\\
    		&\prod\limits_{k_1<k_2}\left(\12\m_{k_1}^2+j_{k_1}-\12\m_{k_2}^2-j_{k_2}+(\m_{k_1}-\m_{k_2})\cdot(\s+\l_6)\right)^2
    	\end{align*}
		To get the lowest order equations which specify $B_0$, let $p_1,...p_5$ be a permutation of $\{1,...,5\}$ and let $\boldsymbol \delta:=((\12)^8)$. Then looking at the coefficients of $t^{2+\s\cdot(2e_{p_1}+e_{p_2}+e_{p_3})}$ gives the equation
		\bea
		\left(1+\sigma_{p_1}+\sigma_{p_2}\right)^2\left(1+\sigma_{p_1}+\sigma_{p_3}\right)^2\left(\sigma_{p_2}-\sigma_{p_3}\right)^2 
		B_0(\s)B_0(\s+e_{p_1}+e_{p_2})B_0(\s+e_{p_1}+e_{p_3})= \notag\\
		\left((\boldsymbol \delta -e_{p_2}-e_{p_3})\cdot\s\right)^2
		\left((\boldsymbol \delta -e_{p_2}-e_{p_3}-e_{p_4}-e_{p_5})\cdot\s\right)^2\left(\sigma_{p_4}+\sigma_{p_5}\right)^2 
		\times \notag\\
		B_0(\s+\boldsymbol \delta+\l)	B_0(\s+\boldsymbol \delta+\l-e_{p_4}-e_{p_5})B_0(\s+e_{p_1}-\l/2) \notag
		\eea
		The solution satisfying the asymptotic behaviour \eqref{asy} is 
		\begin{equation*}
		B^{[E_6]}_0=\prod\limits_{i<j=1}^5 { 1 \over G(1\pm\sigma_i\pm\sigma_j)
		}\prod\limits_{\substack{\varepsilon_i=\pm 1 \\ \prod_{i=1}^8\varepsilon_i = 1, \\ \varepsilon_6=\varepsilon_7=\varepsilon_8}} { 1 \over G(1+{1 \over 2}\sum\limits_{i=1}^8\varepsilon_i \sigma_i) }
		\end{equation*}
		We also solved the recurrence relation arising from \eqref{E6} up to three instantons. For one instanton, our results agree with the ones of \cite{Keller:2011ek}, and for two instantons they agree with the blowup formula. Three instantons proved to be too computationally intensive to check using the blowup formula, however it obeys the expected large-$\s$ limit described in appendix \ref{appendix:deepcoulomb}.
		The one instanton contribution follows most easily by looking at the coefficients of $t^{2+e_{p_1}+e_{p_2}}$, where we obtain
		\begin{gather*}
		3!(\sigma_{p_1}+\sigma_{p_3})^2(1+\sigma_{p_1}+\sigma_{p_2})^2 B_0(\s)^2 B_0(\s+e_{p_1}+e_{p_2}) Z_1(\s)\\
		=\sum_{\substack{\n_1+\n_2+\n_3+3\l
				\\=e_{p_1}+e_{p_2}
				\\\12\n_k^2+\l\cdot\n_k=0
		}}\prod_{i<j}^3 (\s\cdot(\n_i-\n_j))^2 \prod_{i=1}^3B_0(\s+\n_i)
		\\
		-\sum_{\substack{\n_1+\n_2 \\=e_{p_1}+e_{p_2}
				\\\n_1^2=\n_2^2=2
		}}B_0(\s) (\s\cdot(\n_1-\n_2))^2 (\s\cdot\n_1)^2 (\s\cdot\n_2)^2 B_0(\s+\n_1)B_0(\s+\n_2)
		\end{gather*}
		The higher instanton expressions are too cumbersome and unenlighting to display 
		here\footnote{These can be provided privatly to any interested reader.}. 
		
		\subsection{\texorpdfstring{$E_7$}{E7}}
		\begin{center}
			\begin{dynkinDiagram}[upside down, ordering=Carter, extended,edge length=1.5cm, indefinite edge/.style={ultra thick,densely dashed}, edge/.style={ultra thick},o/.style={ultra thick,fill=white,draw=black}, root radius=.2cm,root/.style={ultra thick,fill=white,draw=black}, arrow width = 0.4cm, arrow style={length=5mm, width=5mm,line width = 1pt}]E{ooooooo}
				\node[left=.3cm,below=.3cm] at (root 0) {$\tau_0$};
				\node[below=.3cm] at (root 1) {$\tau_1$};
				\node[below=.3cm] at (root 2) {$\tau_2$};
				\node[below=.3cm] at (root 3) {$\tau_3$};
				\node[left=.3cm,below=.3cm] at (root 4) {$\tau_4$};
				\node[below=.3cm] at (root 5) {$\tau_5$};
				\node[below=.3cm] at (root 6) {$\tau_6$};
				\node[below=.3cm] at (root 7) {$\tau_7$};
			\end{dynkinDiagram}
		\end{center}
		The roots of $E_7$ are obtained by projecting the $E_8$ ones to have the last two coordinates equal, $(x_1,...,x_5,x_6,x_7,x_7)$. 
		$E_7$ has two minuscule coweights $\l_0=0$ and $\l_1=(1,0^5,(-\frac{1}{2})^2)$. The Dynkin diagram exhibits an outer $\mathbb{Z}_2$ symmetry. For this exceptional algebra we obtain the $\tau$-system 
		\begin{gather}
		\tau_7=-t^{-\frac{1}{18}}D^{2}(\tau_0),\quad\tau_2= -t^{-\frac{1}{18}}D^{2}(\tau_1)\\
		\tau_6=-t^{-\frac{1}{6}}\tau_0^{-1}D^{4}(\tau_0),\quad\tau_3=-t^{-\frac{1}{6}}\tau_1^{-1}D^{4}(\tau_1)\\\label{E7}
		\tau_2^{-1}D^{2}(\tau_3)=-t^{-\frac{1}{18}}\tau_4=\tau_7^{-1}D^{2}(\tau_6)
		\end{gather}
		When we rewrite \eqref{E7} in terms of the single equation, the powers of $t^{\frac{1}{18}}$ drop out to give
		\begin{equation}\nonumber
		{1\over D^{2}(\tau_0)}D^2({D^4(\tau_0)\over\tau_0})={1\over D^{2}(\tau_1)}D^2({D^4(\tau_1)\over\tau_1})
		\end{equation}
		Here we recognize an operator defined in \eqref{Yoperators}, which enables us to write
		\begin{equation}\nonumber
		Y^4(f)={1\over D^{2}(f)}D^2({D^4(f)\over f}) \quad\Rightarrow\quad Y^4(\tau_0)=Y^4(\tau_1), \quad
		\end{equation}
		The Kiev Ansatz we insert likewise has $\s,\e\in\mathbb{C}^8$, but with the last two components restricted to be same. The equation to be solved becomes
		\begin{align*}
    		&\sum_{\substack{ \n_{1,2,3,4}\in Q \\ i_{1,2,3,4}\in{\mathbb N}}} \prod\limits_{k=1}^4 e^{2\pi\I \e\cdot\n_k }t^{\12(\s+\n_k)^2+i_k} B_0(\s+\n_k) Z_{i_k}(\s+\n_k) \notag
    		\\
    		&\prod\limits_{k_1<k_2}\left(\12\n_{k_1}^2+i_{k_1}-\12\n_{k_2}^2-i_{k_2}+(\n_{k_1}-\n_{k_2})\cdot\s\right)^2 \notag
    		\\=
    		&\sum_{\substack{ \m_{1,2,3,4}\in Q \\ j_{1,2,3,4}\in{\mathbb N}}} t^3\cdot \prod\limits_{k=1}^4 e^{2\pi\I \e\cdot\m_k }t^{\12(\s+\m_k)^2+\l_1\cdot(\s+\m_k)+i_k} B_0(\s+\m_k+\l_1) Z_{j_k}(\s+\m_k+\l_1) \notag
    		\\
    		&\prod\limits_{k_1<k_2}\left(\12\m_{k_1}^2+j_{k_1}-\12\m_{k_2}^2-j_{k_2}+(\m_{k_1}-\m_{k_2})\cdot(\s+\l_1)\right)^2
    	\end{align*}
		With regards to the linear and quadratic constraints obtained from comparing exponents of $t$, $\{t^{\sigma_i}\}_i$, this is similar to $C_4$. In, $\l_1^2=\frac{3}{2}$, so in the analogue of \eqref{C4sqr} we end up with $\frac{4}{2}\l_1^2=3$ in pure powers of $t$. Likewise, we have even powers of $\tau_1$, and $2\l_1\in Q$. Both of these lead to well defined analogues of \eqref{C4sqr} and \eqref{C4lin}. The lowest possible order in $t$ is $t^3$. If we pick $p_1,...,p_6$ to be a permutation of $\{1,...,6\}$, looking at powers of $t^{3+\s\cdot(2 e_{p_1}+2e_{p_2}+2e_{p_3})}$ we get
		\begin{gather*}
		B_0(\s)\prod\limits_{i<j}^3 (1+\sigma_{p_i}+\sigma_{p_j})^2(\sigma_{p_i}-\sigma_{p_j})^2 B_0(\s+e_{p_i}+e_{p_j})\\
		=B_0(\s+\boldsymbol{\delta}-e_1-e_{p_4}-e_{p_5}-e_{p_6})\prod\limits_{i<j}^3(-\delta_{i,1}+\sigma_{p_i}\pm\sigma_{p_j})^2\prod_{i=1}^3B_0(\s+\boldsymbol{\delta}-e_1-e_{p_{i+3}})
		\end{gather*}
		with 
		$\boldsymbol \delta:=((\12)^8)$ as above.
		 Clearly, the only lattice points satisfying the quadratic constraint while summing up to $2(e_{p_1}+e_{p_2}+e_{p_3})$ are $e_{p_1}+e_{p_2},e_{p_1}+e_{p_3},e_{p_2}+e_{p_3}$ and zero, while the ones on the shifted lattice, which can be inferred from the above equation, are similarly unique up to permutation. The solution satisfying the asymptotic behaviour \eqref{asy} is
		\vspace*{-0.3cm}
		\begin{equation*}
	    B^{[E_7]}_0	=\prod\limits_{i<j=1}^6 { 1 \over G(1\pm\sigma_i\pm\sigma_j)
		}\prod\limits_{\substack{\varepsilon_i=\pm 1 \\ \prod_{i=1}^8\varepsilon_i = 1, \\ \varepsilon_7=\varepsilon_8}} { 1 \over G(1+{1 \over 2}\sum\limits_{i=1}^8\varepsilon_i \sigma_i) }
		\end{equation*}
		The one instanton contribution follows most easily by looking at the coefficients of $t^{3+2\sigma_{p_1}+\sigma_{p_2}+\sigma_{p_3}}$, where we obtain
		\begin{gather*}
		4!(\sigma_{p_2}-\sigma_{p_3})^2(1+\sigma_{p_1}+\sigma_{p_2})^2(\sigma_{p_1}+\sigma_{p_2})^2(1+\sigma_{p_1}+\sigma_{p_3})^2(\sigma_{p_1}+\sigma_{p_3})^2
	    \\B_0(\s)^2B_0(\s+e_{p_1}+e_{p_2})B_0(\s+e_{p_1}+e_{p_3}) Z_1(\s)\\
		=\sum_{\substack{\n_1+\n_2+\n_3+\n_4+4\l
				\\=2e_{p_1}+e_{p_2}+e_{p_3}
				\\\12\n_k^2+\l\cdot\n_k=0
		}}\prod_{i<j}^4 (\s\cdot(\n_i-\n_j))^2 \prod_{i=1}^4B_0(\s+\n_i)
	    \\
		-\sum_{\substack{\n_1+\n_2+\n_3 \\=2e_{p_1}+e_{p_2}+e_{p_3}
		\\\n_1^2=\n_2^2=\n_3^2=2
	    }}B_0(\s)\prod_{i<j}^3 (\s\cdot(\n_i-\n_j))^2\prod_{i=1}^3 (\s\cdot\n_i)^2 B_0(\s+\n_i)
		\end{gather*}
		This can be compared with the general one instanton term, most easily when we specialize all variables except one; for example, leaving intact $\sigma_7$ yields a ratio of a degree 50 and a degree 66 polynomial in $\C[\sigma_7]$. Comparing other powers, i.e. $t^{3+e_{p_1}+e_{p_2}}$ and $t^{3}$ yields different expressions for $Z_1(\s)$. 
		To obtain the two instanton term, we can look at $t^{4+2e_{p_1}+e_{p_2}+e_{p_3}}$. Similarly to the previous case, we obtain a ratio of a degree 166 to one of 198 in $\C[\sigma_7]$. The large-$\sigma$ limit conforms to the expected limit from appendix \ref{appendix:deepcoulomb}.

		\subsection{\texorpdfstring{$E_8$}{E8}}
		\begin{center}
			\begin{dynkinDiagram}[upside down, ordering=Carter, extended,edge length=1.5cm, indefinite edge/.style={ultra thick,densely dashed}, edge/.style={ultra thick},o/.style={ultra thick,fill=white,draw=black}, root radius=.2cm,root/.style={ultra thick,fill=white,draw=black}, arrow width = 0.4cm, arrow style={length=5mm, width=5mm,line width = 1pt}]E{oooooooo}
				\node[below=.3cm] at (root 0) {$\tau_0$};
				\node[below=.3cm] at (root 1) {$\tau_1$};
				\node[below=.3cm] at (root 2) {$\tau_2$};
				\node[below=.3cm] at (root 3) {$\tau_3$};
				\node[below=.3cm] at (root 4) {$\tau_4$};
				\node[left=.3cm,below=.3cm] at (root 5) {$\tau_5$};
				\node[below=.3cm] at (root 6) {$\tau_6$};
				\node[below=.3cm] at (root 7) {$\tau_7$};
				\node[below=.3cm] at (root 8) {$\tau_8$};
			\end{dynkinDiagram}
		\end{center}
		For the exceptional algebra $E_8$ we obtain the system
		\begin{align}
		\label{E81}
		Y^6(\tau_0)&=Y^3(\tau_8)\\
  \label{E82}
		\tau_6 D^2(\tau_8)&=Y^7(\tau_0)\\
  \label{E83}
		D^2(\tau_6)&=Y^6(\tau_0)
		\end{align}
		Here, $\tau_8$ needs to be determined from \eqref{E81}, and then fed into \eqref{E83}, once $\tau_6$ has been eliminated using \eqref{E82}. As the algebra with the largest root system, it was not practical to explicit calculations for the above $E_8$ system.

		\subsection{\texorpdfstring{$G_2$}{G2}}
		\begin{center}
			\begin{dynkinDiagram}[upside down, reverse arrows, extended,edge length=1.5cm, indefinite edge/.style={ultra thick,densely dashed}, edge/.style={ultra thick},o/.style={ultra thick,fill=white,draw=black}, root radius=.1cm,root/.style={ultra thick,fill=white,draw=black}, arrow width = 0.4cm, arrow style={length=5mm, width=8mm,line width = 1pt}]G{oo}
				\node[below=.2cm] at (root 0) {$\tau_0$};
				\node[below=.2cm] at (root 1) {$\tau_1$};
				\node[below=.2cm] at (root 2) {$\tau_2$};
			\end{dynkinDiagram}
		\end{center}
		\vspace*{-0.2cm}
		$G_2$ is a non-simply laced exceptional algebra. As can be seen from the (dual) extended Dynkin diagram, eliminating the node corresponding to $\tau_2$ leaves us with a copy of $A_2$, which is a subalgebra which we previously embedded into a hyperplane orthogonal to $(1,1,1)$ in $\mathbb{R}^3$. We will use the same embedding for $G_2$, with $\sigma_1+\sigma_2+\sigma_3=0$. Besides the $6$ roots of $A_2$, $G_2$ has $6$ other roots of the form $e_{p_1}+e_{p_2}-2e_{p_3}$ for $p_{1,2,3}$ permutations of $\{1,2,3\}$. In the normalization where $G_2$'s longest roots have length 2, the coroot lattice is the span $Q^\vee=\Z\frac{1}{\sqrt3}(-2,1,1)\oplus\Z\sqrt3(1,-1,0)$ - we are not aware of a simpler definition. The $\tau$-system is
		\begin{align}
		    \label{G2first}
		    D^2(\tau_0)&=-t^{\frac{1}{6}}\tau_1 \\
		    \label{G2mid}
		    D^2(\tau_1)&=-t^{\frac{1}{6}}\tau_0\tau_2 \\
		    \label{G2last}
		    D^2(\tau_2)&=-3t^{\frac{1}{6}}\tau_1 
		\end{align}
		By using \eqref{G2first} to eliminate $\tau_1$ from \eqref{G2mid} and then using \eqref{G2mid} to eliminate $\tau_2$ from \eqref{G2last}, the $\tau$-system reduces to the single equation 
		\begin{equation}
		D^2(\tau_0^{-1}D^4(\tau_0))=3t(D^2(\tau_0))^3\, 
		\end{equation}
		which can be simplified to
		\begin{equation}\label{G2}
		Y^4(\tau_0)=3t(D^2(\tau_0))^2\, 
		\end{equation}
		since the Kiev Ansatz \eqref{Kiev} implies $D^2(\tau_0)\neq0$.  We insert
		\begin{equation}\nonumber
		\tau_0(\s,\e|t)=\sum\limits_{\n\in Q^\vee}e^{2\pi\I \e\cdot\n}\left({-t\over 3}\right)^{\12(\s+\n)^2}B_0\left(\s+\n|{-t\over 3}\right)\notag
		\end{equation}
		and after a rescaling $t\mapsto -3 t$ we obtain the equation
		\begin{gather}
		\sum_{\substack{ \n_{1,2,3,4}\in Q^\vee\\ i_{1,2,3,4}\in{\mathbb N_0}}} \prod\limits_{k=1}^4 e^{2\pi\I \e\cdot\n_k }t^{\12(\s+\n_k)^2+i_k} B_0(\s+\n_k) Z_{i_k}(\s+\n_k) \notag
		\\
		\Bigg(\frac{1}{4!}\prod\limits_{k_1<k_2}(\12\n_{k_1}^2+i_{k_1}-\12\n_{k_2}^2-i_{k_2}+(\n_{k_1}-\n_{k_2})\cdot\s)^2\notag
		\\
		-\frac{9}{4}t(\12\n_{1}^2+i_{1}-\12\n_{2}^2-i_{2}+(\n_{1}-\n_{2})\cdot\s)^2\notag
		\\
		(\12\n_{3}^2+i_{3}-\12\n_{4}^2-i_{4}+(\n_{3}-\n_{4})\cdot\s)^2\Bigg)=0 \, .
		\label{G2explicit}
		\end{gather}
		The coefficients of  $t^{3+\s\cdot({4\over\sqrt{3}},-{2\over\sqrt{3}},-{2\over\sqrt{3}})}$ are the lowest order powers which give the functional equation for $B_0(\s)$. Instead of a quartic relation we find that it simplifies to the following quadratic one
		\begin{gather*}
		\left(\frac{2 \sigma_1-\sigma_2-\sigma_3}{\sqrt{3}}+1\right)^2 B_0(\s) B_0\left(\s+\frac{1}{\sqrt3}(2,-1,-1)\right)
		=\left(\frac{\sigma_2-\sigma_3}{\sqrt{3}}\right)^2 
		\left(\frac{\sigma_1+\sigma_2-2 \sigma_3}{\sqrt{3}}\right)^2 
		\\\times
		\left(\frac{\sigma_1-2 \sigma_2+\sigma_3}{\sqrt{3}}\right)^2 
		\left(\frac{\sigma_1+\sigma_2-2 \sigma_3}{\sqrt{3}}+1\right)^2 
		\left(\frac{\sigma_1-2 \sigma_2+\sigma_3}{\sqrt{3}}+1\right)^2 
		\\ \times
		B_0\left(\s+\frac{1}{\sqrt3}(1,-2,1)\right)B_0\left(\s+\frac{1}{\sqrt3}(1,1,-2)\right)
		\end{gather*}
		By imposing \eqref{asy}, these are solved by $B_0^{[G_2]}(\s)=$
		\begin{equation*}
		\prod_{i<j}^3{1\over G(1\pm{1\over\sqrt{3}}(\sigma_i-\sigma_j))}\prod_{\substack{ijk\\ cyclic}}^3{1\over G(1\pm{1\over\sqrt{3}}(2\sigma_i-\sigma_j-\sigma_k))}
		\end{equation*}
		However, such a simplification doesn't apply to the higher orders or different powers of $t$, $\{t^{\sigma_i}\}$. The 1-instanton contribution is obtained by considering the coefficient of the next order $t^{3+\s\cdot(\sqrt{3},0,-\sqrt{3})}$ term: curiously, all $B_0(\s)$ factors drop out and we obtain just 
		\begin{equation}\notag
		Z_1(\s)^{[G_2]}=-\frac{486}{(\sigma_1+\sigma_2-2\sigma_3)^2 (\sigma_1-2\sigma_2+\sigma_3)^2 (-2\sigma_1+\sigma_2+\sigma_3)^2}\stackrel{\sigma_3=-\sigma_1-\sigma_2}{=}-\frac{3}{2\sigma_1^2\sigma_2^2(\sigma_1+\sigma_2)^2}
		\end{equation}
		This expression is in agreement with $(4.39)$ in \cite{Keller:2011ek} when $Q=0$ and their $a_1 = (\sigma_1-\sigma_2)/\sqrt{6}$, $a_2 = (\sigma_1+\sigma_2)/\sqrt{3}$ as well as with the blowup-formula from \ref{appendix:instanton_counting}. The next order in $t$, $t^{4+\s\cdot(\sqrt{3},0,-\sqrt{3})}$, gives the 2-instanton term $Z_2(\s)^{[G_2]}\vert_{\sigma_3=-\sigma_1-\sigma_2}=$
		\begin{equation}\notag		
		\frac{3 \left(9 \sigma_1^4 \left(6 \sigma_2^2+1\right)+18 \sigma_1^3 \left(6 \sigma_2^3+\sigma_2\right)+3 \sigma_1^2 \left(18 \sigma_2^4+9 \sigma_2^2-2\right)+6 \sigma_1 \sigma_2 \left(3 \sigma_2^2-1\right)+\left(1-3 \sigma_2^2\right)^2\right)}{\sigma_1^2 \left(1-3 \sigma_1^2\right)^2 \sigma_2^2 \left(1-3 \sigma_2^2\right)^2 (\sigma_1+\sigma_2)^2 \left(1-3 (\sigma_1+\sigma_2)^2\right)^2}
		\, 
		\end{equation}
		which agrees with the expression obtained from the blowup formula of appendix \ref{appendix:instanton_counting}. $Z_3(\s)$ can be obtained by looking at $t^{5+\s\cdot(4,5,-1)/\sqrt{3}}$, although it is much to cumbersome to display. At this point, comparison with the blowup formula again becomes impossible, and we have to be content with checking that the large-$\s$ limit of appendix \ref{appendix:deepcoulomb} is correct.
		
		\subsection{\texorpdfstring{$F_4$}{F4}}
		\begin{center}
			\begin{dynkinDiagram}[upside down,reverse arrows, ordering=Carter, extended,edge length=1.5cm, indefinite edge/.style={ultra thick,densely dashed}, edge/.style={ultra thick},o/.style={ultra thick,fill=white,draw=black}, root radius=.2cm,root/.style={ultra thick,fill=white,draw=black}, arrow width = 0.4cm, arrow style={length=5mm, width=5mm,line width = 1pt}]F{oooo}
				\node[left=.3cm,below=.3cm] at (root 0) {$\tau_0$};
				\node[below=.3cm] at (root 1) {$\tau_1$};
				\node[below=.3cm] at (root 2) {$\tau_2$};
				\node[below=.3cm] at (root 3) {$\tau_3$};
				\node[below=.3cm] at (root 4) {$\tau_4$};
			\end{dynkinDiagram}
		\end{center}
		$F_4$ is another non-simply laced unimodular exceptional algebra. As such, we have to express the system of equations in terms of the single tau function $\tau_0$ associated to the extended node. From the system 
		\begin{gather*}
		\label{F4system}
		D^2(\tau_0)=-t^{1/9}\tau_1,\quad
		D^2(\tau_1)=-t^{1/9}\tau_0\tau_2,\quad
		D^2(\tau_2)=-t^{1/9}\tau_1\tau_3\\
		D^2(\tau_3)=-t^{1/9}\tau_2^2\tau_4,\quad
		D^2(\tau_4)=-t^{1/9}\tau_3
		\end{gather*}
		we obtain the single equation 
		\begin{equation}
		\label{F4}
		D^2({Y^5(\tau_0)\over Y^3(\tau_0)})=-8 t^{2}Y^4(\tau_0)
		\end{equation}
	Like for $E_8$, we leave it as it is, being beyond our computational power.
		\color{black}
		
		\section{Twisted affine Lie algebras: radial Bullough-Dodd and \texorpdfstring{$BC_1$}{BC1}}
		\color{black}
		
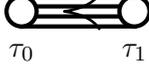
    	\begin{center}
    		\begin{dynkinDiagram}[upside down, ordering=Carter, extended,edge length=1.5cm, indefinite edge/.style={ultra thick,densely dashed}, edge/.style={ultra thick},o/.style={ultra thick,fill=white,draw=black}, root radius=.2cm,root/.style={ultra thick,fill=white,draw=black}, arrow width = 0.4cm, arrow style={length=5mm, width=5mm,line width = 1pt}]A[2]{o}
    			\node[below=.3cm] at (root 0) {$\tau_0$};
    			\node[below=.3cm] at (root 1) {$\tau_1$};
    		\end{dynkinDiagram}
    	\end{center}
    	We consider the \textit{twisted} affine Lie algebra, called either $A^{(2)}_{2n}$ or $BC_n$, with roots of three different lengths inherited from a folding of affine $D_{2n+2}$; the roots are $\pm e_k$, $\pm e_j\pm e_k$ as well as $\pm 2 e_k$ of lengths $1,2,4$ respectively\footnote{Note that the $BC$ nomenclature refers to both sets of roots $\pm e_k$ and $\pm 2 e_k$, which are peculiar to algebras of $B_n$ and $C_n$ respectively, being present along with the usual roots of $D_n$ algebras.}. The simplest case $n=1$ is slightly exceptional in this regard. Indeed, it comes from a quotient of affine $D_4$ by its order 4 automorphism and possessing no middle roots. It gives us
    	\begin{align*}
    		D^2(\tau_0)&=-\12 t^{1/3}\tau_1\\
    		D^2(\tau_1)&=-2t^{1/3}\tau_0^4
    	\end{align*}
    	We redefine $t\mapsto 3 2^{-4} t^2$ from which we obtain the single equation 
    	\begin{equation}
    		\label{A22}
    		Y^3(\tau_0)=-6 t^2  \tau_0^3
    	\end{equation}
    	suitable for inserting an Ansatz analogous to the one used for the $A_1$ case,
    	\begin{equation}\label{BD_kiev}
    		\tau_0(\sigma,\eta|t)=\sum_{n\in \Z, i\in \N_0}e^{2\pi\I \eta n } t^{(\sigma+n)^2+i}B_0(\sigma+n)Z_i(\sigma+n),
    	\end{equation}
    	yielding from \eqref{A22} the equation
    	\begin{gather}\label{BD}
    		\sum_{\substack{ n_{1,2,3}\in \Z \\ i_{1,2,3}\in{\mathbb N}}} \prod\limits_{k=1}^3 e^{2\pi\I \eta n_k }t^{(\sigma+n_k)^2+i_k} B_0(\sigma+n_k) Z_{i_k}(\sigma+n_k) \notag
    		\\
    		\Bigg(\prod\limits_{k_1<k_2}(n_{k_1}^2+i_{k_1}-n_{k_2}^2-i_{k_2}+2(n_{k_1}-n_{k_2})\sigma)^2 +6t^2\Bigg)=0 \, .
    	\end{gather}
    	The lowest order terms are the ones proportional to $t^{2}$. For terms proportional to the $6t^2$, the solution has no shifts or instanton numbers, while for rest of the the only nonvanishing possibilities are $n_1=1,n_2=-1,n_3=0$ and permutations. After a cancellation of a $B_0(\sigma)$ factor, this leads to 
    	\begin{equation}\label{BD_1loop_eqn}
    		4 \sigma ^2 \left(1-4 \sigma ^2\right)^2 B_0(\sigma\pm 1)=B_0(\sigma)^2
    	\end{equation}
    	At this point we have to discuss what kind of asymptotics would be suitable. Notice that we have the roots $\pm 2$. If we identify these with the ones of the usual SU(2) adjoint representation, then the roots $\pm 1$ correspond to the fundamental representation. Both representations are obtained from the folding of a pure $D_4$ Super Yang-Mills theory. As the parent theory has no mass parameters, it is natural to consider asymptotic conditions
    	corresponding to an SU(2) with one massless fundamental flavor, i.e.
    	\begin{equation}\nonumber
    		\log (B_0)\sim
    		\frac{1}{4}   (\sigma)^2  \log\left(\sigma \right)^2 
    		-\frac{1}{4}  (2\sigma)^2  \log\left(2\sigma \right)^2 
    	\end{equation}
    	We find that the solution to equation \eqref{BD_1loop_eqn} with the appropriate asymptotics is
    	\begin{equation}\label{BD_1loop}
    		B_0(\sigma)={ G(1\pm\sigma) \over G(1\pm 2\sigma)}
    	\end{equation}
    	
    	Further, by looking at $t^{2k+2\sigma}$ terms in \eqref{BD}, we find that there is a unique term proportional to 
    	\begin{equation}
    		(2k-1)^2(1-2k+2\sigma)^2(1+2\sigma)^2B_0(\sigma+1)B_0(\sigma)Z_{2k-1}(\sigma)
    	\end{equation}
        which comes from a factor from the Kiev Ansatz with $t^{1+2\sigma}$, one with no shifts, and another with only an instanton contribution $t^{2k-1}$. All the other terms are necessarily combinations of terms proportional to $Z_{2k'-1}$ with $k'<k$. To get $t^{2k+2\sigma}$ have to solve
    	\begin{align}
    		&n_1+n_2+n_3 = 1\\
    		&n_1^2+n_2^2+n_3^2+i_1+i_2+i_3+(2) = 2k
    	\end{align}
        The first equation, however, implies $n_1^2+n_2^2+n_3^2$ is odd, and so one of the $i$'s always has to be odd and accordingly $Z_{2k-1}=0$. Indeed, for $k=1$ there is only one such term possible and we find that it has to vanish due to \eqref{BD}, so by induction we can conclude that 
    	\begin{equation}\nonumber
    		Z_{odd}(\sigma)=0.
    	\end{equation}
    	For the rest we find
    	\begin{align*}
    		Z_2(\sigma)&= -\frac{3}{2^2\left(1-4 \sigma ^2\right)^2}\\
    		Z_4(\sigma)&= \frac{9 \left(4 \sigma ^2+1\right)}{2^7 \sigma ^2 \left(1-4 \sigma ^2\right)^2\left(9-4 \sigma ^2\right)^2 }\\
    		Z_6(\sigma)&=-\frac{576 \sigma ^6-2160 \sigma ^4+5324 \sigma ^2+75}{2^9 \sigma ^2 \left(1-4 \sigma ^2\right)^4 \left(9-4 \sigma ^2\right)\left(25-4 \sigma ^2\right) }\\
    		Z_8(\sigma)&= \frac{3 \left(4608 \sigma ^{10}-78336 \sigma ^8+482560 \sigma ^6-615824 \sigma ^4+243742 \sigma ^2+62475\right)}{2^{16} \sigma ^2 \left(1-\sigma ^2\right)^2\left(1-4 \sigma ^2\right)^4 \left(9-4 \sigma ^2\right)^2 \left(25-4 \sigma ^2\right)^2 \left(49-4 \sigma ^2\right)^2} 
    	\end{align*}
    	\color{black}
    	We can recognize here exactly the SU(2) Nekrasov functions with one massless flavor in the fundamental representation. 
    	Note that, starting from PIII$_2$ in the form
    	\begin{equation}\nonumber
    		\ddot{q} = \frac{\dot{q}^2}{q(t)}-\frac{\dot{q}}{t}-\frac{(1-2 \theta ) \dot{q}^2}{t}+q^3-\frac{2}{t^2}
    	\end{equation}
    	and setting $q=t^{-2/3}\exp{w},\theta=1/2$ gives us
    	\begin{equation}\nonumber
    		\partial_{\log{t}}^2 w = t^{2/3}(e^{2 w}-2 e^{-w})
    	\end{equation}
    	Setting $w = a + 2(\sigma-1/6)\log{t}+X(t)$, since $\rho/h^\vee = 1/6$, gives us 
    	\begin{equation}\nonumber
    		\partial_{\log{t}}^2 X = e^{2 a} t^{4\sigma}e^{2 X}-2 e^{a} t^{1-2\sigma} e^{-X}
    	\end{equation}
    	This cannot be obtained by directly applying \eqref{Qsystem} to this affine root system. Instead we must start from $D_4$ and require a solution of the form $\phi_1=\phi_4=0,\phi_3=-\phi_2$ 
        Let us comment on other interesting directions to investigate further and look at the equation for the surviving degree of freedom. This is exactly the folding which gives the diagram $BC_1$ depicted at the beginning of this subsection. 
    	Solving the equation we find 
    	\begin{equation}\nonumber
    		X(\sigma,a|t) = 2\log{\tau_0(\sigma,\tilde{a}|\I t) }-\log{\tau_1(\sigma,a| t)}
    	\end{equation}
    	where $\partial_{\log{t}}^2\log\tau_1(\sigma,a| t)=e^{2 a} t^{4\sigma}e^{2 X}$, normalized 
    	such that $\tau_1(\sigma,a|0)=1$,
    	$\tau_0$ is \eqref{BD_kiev} with perturbative term \eqref{BD_1loop}, first four instanton terms we found and the initial condition
    	\begin{equation}\nonumber
    		\tilde{a}=a-\log \left(\frac{\Gamma (1-2 \sigma )^2 \Gamma (\sigma )}{\Gamma (2 \sigma )^2 \Gamma (1-\sigma ) }\right)-i \pi  \sigma +\frac{i \pi }{2}
    	\end{equation}	   
    	Note that this is different from the tau function defined in $(2.25)$, $(2.29)$ of \cite{Gamayun:2013auu}.

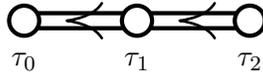
    	\begin{center}
    		\begin{dynkinDiagram}[upside down, ordering=Carter, extended,edge length=1.5cm, indefinite edge/.style={ultra thick,densely dashed}, edge/.style={ultra thick},o/.style={ultra thick,fill=white,draw=black}, root radius=.2cm,root/.style={ultra thick,fill=white,draw=black}, arrow width = 0.4cm, arrow style={length=5mm, width=5mm,line width = 1pt}]A[2]{oo}
    			\node[below=.3cm] at (root 0) {$\tau_0$};
    			\node[below=.3cm] at (root 1) {$\tau_1$};
    			\node[below=.3cm] at (root 2) {$\tau_2$};
    		\end{dynkinDiagram}
    	\end{center}
    	The slightly more general case of $n=2$ gives us the system
    	\begin{align*}
    		\label{BC2system}
    		D^2(\tau_0)&=-\12 t^{1/5}\tau_1\\
    		D^2(\tau_1)&=-t^{1/5}\tau_0^2\tau_2\\
    		D^2(\tau_2)&=-2 t^{1/5}\tau_1^2
    	\end{align*}
    	from which we obtain the single equation 
    	\begin{equation}
    		\label{BC2}
    		\tau_0^2 Y^4(\tau_0) -(Y^3(\tau_0))^2=-\12 t  \tau_0^4 D^2(\tau_0)
    	\end{equation}
    	The lattice is $Q^\vee = \Z^2$, rescaled by a factor of $\sqrt{2}$, as the underlying finite root system is $C_2$. Examining the lowest order terms we find that from the Ansatz, where the rescaling is taken care of by fractional $t$, 
    	\begin{equation}\nonumber
    		\tau_0(\s,\e|t) = \sum\limits_{\n\in \Z^2,\,i\in \N_0/2}e^{2\pi\sqrt{-1}\n\cdot\e}
    		t^{\12(\s+\n)^2+i}B_0(\s+\n)Z_{2i}(\s+\n)
    	\end{equation}
    	we find that $Z_1(\s)$ should vanish again. Looking at the lowest order gives us the equation 
    	\begin{gather*}
    		16 \left(2 \sigma _1+1\right){}^2 \left(\sigma _1-\sigma _2\right){}^2 \sigma _2^2 \left(\sigma _1+\sigma _2\right){}^2 \left(2 \sigma _2-1\right){}^2 \left(2 \sigma _2+1\right){}^2
    		\\ \times B_0(\s)B_0(\s+e_1)B_0(\s+e_2)B_0(\s-e_2)
    		\\
    		=-\left(2 \sigma _1+1\right){}^2 B_0(\s)^3 B_0(\s+e_1)
    	\end{gather*}
    	with the solution 
    	\begin{equation}\nonumber
    		B_0(\s)={ G(1\pm \sigma_1) G(1\pm \sigma_2) \over G(1\pm 2\sigma_1) G(1\pm 2\sigma_2) G(1\pm \sigma_1\pm \sigma_2)}
    	\end{equation}
    	and further we find
    	\begin{equation}\nonumber
    		Z_2(\s)=\frac{24 \sigma _1^4-32 \sigma _2^2 \sigma _1^2-4 \sigma _1^2+24 \sigma _2^4-4 \sigma _2^2+1}{2 \left(2 \sigma _1-1\right){}^2 \left(2 \sigma _1+1\right){}^2 \left(\sigma _1-\sigma _2\right){}^2 \left(\sigma _1+\sigma _2\right){}^2 \left(2 \sigma _2-1\right){}^2 \left(2 \sigma _2+1\right){}^2}
    	\end{equation}
    	
    	We do not have at present a clear 4D gauge theory interpretation for this case.

	    \color{black}
	    
	    \section{\texorpdfstring{$SU(2)^{n}$}{SU(2)n} linear quiver gauge theories} 
	    In this section we will be focusing on the case of linear $SU(2)^{\times n}$ quivers in the pure, non-conformal case. Our proposal is the following modification of the $SU(2)$ system
	    \begin{align}\label{quiver}
	    \tau_{0}^2 \partial_{\log{t_1}}\partial_{\log{t_n}}\log\tau_{0}&=-t_1^{1/4}\cdots t_n^{1/4}\tau_{1}^2 \\
	    \tau_{1}^2 \partial_{\log{t_1}}\partial_{\log{t_n}}\log\tau_{1}&=-t_1^{1/4}\cdots t_n^{1/4}\tau_{0}^2
	    \end{align}
		Notice that, in the case $n>2$, the system of equations explicitely describes only the dynamics associated to the irregular punctures moduli. As we will see in the following, the dependence on the moduli of the regular punctures is uniquely fixed by suitable asymptotic conditions on the solutions. These are obtained in the limiting cases of identity punctures leading to trivial monodromy or degenerating limits dividing the punctured Riemann sphere into disconnected components.
		
		 We solve to above equations in terms of the following generalised $SU(2)$ quiver Kiev Ansatz
		\begin{equation}\nonumber
		\tau_j(\{\sigma_k\},\{\eta_k\}|t_1,...,t_n)=\sum\limits_{\substack{n_1,...,n_k\in\frac{j}{2}+\Z\\ i_1,...,i_n\in\N_0}}\prod\limits_{l=1}^n\left(e^{2\pi\I\eta_l\cdot n_l}t_l^{(\sigma_l+n_l)^2+i_l}\right)B_0(\{\sigma_k+n_k\})Z_{i_1,...,i_n}(\{\sigma_k+n_k\})
		\end{equation}
	    where $Z_{0,...,0}\equiv 1$. Notice that the shift is simultaneous in all the lattices as there is no mixing. In the next subsection, by imposing appropriate asymptotic conditions, we will find that $B_0(\{\sigma_k+n_k\})=B^{\text{quiver}}(\{\sigma_k+n_k\})$ where
	    \begin{equation}\nonumber
	    B^{\text{quiver}}(\{\sigma_k\})={\prod\limits_{i=1}^{n-1} G(1+m_{i,i+1}+\sigma_i\pm\sigma_{i+1})G(1+m_{i,i+1}-\sigma_i\pm\sigma_{i+1})
	    	 \over 
	    \prod\limits_{k=1}^n  G(1\pm2\sigma_k) }
	    \end{equation}
		where $m_{i,j}\in\C$ are arbitrary bifundamental masses. We conjecture that these one-loop terms, along with the recursion relations arising from \eqref{quiver} and suitable additional constraints, lead to the identification
		\begin{equation}\nonumber
	    Z_{i_1,...,i_n}(\{\sigma_k\})=\sum_{\substack{(\vec{Y}_1,...,\vec{Y}_n)\\ |Y_{k,1}|+|Y_{k,2}|=i_k}} 
	    {
	    \prod_{i=1}^{n-1}Z_{bifund.}(\sigma_{i},\vec{Y}_{i},\sigma_{i+1},\vec{Y}_{i+1},m_{i,i+1})
	    \over
	    \prod_{i=1}^{n}Z_{bifund.}(\sigma_{i},\vec{Y}_{i},\sigma_{i},\vec{Y}_{i},0)
	    }
		\end{equation}
where $Z_{bifund.}$ is defined in Appendix B.

		\subsection{One-loop normalisation}
		Examining the $\prod_l t_l^{1+2\sigma_l}$ term in \eqref{quiver} gives, for general $n$,
		\begin{gather}\nonumber
		(1+2\sigma_1)(1+2\sigma_n)\sum\limits_{p_1,...,p_n\in\{0,1\}}{(-1)^{1+p_1+p_n}\over 2}B_0(\{\sigma_k+p_k\})B_0(\{\sigma_k+1-p_k\})\\ \label{quiver1loop} =-B_0\left(\{\sigma_k+\12\}\right)^2
		\end{gather}
		We prove this by induction. First, we need the auxiliary result that, for $k_1^+\leq k_1^-\leq k_2^+\leq k_2^-\leq...\leq k_l^+\leq k_l^-$, if we denote 
		\tiny
		\begin{align*}
		\s_+ &:= (\sigma_1,...,\sigma_{k_1^+-1},\sigma_{k_1^+}+1,...,\sigma_{k_1^-}+1,\sigma_{k_1^-+1},...,\sigma_{k_2^+-1},\sigma_{k_2^+}+1,...,\sigma_{k_2^-}+1,\sigma_{k_2^-+1},...,\sigma_{k_l^+-1},\sigma_{k_l^+}+1,...,\sigma_{k_l^-}+1,\sigma_{k_l^-+1},...,\sigma_n)\\
		\s_- &:= (\sigma_1+1,...,\sigma_{k_1^+-1}+1,\sigma_{k_1^+},...,\sigma_{k_1^-},\sigma_{k_1^-+1}+1,...,\sigma_{k_2^+-1}+1,\sigma_{k_2^+},...,\sigma_{k_2^-},\sigma_{k_2^-+1}+1,...,\sigma_{k_l^+-1}+1,\sigma_{k_l^+},...,\sigma_{k_l^-},\sigma_{k_l^-+1}+1,...,\sigma_n+1)
		\end{align*}
		\normalsize
		then it follows that
		\begin{align*}\nonumber
		{ 
			B^{\text{quiver}}(\s_+)B^{\text{quiver}}(\s_-)
			\over 
		   -B^{\text{quiver}}\left(\{\sigma_k+\12\}\right)^2
		} 
		&={\prod_{i=1}^{n-1}(1+\sigma_i+\sigma_{i+1}\pm m_{i,i+1})
			\over
		\prod_{i} (1+2\sigma_i)^2}\\
	&\times\prod_{q=1}^l 
	{ (\sigma_{k_q^+-1}-\sigma_{k_q^+}\pm m_{k_q^+-1,k_q^+})(\sigma_{k_q^-}-\sigma_{k_q^-+1}\pm m_{k_q^-,k_q^-+1})\over  (1+\sigma_{k_q^+-1}+\sigma_{k_q^+}\pm m_{k_q^+-1,k_1^+})(1+\sigma_{k_q^-}+\sigma_{k_q^-+1}\pm m_{k_q^-,k_q^-+1})
	}
		\end{align*}
		\normalsize
		Next we need another auxiliary result which we use to tame the summation in \eqref{quiver1loop}. Namely, for $n\geq 3$
		\begin{align*}
		\nonumber
		&{ -(\sigma_{1}-\sigma_{2}\pm m_{1,2})\over  (1+\sigma_{1}+\sigma_{2}\pm m_{1,2})
		}
	\Big(
	1
	+\sum_{i_1=2}^{n-1} 
	{ -(\sigma_{i_1}-\sigma_{i_1+1}\pm m_{i_1,i_1+1})\over  (1+\sigma_{i_1}+\sigma_{i_1+1}\pm m_{i_1,i_1+1})
	}
\\\nonumber
&+\sum_{i_1=2}^{n-1} 
{ -(\sigma_{i_1}-\sigma_{i_1+1}\pm m_{i_1,i_1+1})\over  (1+\sigma_{i_1}+\sigma_{i_1+1}\pm m_{i_1,i_1+1})
}
\sum_{i_2=i_1+1}^{n-1} { -(\sigma_{i_2}-\sigma_{i_2+1}\pm m_{i_2,i_2+1})\over  (1+\sigma_{i_2}+\sigma_{i_2+1}\pm m_{i_1,i_2+1})
}
\\\nonumber
&+\sum_{i_1=2}^{n-1} 
{ -(\sigma_{i_1}-\sigma_{i_1+1}\pm m_{i_1,i_1+1})\over  (1+\sigma_{i_1}+\sigma_{i_1+1}\pm m_{i_1,i_1+1})
}
\sum_{i_2=i_1+1}^{n-1} { -(\sigma_{i_2}-\sigma_{i_2+1}\pm m_{i_2,i_2+1})\over  (1+\sigma_{i_2}+\sigma_{i_2+1}\pm m_{i_1,i_2+1})
}
\sum_{i_3=i_2+1}^{n-1} { -(\sigma_{i_3}-\sigma_{i_3+1}\pm m_{i_3,i_3+1})\over  (1+\sigma_{i_3}+\sigma_{i_3+1}\pm m_{i_3,i_3+1})
}
\\\nonumber
    & + ... + \prod_{i=2}^{n-1}{ -(\sigma_{i}-\sigma_{i+1}\pm m_{i,i+1})\over  (1+\sigma_{i}+\sigma_{i+1}\pm m_{i,i+1})
    }\Big)
\\ &= { (\sigma_1-\sigma_2\pm m_{1,2})	(1+2\sigma_2)(1+2\sigma_n)\prod_{i=3}^{n-1} (1+2\sigma_i)^2 
	\over 
	\prod_{i=1}^{n-1}(1+\sigma_i+\sigma_{i+1}\pm m_{i,i+1})
}
	\end{align*}
    This also follows from induction starting from $n=3$, for which
    \begin{equation}\nonumber
    -1 +{(\sigma_2-\sigma_3\pm m_{2,3}) 
    \over (1+\sigma_2+\sigma_3\pm m_{2,3})
} 
= -{(1+2\sigma_2)(1+2\sigma_3) 
	\over (1+\sigma_2+\sigma_3\pm m_{2,3})
}  
    \end{equation}
    by iterating the identity we get
    \begin{align*}\nonumber
    &{ (\sigma_1-\sigma_2\pm m_{1,2})	(1+2\sigma_2)(1+2\sigma_{n-1})\prod_{i=3}^{n-2} (1+2\sigma_i)^2 
    	\over 
    	\prod_{i=1}^{n-2}(1+\sigma_i+\sigma_{i+1}\pm m_{i,i+1})
    }
\\ \nonumber
&+{ -(\sigma_{n-1}-\sigma_{n}\pm m_{n-1,n})\over  (1+\sigma_{n-1}+\sigma_{n}\pm m_{n-1,n})
}
\Big(
1
+\sum_{i_1=2}^{n-2} 
{ -(\sigma_{i_1}-\sigma_{i_1+1}\pm m_{i_1,i_1+1})\over  (1+\sigma_{i_1}+\sigma_{i_1+1}\pm m_{i_1,i_1+1})
}+...\Big)
\\\nonumber
 &={ (\sigma_1-\sigma_2\pm m_{1,2})	(1+2\sigma_2)(1+2\sigma_{n-1})\prod_{i=3}^{n-2} (1+2\sigma_i)^2 
 	\over 
 	\prod_{i=1}^{n-2}(1+\sigma_i+\sigma_{i+1}\pm m_{i,i+1})
 }\left(1+{ -(\sigma_{n-1}-\sigma_{n}\pm m_{n-1,n})\over  (1+\sigma_{n-1}+\sigma_{n}\pm m_{n-1,n})
}\right)
\\
&= { (\sigma_1-\sigma_2\pm m_{1,2})	(1+2\sigma_2)(1+2\sigma_n)\prod_{i=3}^{n-1} (1+2\sigma_i)^2 
	\over 
	\prod_{i=1}^{n-1}(1+\sigma_i+\sigma_{i+1}\pm m_{i,i+1})
}
    \end{align*}
    which is what we wanted. Using both results, \eqref{quiver1loop} becomes equivalent to the following identity after some reorganizing,
    \begin{align*}
    \nonumber
    &1+{ (\sigma_1-\sigma_2\pm m_{1,2})	(1+2\sigma_2)(1+2\sigma_n)\prod_{i=3}^{n-1} (1+2\sigma_i)^2 
    	\over 
    	\prod_{i=1}^{n-1}(1+\sigma_i+\sigma_{i+1}\pm m_{i,i+1})
    } 
	\\ \nonumber
	&+{ (\sigma_2-\sigma_3\pm m_{2,3})	(1+2\sigma_3)(1+2\sigma_n)\prod_{i=4}^{n-1} (1+2\sigma_i)^2 
		\over 
		\prod_{i=2}^{n-1}(1+\sigma_i+\sigma_{i+1}\pm m_{i,i+1})
	} 
\\ \nonumber
&+...+{ (\sigma_{n-2}-\sigma_{n-1}\pm m_{n-2,n-1})	(1+2\sigma_{n-2})(1+2\sigma_n) 
	\over 
	(1+\sigma_{n-2}+\sigma_{n-1}\pm m_{n-2,n-1})(1+\sigma_{n}+\sigma_{n-1}\pm m_{n,n-1})
} - { (\sigma_{n}-\sigma_{n-1}\pm m_{n,n-1})
\over 
(1+\sigma_{n}+\sigma_{n-1}\pm m_{n,n-1})
}
\\
&= { 	(1+2\sigma_1)(1+2\sigma_n)\prod_{i=2}^{n-1} (1+2\sigma_i)^2 
	\over 
	\prod_{i=1}^{n-1}(1+\sigma_i+\sigma_{i+1}\pm m_{i,i+1})
}
    \end{align*}
    
    \subsection{Instanton terms}
    The one instanton contributions can be obtained in a similar way. Let us focus on the detailed analysis of the simplest cases, starting with that of $n=2$. To obtain the instanton terms, we need to impose correct boundary conditions.
    \begin{itemize}
    	\item The first condition is the one corresponding to the identity puncture. This implies that when $m_{1,2}=0$, by setting $\sigma_1=\sigma_2$  
    	kills  off-diagonal terms in the expansion of the quiver tau function. Once we equate $t_1=t_2$, the solution has to equal that of pure SU(2) as the equation it solves is the same. \footnote{In the D-brane language, this condition can be seen most easily from the Hanany-Witten brane setup - this is the point at which the theory touches the Higgs branch and the gauge group gets broken down to the diagonal. There is no bifundamental, and the branes are fixed to move in unison, ignoring the intermediate NS5 brane.} A detailed proof of this is provided in Appendix B. 
    	\item The second condition is the one associated to the dividing degeneration limit. This is obtained by sending $m_{1,2}\to \infty$ while scaling $t_{1,2}\mapsto t_{1,2}/m_{1,2}^2$, inducing the factorization
        	\begin{equation}\nonumber
        	Z_{k_1,k_2}(\sigma_1,\sigma_2) \left({t_1 \over m_{1,2}}\right)^{k_1}\left({t_2 \over m_{1,2}}\right)^{k_2}\to Z_{k_1}(\sigma_1)Z_{k_2}(\sigma_2)t_1^{k_1} t_2^{k_2}
        	\end{equation}
    	This is consistent with \eqref{quiver}. Indeed, under the scaling itself, the RHS goes to zero as $1/m_{1,2}$, while, the LHS automatically vanishes if the tau function factorizes.
    \end{itemize}  
    The equation for $n=2$ yields two bilinear equations related one to the other by $\sigma_i\mapsto\sigma_i+1/2$ symmetry:
    \begin{align*}
	&\sum_{\substack{\n,\m\in \mathbb{Z}^2\\i_1,i_2\geq 0}}e^{2\pi\sqrt{-1}(\n+\m)\cdot\e}
	t_1^{(\sigma_1+n_1)^2+i_1+(\sigma_2+n_2)^2+i_2}t_2^{(\sigma_1+m_1)^2+j_1+(\sigma_2+m_2)^2+j_2}\notag \\
	&\times \left(i_1-i_2+(n_1-n_2)(n_1+n_2+2\sigma_1)\right)\left(j_1-j_2+(m_1-m_2)(m_1+m_2+2\sigma_2)\right) \notag \\
	&\times B_0(\sigma_1+n_1,\sigma_2+m_1)B_0(\sigma_1+n_2,\sigma_2+m_2)Z_{i_1,j_1}(\sigma_1+n_1,\sigma_2+m_1)Z_{i_2,j_2}(\sigma_1+n_2,\sigma_2+m_2) \notag \\
	&=-\sum_{\substack{\n,\m\in \mathbb{Z}^2\\i_1,i_2\geq 0}}e^{2\pi\sqrt{-1}(\n+\m)\cdot\e}
	t_1^{1+(\sigma_1+n_1)^2+\sigma_1+n_1+i_1+(\sigma_2+n_2)^2+\sigma_2+n_2+i_2}t_2^{1+(\sigma_1+m_1)^2+\sigma_1+m_1+j_1+(\sigma_2+m_2)^2+\sigma_2+m_2+j_2}\notag \\
	&\times B_0(\sigma_1+n_1+\12,\sigma_2+m_1+\12)B_0(\sigma_1+n_2+\12,\sigma_2+m_2+\12)\\
	&\times Z_{i_1,j_1}(\sigma_1+n_1+\12,\sigma_2+m_1+\12)Z_{i_2,j_2}(\sigma_1+n_2+\12,\sigma_2+m_2+\12)
	\end{align*}
	The $t_1 t_2^{1+2\sigma_2}$ term gives 
    \begin{equation}\nonumber
        Z_{1,0}(\sigma_1,\sigma_2+1)-Z_{1,0}(\sigma_1,\sigma_2)=-{2 B_0(\sigma_1+\12,\sigma_2+\12) B_0(\sigma_1-\12,\sigma_2+\12)
        \over
        (1+2\sigma_2)B_0(\sigma_1,\sigma_2)B_0(\sigma_1,\sigma_2+1)
    }=-{ 1+2\sigma_2\over 2\sigma_1^2}
    \end{equation}
    once we put $B_0=B^{\text{quiver}}$. The unique solution satisfying the boundary conditions is
    \begin{equation}\nonumber
        Z_{1,0}(\sigma_1,\sigma_2)={m_{1,2}^2+\sigma_1^2-\sigma_2^2\over 2\sigma_2^2}
    \end{equation}
    Note also the obvious symmetry $Z_{i,j}(\sigma_1,\sigma_2)=Z_{j,i}(\sigma_2,\sigma_1)$ which leads to $Z_{0,1}$. Finally, the $t_1^{1+2\sigma_1} t_2^{1+2\sigma_2}$ term gets us
     \begin{align*}
        &Z_{1,1}(\sigma_1,\sigma_2)-Z_{1,0}(\sigma_1,\sigma_2)Z_{0,1}(\sigma_1,\sigma_2)
        \\
        & =2{ B_0(\sigma_1+\12,\sigma_2-\12) B_0(\sigma_1-\12,\sigma_2+\12)+B_0(\sigma_1+\12,\sigma_2+\12) B_0(\sigma_1-\12,\sigma_2-\12)
        	\over
        	B_0(\sigma_1,\sigma_2)^2
        }\\
    	&=-{ m_{1,2}^2-\sigma_1^2-\sigma_2^2\over 4\sigma_1^2\sigma_2^2}
    \end{align*}
    giving the correct mixed 2-instanton term
    \begin{equation}\nonumber
        Z_{1,1}(\sigma_1,\sigma_2)={ m_{1,2}^2(m_{1,2}^2-1)+\sigma_1^2+\sigma_2^2-(\sigma_1^2-\sigma_2^2)^2
        	\over 4\sigma_1^2\sigma_2^2}
    \end{equation}
    In general, we find that the coefficients of the form $Z_{k,0}(\s)$ satisfy simple recurrence relations, while mixed terms are determined uniquely. We checked agreement with instanton counting up to $Z_{3,3}(\s)$. 
    \\
    \noindent
    For the $SU(2)^3$ quiver, the boundary conditions are the logical extension of the above. 
        \begin{itemize}
        	\item When $m_{1,2}=0$, setting $\sigma_1=\sigma_2$ has to kill all terms $Z_{k_1,k_2,k_3}$ with $k_1\neq k_2$. Further, if we put $t_{1}\mapsto t_{1}^{1/2}$ and $t_{2}\mapsto t_{1}^{1/2}$, then
        	\begin{equation}\nonumber
        	\left(Z_{k_1,k_2,k_3}(\sigma_1,\sigma_1,\sigma_3)|_{m_{1,2}=0}\right) t_1^{k_1}t_2^{k_2}t_3^{k_3}= Z_{k_1,k_3}(\sigma_1,\sigma_3) t^{k_1}t_3^{k_3}
        	\end{equation}
        	where on the RHS we have the partition functions of the $SU(2)^2$ quiver. Similar considerations apply if $m_{2,3}=0$ and $\sigma_2=\sigma_3$. Clearly, at this point previous considerations apply, and we are free to set the remaining mass to zero and reach pure $SU(2)$ again. 
        	\item We can decouple bifundamental hypermultiplets individually. For instance, by sending $m_{1,2}\to \infty$ while scaling $t_{1,2}\mapsto t_{1,2}/m_{1,2}^2$, we obtain the factorization
        	\begin{equation}\nonumber
        	Z_{k_1,k_2,k_3}(\sigma_1,\sigma_2,\sigma_3) \left({t_1 \over m_{1,2}}\right)^{k_1}\left({t_2 \over m_{1,2}}\right)^{k_2}t_3^{k_3}\to Z_{k_1}(\sigma_1)Z_{k_2,k_3}(\sigma_2,\sigma_3)t_1^{k_1} t_2^{k_2}t_3^{k_3}
        	\end{equation}
        	Again, under any factorization, the LHS of \eqref{quiver} vanishes. 
        \end{itemize}  
    	Noting the symmetry "$1\leftrightarrow3$", we list in the Appendix \ref{E} the first several equations and their solutions, which are the same as those obtained by instanton counting as in \ref{appendix:instanton_counting}. Indeed the calculations are analogous to the ones already presented for the quiver with two nodes. 
    	
		\appendix
		
		\color{black}
	    \section{The operators \texorpdfstring{$Y^n$}{Yn}}
	    \label{appendix:Y}
		In the main text we made frequent use of the operators recursively defined as
		\begin{align}\label{Yoperators}
		Y^1(f)&=f 
		\\
		Y^2(f)&=D^2(f)
		\\
		Y^n(f)&=(Y^{n-2}(f))^{-1}D^2(Y^{n-1}(f)),\quad n\geq 2
		\end{align}
		Besides being shorthands, their utility consists in the property that they act on a formal power series as 
		\begin{equation}\label{Yn}
		Y^n(\sum_i y_i t^{x_i})=\sum_{i_1,...,i_n}\prod_{j=1}^n y_{i_j} t^{x_{i_j}}\prod\limits_{k<l=1}^n (x_{i_k}-x_{i_l})^2 
		\end{equation}
		It is easy to calculate by hand that is true for $n=2$. For $n>2$, suppose we only have a total of $n$ distinct exponents $x_{i_n}$ in the sum $\sum_i y_i t^{x_{i}}$. Then if we assume  \eqref{Yn} holds for $n-1$, $Y^{n-1}$ will by assumption have only ${}_{n}C_{n-1}=n$ terms, differring by one pair of indices. Without loss of generality, consider two terms with the last index labelled differently. That is, let $x_{\text{same}}:=\sum_{k=1}^{n-2}x_{i_k}$ and $x_{i_{n-1}}\neq x_{i_{n}}$. By assumption of the induction we have that in applying $Y^{n-1}$ to $\sum_i y_i t^{x_i}$ we end up with two different terms
		\begin{equation}\nonumber
		   c_1\, t^{x_{\text{same}}+x_{n-1}},c_2\, t^{x_{\text{same}}+x_{n}} \in Y^{n-1}(\sum_i y_i t^{x_i})
		\end{equation}
		where the coefficients $c_{1,2}$ are
		\begin{align*}
		    c_1 &= y_{i_{n-1}}\prod_{k=1}^{n-2}(x_{i_k}-x_{i_{n-1}})^2\cdot c_{\text{same}}\\
		    c_2 &= y_{i_{n}}\prod_{k=1}^{n-2}(x_{i_k}-x_{i_{n}})^2\cdot c_{\text{same}}
	    \end{align*}
	    where 
	    \begin{equation}\nonumber
		    c_{\text{same}} = \prod_{j=1}^{n-2} y_{i_j} \prod\limits_{k<l=1}^{n-2} (x_{i_k}-x_{i_l})^2
		\end{equation}
		comes from the exponents purely inside $x_{\text{same}}$. Considering the application of $D^2$ to just those two terms we obtain
		\begin{equation}\nonumber
		    D^2\left(c_1c_{\text{same}}t^{x_{\text{same}}+x_{i_{n-1}}} + c_2c_{\text{same}}t^{x_{\text{same}}+x_{i_{n}}}\right) = c_1 c_2 c_{\text{same}}^2 (x_{i_{n-1}}-x_{i_n})^2 t^{2 x_{\text{same}}+x_{i_{n-1}}+x_{i_{n}}}
		\end{equation}
		which we can write as
		\begin{gather*}
		    \prod_{j=1}^{n-2} y_{i_j} t^{x_{i_j}}\prod\limits_{k<l=1}^{n-2} (x_{i_k}-x_{i_l})^2  \cdot \prod_{j=1}^n y_{i_j} t^{x_{i_j}}\prod\limits_{k<l=1}^n (x_{i_k}-x_{i_l})^2 
		    \\
		    = \prod_{j=1}^{n-2} y_{i_j} t^{x_{i_j}}\prod\limits_{k<l=1}^{n-2} (x_{i_k}-x_{i_l})^2  \cdot Y^n(\sum_i y_i t^{x_i})
		\end{gather*}
		and in the last line we've used \eqref{Yn} in this particular case of only $n$ distinct exponents. The term in front explicitly lacks the pair of indices we chose. Now if we extend by linearly to all such pairs, we see that we have shown that $D^2(Y^{n-1}(f)) =Y^{n-2}(f)Y^n(f)$. But now we are done since we can reduce the general case of more than $n$ distinct exponents to this one by multilinearity. 
		
		\color{black}
		\section{Instanton counting}
		\label{appendix:instanton_counting}  \color{black}
		Here we review the Nekrasov partition function of interest for this work 
		\cite{Flume:2002az,Bruzzo:2002xf}. Given two partitions $Y_1=(k_1\geq k_2\geq...\geq k_l>0)$, $Y_2=(\tilde{k}_1\geq \tilde{k}_2\geq...\geq \tilde{k}_{\tilde{l}}>0)$ and a cell $c=(i,j)\in Y_1$ we define the auxiliary functions 
		\begin{gather*}
		\phi(a,c)=a +\ea(i-1)+\eb(j-1) \\
		\xi(a,b,Y_1,Y_2,c)=a-b + \ea(\text{leg}(c,Y_1)+1)-\eb(\text{arm}(c,Y_2))
		\end{gather*}  
		the last of which, a deformed hook length, uses $\text{arm}(c,Y)=k_i-j$, $\text{leg}(c,Y^t)=k_j-i$, and finally set
		\begin{equation}\nonumber
        E(a,b,Y_1,Y_2)=\prod_{c\in Y_1}\xi(a,b,Y_1,Y_2,c)\left(\ea+\eb-\xi(a,b,Y_1,Y_2,c)\right)	\end{equation} 
		\subsection{The classical gauge groups \texorpdfstring{$SU(n)$}{SU(n)}, \texorpdfstring{$SO(2n+\chi)$}{SO(2n+chi)}, \texorpdfstring{$Sp(n)$}{Sp(n)}}
		In the following we consider $n$ partitions $(Y_1,...,Y_n)=\vec{Y}$ with the total number of boxes $k$. Then the equivariant volume of the $k$ instanton moduli space for $U(n)$ is given by
		\begin{gather}\label{AnCountDiagram}
	Z^{SU(n)}(\vec Y ) =  \prod_{i,j=1}^{n}\left(E(\sigma_i,\sigma_j,Y_i,Y_j)\right)^{-1}\\
Z_k^{SU(n)}=\sum_{|\vec{Y}|=k}Z^{SU(n)}(\vec Y ) \label{AnCount}
		\end{gather} 
		and $SU(n)$ is obtained by restricting to the $\sum_k\sigma_k=0$ slice. 
		For the orthogonal groups, if $\chi\in\{0,1\}$,
		\begin{equation}\label{so}
		Z_k^{SO(2n+\chi)}=\sum_{|\vec{Y}|=k}\prod_{i=1}^{n}
		{
			\prod_{c\in Y_i}4(\phi(\sigma_i,c))^{2\chi}(4\phi(\sigma_i,c)^2-1)^2
			\over
		    \prod_{j=1}^n E(\sigma_i-\sigma_j,Y_i,Y_j)^2 E(-\sigma_i-\sigma_j,Y_i^t,Y_j) E(\sigma_i+\sigma_j,Y_i,Y_j^t)
		}
		\end{equation}
		The combinatorial expressions for $Sp(n)$ are more involved. Namely, one has to multiply \eqref{so} by extra factors depending just on the $\Omega$-background parameters\footnote{In the brane realisation of instanton counting these are usually dubbed \textit{fractional instantons}, stuck at the orientifold plane \cite{Fucito2004}.}. A combinatorial solution is proposed in \cite{Keller:2011ek}, the issue can also be approached using Jeffrey-Kirwan residues as in \cite{Nakamura2015}. For the self-dual background we can simplify the latter procedure via a $\epsilon_{1,2}=\pm1\mp i0$ prescription which renders the pole structure easier to handle. Namely, in that case we need to define two-indexed functions $Z_{2k,l}^{Sp(n)}$ such that 
		\begin{gather}\label{spcounting}
		Z_{2k,0}^{Sp(n)}=\sum_{|\vec{Y}|=k}\prod_{i=1}^{n}
		\left(
		\prod_{c\in Y_i}	4(\phi(\sigma_i,c))^{2}(4\phi(\sigma_i,c)^2-1)^2 \right)^{-1}
			\\ \nonumber
			\times \left(
			\prod_{j=1}^n E(\sigma_i-\sigma_j,Y_i,Y_j)^2 E(-\sigma_i-\sigma_j,Y_i^t,Y_j) E(\sigma_i+\sigma_j,Y_i,Y_j^t)
		\right)^{-1}
		\end{gather}
		and then the fractional instanton contributions are given as 
		\begin{align*}
		Z_{2k,1}^{Sp(n)}&=\frac{1}{2}Z_{2k,0}^{Sp(n)}\prod_{i=1}^{n}\frac{1}{-\sigma_i^2}\sum_{|\vec{Y}|=k}\prod_{c\in Y_i} { \phi(\sigma_i,c)^4 \over (\phi(\sigma_i,c)^2-1)^2}
		\\
		Z_{2k,2}^{Sp(n)}&=\frac{1}{8}Z_{2k,0}^{Sp(n)}\prod_{i=1}^{n}\frac{1}{(\sigma_i^2-1/4)^2}\sum_{|\vec{Y}|=k}\prod_{c\in Y_i} { (\phi(\sigma_i,c)^2-1/4)^2 \over (\phi(\sigma_i,c)^2-9/4)^2}
		\\
		Z_{2k,3}^{Sp(n)}&=\frac{1}{144}Z_{2k,0}^{Sp(n)}\prod_{i=1}^{n}\frac{1}{(-\sigma_i^2)(\sigma_i^2-1/4)^2}\sum_{|\vec{Y}|=k}\prod_{c\in Y_i} { \phi(\sigma_i,c)^4  (\phi(\sigma_i,c)^2-1/4)^2 \over (\phi(\sigma_i,c)^2-1)^2 (\phi(\sigma_i,c)^2-9/4)^2}
		\\
		&+\frac{1}{72}Z_{2k,0}^{Sp(n)}\prod_{i=1}^{n}\frac{1}{(-\sigma_i^2)(\sigma_i^2-1)^2}\sum_{|\vec{Y}|=k}\prod_{c\in Y_i} {  (\phi(\sigma_i,c)^2-1)^2 \over (\phi(\sigma_i,c)^2-4)^2 }
		\end{align*}
		where the summands in the last expressions are due to $V=T^{1/2}+1+T^{-1/2}$, $V_2=T+1+T^{-1}$ and $V=T^{1}+1+T^{-1}$, $V_2=T^2+1+T^{-2}$ contributions to be put in the character $(4.16)$ of \cite{Fucito2004}, which can be continued further easily. This finally enables one to compute 
		\begin{equation}\nonumber
		Z^{Sp(n)}_k =\sum_{2m+l=k}Z^{Sp(n)}_{2m,l}
		\end{equation} 
		and it agrees with appendix B of \cite{Marino:2004cn}. Further, we can add fundamental matter by adding a factor of
		\begin{equation}\nonumber
		\prod_{i=1}^{N_f}\prod_{j=1}^{n}\prod_{c\in Y_j}\left(\phi(\sigma_j,c)^2-m_i^2\right)
		\end{equation}
		in the numerators.
		
		\color{black}
		
		\subsection{\texorpdfstring{$SU(2)$}{SU(2)} with fundamental matter}
		Given the partitions $Y_{1,2},W_{1,2}$ we can define 
		\begin{align*}
		&Z_{bifund.}(a_1,a_2,Y_1,Y_2,b_1,b_2,W_1,W_2,m)=\\
		&\prod_{i,j=1}^2 \prod_{c\in Y_i}\left(\xi(a_i-b_j,Y_i,W_j,c)-m\right)
		\prod_{c\in W_j}\left(\ea+\eb-\xi(b_j-a_i,W_j,Y_i,c)-m\right)
		\end{align*}  
		Further we define, 
		\begin{align*}
		Z_{adj.}(a_1,a_2,Y_1,Y_2)&=Z_{bifund.}(a_1,a_2,Y_1,Y_2,a_1,a_2,Y_1,Y_2,0)^{-1}
		\\Z_{fund.}(a_1,a_2,Y_1,Y_2,m)&=\prod_{i,j=1}^2 \prod_{c\in Y_i}\left(\phi(a_i,c)+m\right)
		\end{align*}
		Then for $SU(2)$ with $N_f$ fundamental flavors we have
		\begin{equation}
		Z_{k}(\sigma)= \sum_{|Y_1|+|Y_2|=k} {\prod_{i=1}^{N_f} Z_{fund.}(\sigma,-\sigma,Y_{1},Y_{2},m_{i}) \over  Z_{adj.}(\sigma,-\sigma,Y_{1},Y_{2}) }
		\end{equation}
		To obtain $U(2)$, replace $(\sigma,-\sigma)$ with $(\sigma_1,\sigma_2)$ in the above. 
		
		\color{black}
		
		\subsection{Linear \texorpdfstring{$SU(2)^n$}{SU(2)n} quivers}
		Here we have to consider $n$ pairs of Young diagrams $\{Y_{i,1},Y_{i,2}\}_{i=1}^{n}$, with the total box number of each pair fixed to $|Y_{i,1}|+|Y_{i,2}|=k_i\geq 0$. Then the $(k_1,...,k_n)$-instanton partition function is given by
		\begin{equation}\nonumber
		Z_{k_1,...,k_n}(\sigma_1,...,\sigma_n)= \sum_{ |Y_{i,1}|+|Y_{i,2}|=k_i}
		{\prod_{i=1}^{n-1} Z_{bifund.}(\sigma_i,-\sigma_i,Y_{i+1,1},Y_{i+1,2},\sigma_{i+1},-\sigma_{i+1},Y_{i+1,1},Y_{i+1,2},m_{i,i+1}) \over \prod_{i=1}^n Z_{adj.}(\sigma_i,-\sigma_i,Y_{i,1},Y_{i,2}) }
		\end{equation}  
		In the main text, several claims about this function were made, and among them, for $n=2$,
		\begin{align}\label{Zbifundcollapse}
		Z_{k_1,k_2}(\sigma,\sigma)\lvert_{m_{1,2}=0}&= \sum_{\substack{|Y_{1}|+|Y_{2}|=k_1 \\ |W_{1}|+|W_{2}|=k_2}} { Z_{bifund.}(\sigma,-\sigma,Y_{1},Y_{2},\sigma,-\sigma,W_{1},W_{2},0) \over  Z_{adj.}(\sigma,-\sigma,Y_{1},Y_{2})Z_{adj.}(\sigma,-\sigma,W_{1},W_{2}) } \\
	\nonumber	&=\delta_{k_1,k_2} Z_{k_1}^{SU(2)}(\sigma)
		\end{align}  
		This can be seen to follow from 
		\begin{equation}\label{Zbifundcollapse2}
		    Z_{bifund.}(\sigma,-\sigma,Y_{1},Y_{2},\sigma,-\sigma,W_{1},W_{2},0) =\delta_{Y_1,W_1}\delta_{Y_2,W_2}Z_{adj.}(\sigma,-\sigma,Y_{1},Y_{2})
		\end{equation}
		which we show to be true in the self-dual case. In the general $\Omega$-background, the equality \eqref{Zbifundcollapse2} is not true. Nevertheless, the full sum \eqref{Zbifundcollapse} seems to be true universally, although this follows from more complicated cancellations. 
		In any case, we are interested only in the self-dual case $\epsilon_1=1$, $\epsilon_2=-1$. With that in mind, we write
		\begin{align*}
		&Z_{bifund.}(\sigma,-\sigma,Y_1,Y_2,\sigma,-\sigma,W_1,W_2,0)=\\
		&\prod_{i=1}^2 \prod_{c\in Y_i}\xi((-1)^{1-i}2\sigma,Y_i,W_{1-i},c)
		\prod_{c\in W_i}\left(-\xi((-1)^{1-i}2\sigma,W_i,Y_{1-i},c)\right) \\
		&\prod_{i=1}^2 \prod_{c\in Y_i}\xi(0,Y_i,W_i,c)
		\prod_{c\in W_i}\left(-\xi(0,W_i,Y_i,c)\right)
		\end{align*}  
		We prove the last line vanishes unless the Young diagrams are equal as $Y_1=W_1$, $Y_2=W_2$. Focusing on just one factor, 
		\begin{equation}\nonumber
		    \xi(0,Y,W,c) = \text{leg}(c,Y) + \text{arm}(c,W) + 1
		\end{equation}
		consider row diagrams $Y=(1^{l_1})$ and $W=(1^{l_2})$ with $l_1\neq\tilde l_2$. WLOG, assume $l_1>\tilde l_2$. In this case for $c=(l_1,1)$ we have
		\begin{equation}\nonumber
		    \begin{rcases}
		    \text{leg}(c,Y) = (1^{l_1})^t_1 - l_1 = l_1-l_1 = 0\\
		    \text{arm}(c,W) = 0 - 1 = -1
		    \end{rcases}
		    \Rightarrow \xi(0,Y,W,c) = 0-1 + 1 = 0 \end{equation}
    However, adding any amount of rows to any of the diagrams after the $j=1$ one doesn't change this calculation. The other case is $l_1=\tilde l_2$. Then, for the same cell, $\text{arm}(c,W) = 1 - 1 = 0$, and $\xi(0,Y,W,c) \neq 0$. In fact, both the arm and the leg lengths have to be positive indefinite, since the cell $c$ is contained within the diagrams $Y$ and $W$, so $\xi(0,Y,W,c) \neq 0$ for the whole row, i.e. $i\in[1,l_1]$, $j=1$. Next, we add rows to both diagrams, and we are in the same situation as before. If the rows are of equal length, the cell will be contained in both diagrams and the relative hook length will never vanish. Otherwise, if the $j$'th row is the first one of unequal length, with lengths $l_1>\tilde l_2$, say, then the relative hook length of the cell $c=(l_1,j)$ vanishes by an analogous calculation. If all rows are equal, the diagrams are obviously the same, and there is no vanishing factor. Along with the trivial equality
		\begin{equation}\nonumber
		    Z_{bifund.}(\sigma,-\sigma,Y_{1},Y_{2},\sigma,-\sigma,Y_{1},Y_{2},0) =Z_{adj.}(\sigma,-\sigma,Y_{1},Y_{2})
		\end{equation}
		this proves our claim.

		\subsection{Universal one instanton formula} 
		It was found in \cite{PhysRevD.16.2967} that an instanton of topological charge 1 may be constructed by means of an $\mathfrak{sl}_2$ triple corresponding to a long root. This was used to calculate the 1-instanton corrections to the Seiberg-Witten curve \cite{Ito1996}. Besides this embedding in the internal degrees of freedom, the instanton has a $\C^2$ of moduli specifying its position, therefore the holomorphic functions on this product space is a $U(1)_{\epsilon_1}\times U(1)_{\epsilon_2}\times W$-module.  It is precisely its character that the $5$ dimensional uplift of the theory will be calculating, and the $4$ dimensional formula may be seen as its "Weyl dimension" analogue, and in 
		\begin{gather}\nonumber
		\Lambda^{2h^\vee}Z_1 = \frac{-1}{\epsilon_1\epsilon_2}\sum_{\b\text{ long}}\frac{1}{(\epsilon_1+\epsilon_2+\b\cdot\A)(\b\cdot\A)\prod\limits_{\a\cdot\b^\vee=1}(\a\cdot\A)}\\ \label{universal1instanton}
  \mapsto \sum_{\b\text{ long}}\frac{1}{(\b\cdot\s)^2\prod\limits_{\a\cdot\b^\vee=1}(\a\cdot\s)}
		\end{gather}
		rewritten in $\epsilon$-units \cite{Keller:2011ek}. Comparisons with ADHM calculations \cite{Nekrasov2004,Marino:2004cn,Fucito2004} tend to reveal some sign differences, e.g. $Z_1^{B_n,C_n}|_{\eqref{universal1instanton}}=  -Z_1^{B_n,C_n}|_{\text{ADHM}}$, which is why the rescalable instanton counting factor of $\Lambda$ ought to be kept in mind. 
		
			\color{black}
		
		\subsection{Blowup equations}
		
		The instanton moduli spaces for exceptional groups lack an ADHM description since their fundamental representation is different from their defining one.
		Thus to compute higher instanton terms one cannot resort to the usual localisation techniques. An alternative approach, besides the one discussed in this paper, is by blowup equations, although these do not give compact expressions such as \eqref{universal1instanton}. Generalizing the 4d expression in \cite{Nakajima:2003pg} to general gauge groups as was done for 5d in \cite{Keller:2011ek}, except noting that in 4d the partition function with flux on the exceptional divisor vanishes -- so $\hat Z_{d=0}=Z$ but $ \hat Z_{d\geq 1}=0$ -- we obtain 
		\begin{gather}\nonumber
		Z_n(\epsilon_1,\epsilon_2,\s)=\frac{1}{n^2 \epsilon_{1}\epsilon_2}\sum\limits_{\substack{\12 \m^2+i_1+i_2=n\\ \m\in Q^\vee,\, i_{1,2}<n}}\frac{\left(\epsilon_1 i_1 + (\epsilon_1+\epsilon_2)i_2+\m\cdot\s+\frac{1}{2}\m^2(2\epsilon_1+\epsilon_2)\right)}{L(\epsilon_1,\epsilon_1+\epsilon_2,\s,\m)}\\
		\nonumber
		\left(\epsilon_1 i_1 + (\epsilon_1+\epsilon_2)(i_2-n)+\m\cdot\s+\frac{1}{2}\m^2(2\epsilon_1+\epsilon_2)\right)\\
		\label{blowup}
  Z_{i_1}(\epsilon_1,\epsilon_2,\s+ \epsilon_1 \m)Z_{i_2}(-\epsilon_2,\epsilon_1+\epsilon_2,\s+ (\epsilon_1+\epsilon_2) \m)
		\end{gather}
		starting from $Z_0(\epsilon_1,\epsilon_2,\s)=1$, where $L(\epsilon_1,\epsilon_2,\s,\m):=\prod_{\a\in R}\ell(\epsilon_1,\epsilon_2,\s,\m,\a)$ and
		\begin{equation}\label{ell_definition}
		\ell(\epsilon_1,\epsilon_2,\s,\m,\a) = \begin{cases}
		\prod\limits_{\substack{i,j\geq0\\ i+j\leq -\m\cdot\a-1}}(-i\epsilon_1-j\epsilon_2 +\m\cdot\s), & \text{if } \m\cdot\a<0,\\
		\prod\limits_{\substack{i,j\geq0\\ i+j\leq \m\cdot\a-2}}((i+1)\epsilon_1+(j+1)\epsilon_2 +\m\cdot\s), & \text{if } \m\cdot\a>1,\\
		1 & \text{otherwise.}
		\end{cases}
		\end{equation}	
		Since we are interested in the self-dual background, we see that naively taking it leads to some singular terms in the summands due to the NS limit getting involved, so care must be taken to first preform the summation and then to take the limit. In particular, one can take $\ea=1+\delta$, $\eb=-1$ and then safely send $\delta\to0$ in the final expression.

		\section{The deep Coulomb approximation, and Lie algebraic theta function identities}
		\label{appendix:deepcoulomb}
		
		Many times in the main text it was useful to check equivariant volume calculations by performing a large $\s$ limit and studying the leading term, especially when we had either nothing to compare with at all or the blowup technique turned out to be too computationally costly. To describe this, we first define the auxiliary function
		\begin{equation}\nonumber
		Z^{dC}_n(\ea,\eb,\s)= \frac{1}{n!} \left( \frac{-1}{\ea\eb}Z_1(1,-1,\s)\right)^n
		\end{equation}
		which we call the deep Coulomb instanton function. This is going to be the large $\s$ limit, in which the only contribution to the equivariant volume given by the $n$-fold symmetric product of one instantons terms and more complicated configurations of Young diagrams are not involved. We claim that
		\begin{equation}\label{deep_limit}
		Z^{dC}_n(\ea,\eb,\s)= \lim_{\gamma\to0} \gamma^{-(2 h^\vee-2) n} Z_n(\ea,\eb,\s/\gamma)
		\end{equation}
		In other words, this is the leading part under the scaling 
		\begin{equation}\nonumber
	    Z_n(\ea,\eb,\s/\gamma)=	 \gamma^{(2 h^\vee-2) n} Z^{dC}_n(\ea,\eb,\s)+\mathcal{O}(\gamma^{(2 h^\vee-2) n+1}) 
		\end{equation}
		Clearly, $Z^{dC}_1(1,-1,\s)=Z_1(1,-1,\s)$. In this case, the subleading terms are absent, and that the scaling is correct can be seen from the universal 1-instanton term, since there are $2 h^\vee -2$ terms in the denominator of \eqref{universal1instanton}. In the refined case $\ea+\eb\neq 0$, this is no longer true, but it's immediate that \eqref{deep_limit} is true. 
		
		We prove the rest by induction, using \eqref{blowup}. In the following we split the sum into one with $\m=0$ and the rest. For $n>1$ write 
		\begin{align*}
		&\lim_{\gamma\to0}{ Z_n(\ea,\eb,\s/\gamma) \over Z^{dC}_n(\ea,\eb,\s/\gamma)}\\
		&=
		\lim_{\gamma\to0}\frac{1}{n^2 \ea\eb}\Big(
		\sum\limits_{\substack{i_1+i_2=n\\ \, i_{1,2}<n}} (\ea i_1 + (\ea+\eb) i_2)(-\eb i_2) { Z_{i_1}(\ea,\eb,\s/\gamma)  Z_{i_2}(-\eb,\ea+\eb,\s/\gamma) \over Z^{dC}_n(\ea,\eb,\s/\gamma)}
		\\
		&+\sum\limits_{\substack{\12 \m^2+i_1+i_2=n\\ \vb 0 \neq\m\in Q^\vee,\, i_{1,2}<n}}\frac{\left(\epsilon_1 i_1 + (\epsilon_1+\epsilon_2)i_2+\m\cdot\s/\gamma+\frac{1}{2}\m^2(2\epsilon_1+\epsilon_2)\right)}{L(\epsilon_1,\epsilon_1+\epsilon_2,\s/\gamma,\m)}\\
		&
		\left(\epsilon_1 i_1 + (\epsilon_1+\epsilon_2)(i_2-n)+\m\cdot\s/\gamma+\frac{1}{2}\m^2(2\epsilon_1+\epsilon_2)\right)\\
		&{ 
		Z_{i_1}(\epsilon_1,\epsilon_2,\s/\gamma+ \epsilon_1 \m)Z_{i_2}(-\epsilon_2,\epsilon_1+\epsilon_2,\s/\gamma+ (\epsilon_1+\epsilon_2) \m)
		\over 
		Z^{dC}_n(\ea,\eb,\s/\gamma)
		}\Big)
		\end{align*}
		Assume \eqref{deep_limit} is true for all $n'< n$. Then the first sum becomes
		\begin{align*}
		&\frac{1}{n^2 \ea\eb}
		\sum\limits_{\substack{i_1+i_2=n\\ \, i_{1,2}<n}} (\ea i_1 + (\ea+\eb) i_2)(-\eb i_2) {(i_1+i_2)!\over i_1! i_2!} \left(-\frac{\ea}{\ea+\eb}\right)^{i_2}
		\\
		&=\frac{1}{n^2 \ea\eb}
		\sum\limits_{i=1}^{n-1} \left( -(n-i)^2 \ea \eb +n(n-i)(\ea+\eb)\right) \binom{n}{i} \left(-\frac{\ea}{\ea+\eb}\right)^{i}
		\\
		&= 1-\frac{\eb^{n-2}\left((n-1)\ea+n\eb\right)}{n (\ea+\eb)^{n-1}}+\frac{(n-1)\eb^{n-2}}{n (\ea+\eb)^{n-2}}=1-\frac{\eb^{n-1}}{n (\ea+\eb)^{n-1}}
		\end{align*}
		In the second sum, all terms except the ones proportional to $\s$ can be ignored, as well as all the shifts. It becomes
		\begin{align*}
	    &\frac{1}{n^2 \ea\eb}\sum\limits_{\substack{\12 \m^2+i_1+i_2=n\\ \vb 0 \neq\m\in Q^\vee,\, i_{1,2}<n}}\left(\m\cdot\s\right)^2\lim_{\gamma\to0} { 
			Z^{dC}_{i_1}(\epsilon_1,\epsilon_2,\s/\gamma)Z^{dC}_{i_2}(-\epsilon_2,\epsilon_1+\epsilon_2,\s/\gamma)
			\over 
			\gamma^2 L(\epsilon_1,\epsilon_1+\epsilon_2,\s/\gamma,\m) Z^{dC}_n(\ea,\eb,\s/\gamma)
		}
		\\
	    &=\frac{1}{n^2 \ea\eb(-Z_1(1,-1,\s))}\sum\limits_{\substack{\12 \m^2+i_1+i_2=n\\ \vb 0 \neq\m\in Q^\vee,\, i_{1,2}<n}}\left(\m\cdot\s\right)^2 \ea\eb  {(i_1+i_2+1)!\over i_1! i_2!} \left(-\frac{\ea}{\ea+\eb}\right)^{i_2}
	    \\
	    &
	    \lim_{\gamma\to0} { 
	    	1
	    	\over 
	        \gamma^{2 h^\vee} L(\epsilon_1,\epsilon_1+\epsilon_2,\s/\gamma,\m)
	    }
		\end{align*}
		The limit is dependent on the incidence properties of the colattice vector with respect to the roots. Note that if $\m^2=2$, $|\{\a\in R|\a\cdot\m=-1\}|=2 h^\vee-4$ which contribute 1 term each in \eqref{ell_definition} and $|\{\a\in R|\a\cdot\m=\pm 2\}|=2$ which contribute 2 terms each. These vectors are the short coroots. For any other nonzero vector, the number of terms is greater than $2h^\vee$. Explicitly,
		\begin{equation}\nonumber
		\lim_{\gamma\to0}\gamma^{2 h^\vee} L(\epsilon_1,\epsilon_1+\epsilon_2,\s/\gamma,\m)=\begin{cases}
		0, & \text{if } \m=\vb 0, \\
		(\m\cdot\s)^4 \prod\limits_{\a\cdot \m=1}(\a\cdot\s), & \text{if } \m\in R^\vee_{\text{short}}, \\
		\infty & \text{otherwise.}
		\end{cases}
		\end{equation}
		Therefore, the sum becomes
		\begin{gather*}
		\frac{1}{n^2(-Z_1(1,-1,\s))}\sum\limits_{\m\in R^\vee_{\text{short}}}{1\over (\m\cdot\s)^2 \prod\limits_{\a\cdot \m=1}(\a\cdot\s)}\sum\limits_{\substack{i_1+i_2=n-1\\ i_{1,2}<n}}  {(i_1+i_2+1)!\over i_1! i_2!} \left(-\frac{\ea}{\ea+\eb}\right)^{i_2}
		\\
		=\frac{\eb^{n-1}}{n (\ea+\eb)^{n-1}}
		\end{gather*}
		using \eqref{universal1instanton}. This proves the claim. Moreover, we can prove quite easily that the full expansion around $\gamma=0$ has to be even,
		\begin{equation}\nonumber
        { Z_n(\ea,\eb,\s/\gamma) \over Z^{dC}_n(\ea,\eb,\s/\gamma)}=1+\sum_{k>1} z_k(\s) \gamma^{2k},
		\end{equation} 
		since this expression is holomorphic and Weyl invariant, and the Weyl group is generated by reflections. This can also be inferred from the blowup formula by a more involved, but straightforward calculation.
	%
		
		\section{The algebras \texorpdfstring{$D_2$}{D2} and \texorpdfstring{$D_3$}{D3}}
		\label{appendix:D23}
		\color{black}
		\subsection{\texorpdfstring{$D_2=A_1\times A_1$}{D2=A1xA1}}\label{sec:D2}
		An interesting thing about \eqref{Drec} is that it generalizes to lower $n$. Explicitly, under the isomorphism 
		\begin{align}\label{D2isom}
		\sigma_1^{[D_2]}&=(\sigma_1+\sigma_2)^{[A_1\times A_1]}\\
		\sigma_2^{[D_2]}&=(\sigma_1-\sigma_2)^{[A_1\times A_1]}
		\end{align}
		we find 
		\begin{equation}\nonumber
		Z_1(\s)^{[D_2]}=Z_1(\sigma_1)^{[A_1]}+Z_1(\sigma_2)^{[A_1]}
		\end{equation}
		\begin{equation}\nonumber
		Z_2(\s)^{[D_2]}=Z_2(\sigma_1)^{[A_1]}+2Z_1(\sigma_1)^{[A_1]}Z_1(\sigma_2)^{[A_1]}+Z_2(\sigma_2)^{[A_1]}
		\end{equation}
		That this continues can be confirmed by the recursion relations or instanton counting. Together with \eqref{D2isom} which splits $Q_{D_2}\cong Q_{A_1}\times Q_{A_1}$ this suggests 
		\begin{equation}\nonumber
		\tau_{0}^{[D_2]}=(\tau_{0}^{[A_1]})^2, \quad \tau_{1}^{[D_2]}=(\tau_{1}^{[A_1]})^2 
		\end{equation}
		Since $D^2(\tau_{0}^{[A_1]})=-t^{1/2}(\tau_{1}^{[A_1]})^2$ and $D^2(\tau_{1}^{[A_1]})=-t^{1/2}(\tau_{0}^{[A_1]})^2$, 
		\begin{equation}\nonumber
		D^2((\tau_{0}^{[A_1]})^2)=2(\tau_{0}^{[A_1]})^2 D^2((\tau_{0}^{[A_1]})^2)=2 t^{-1/2}D^2((\tau_{1}^{[A_1]})^2)t^{1/2}\tau_{1}^{[A_1]}=D^2((\tau_{1}^{[A_1]})^2)
		\end{equation}
		So \eqref{Dtau} is valid in this case as well. From the isomonodromic viewpoint, a linear quiver such as $A_1\times A_1$ corresponds to the degeneration of the sphere with 5 points where we have two complex deformations, the Garnier system. Up to now, we have not considered masses. However, due to this identification, we expect an identification between $D_n$ with one fundamental flavor and the $SU(2)\times SU(2)$ quiver with one bifundamental. Indeed, we find
		\begin{equation}\nonumber
		    \sum\limits_{i\geq0} Z_i^{[D_2]}(b_1-b_2,b_1+b_2,m) t^i = e^{4t}\sum\limits_{i\geq0} Z_i^{[A_1\times A_1]}(b_1,b_2,m) (-1)^it^i 
		\end{equation}
		This agrees with \cite{Hollands:2011zc}.

		\subsection{\texorpdfstring{$D_3=A_3$}{D3=A3}}\label{sec:D3}	
		Paralleling the previous discussion, there is a linear isomorphism of $D_3$ and $A_3$. Their extended root systems are the same, and from \eqref{tausystem} for $D_3$ we can obtain \eqref{Dtau}, \eqref{Drec} since
		\begin{equation}\nonumber
		D^2(\tau_0)=-t^{1/4}\tau_2\tau_3 = D^2(\tau_1)
		\end{equation} 
		so the equations are likewise the same.

		\section{Three node quiver calculations}\label{E}
    We present in brief the lowest order calculations for the $SU(2)^{3}$ quiver. The $t_1^{1+2\sigma_1}t_2^{1+2\sigma_2}t_3^{1+2\sigma_3}$ term leads to the one-loop term already discussed in general in the main text. Next, the $Z_{a,b,c}$ coefficients with positive integers $a^2+b^2+c^2=1$ are accessed by looking at $t_1^{1+2(1-a)\sigma_1}t_2^{1+2(1-b)\sigma_2}t_3^{1+2(1-c)\sigma_3}$ terms. This leads to
    \begin{gather}\nonumber
    	\left(-2 m_{2,3}^2+2 \sigma _2 \left(\sigma _2+1\right)+2 \sigma _3 \left(\sigma _3+1\right)+1\right) Z_{1,0,0}\left(\sigma _1,\sigma _2,\sigma _3\right)
    	\\ \nonumber
    	=\left(\left(\sigma _2-\sigma _3\right){}^2-m_{2,3}^2\right) \left(2 Z_{1,0,0}\left(\sigma _1,\sigma _2,\sigma _3+1\right)-Z_{1,0,0}\left(\sigma _1,\sigma _2+1,\sigma _3\right)\right)
    	\\ \nonumber
    	+\left((1+\sigma _2+\sigma _3)^2-m_{2,3}^2\right) Z_{1,0,0}\left(\sigma _1,\sigma _2+1,\sigma _3+1\right)+\frac{\left(2 \sigma _3+1\right) \left(2 \sigma _2+1\right){}^2}{2 \sigma _1^2}
    	\\ 
    	\Rightarrow Z_{1,0,0}(\sigma_1,\sigma_2,\sigma_3)=\frac{m_{1,2}^2+\sigma _1^2-\sigma _2^2}{2 \sigma _1^2},
    	\end{gather}
    an analogous equation for $Z_{0,0,1}$ which we omit, and finally
    	\begin{gather}
     \nonumber
    	Z_{0,1,0}\left(\sigma _1,\sigma _2,\sigma _3\right)+Z_{0,1,0}\left(\sigma _1+1,\sigma _2,\sigma _3+1\right)
    	\\ \nonumber
    	=Z_{0,1,0}\left(\sigma _1,\sigma _2,\sigma _3+1\right)+Z_{0,1,0}\left(\sigma _1+1,\sigma _2,\sigma _3\right)+\frac{\left(2 \sigma _1+1\right) \left(2 \sigma _3+1\right)}{2 \sigma _2^2}
    	\\ 
    	\Rightarrow Z_{0,1,0}(\sigma_1,\sigma_2,\sigma_3)=\frac{-4 \sigma _2^2 m_{1,2} m_{2,3}+m_{2,3}^2 \left(m_{1,2}^2-\sigma _1^2+\sigma _2^2\right)+\left(\sigma _2^2-\sigma _3^2\right) \left(m_{1,2}^2-\sigma _1^2+\sigma _2^2\right)}{2 \sigma _2^2}
    	\end{gather}
     The $Z_{a,b,c}$ coefficients with positive integers $a^2+b^2+c^2=2$ are likewise accessed by looking at $t_1^{1+2(1-a)\sigma_1}t_2^{1+2(1-b)\sigma_2}t_3^{1+2(1-c)\sigma_3}$ terms. They feature the terms we calculated in the previous step. 
    	\begin{gather}
     \nonumber
    	\left(Z_{0,1,0}\left(\sigma _1,\sigma _2,\sigma _3\right)+Z_{0,1,0}\left(\sigma _1,\sigma _2,\sigma _3+1\right)\right) Z_{1,0,0}\left(\sigma _1,\sigma _2,\sigma _3\right)+Z_{1,1,0}\left(\sigma _1,\sigma _2,\sigma _3\right)
    	\\ \nonumber
    	=\frac{\left(2 \sigma _3+1\right) \left(-m_{1,2}^2+\sigma _1^2+\sigma _2^2\right)}{4 \sigma _1^2 \sigma _2^2}+2 Z_{0,1,0}\left(\sigma _1,\sigma _2,\sigma _3\right) Z_{1,0,0}\left(\sigma _1,\sigma _2,\sigma _3+1\right)+Z_{1,1,0}\left(\sigma _1,\sigma _2,\sigma _3+1\right)
    	\\ \nonumber 
    	\Rightarrow Z_{1,1,0}(\sigma_1,\sigma_2,\sigma_3)=-\frac{4 \sigma _2^2 m_{1,2} m_{2,3} \left(m_{1,2}^2+\sigma _1^2-\sigma _2^2\right)-m_{1,2}^2 \left(m_{1,2}^2-1\right) \left(m_{2,3}^2+\sigma _2^2-\sigma _3^2\right)}{4 \sigma _1^2 \sigma _2^2}
    	\\
    	-\frac{\left(\sigma _1^4-\left(2 \sigma _2^2+1\right) \sigma _1^2+\sigma _2^4-\sigma _2^2\right) \left(m_{2,3}^2+\sigma _2^2-\sigma _3^2\right)}{4 \sigma _1^2 \sigma _2^2}
    	\end{gather}
     as well as an analogous equation for $Z_{0,1,1}$, and  
    	\begin{gather} \nonumber
    	2 Z_{1,0,1}\left(\sigma _1,\sigma _2,\sigma _3\right)+Z_{1,0,1}\left(\sigma _1,\sigma _2+1,\sigma _3\right) 
    	\\ \nonumber
    	=Z_{0,0,1}\left(\sigma _1,\sigma _2+1,\sigma _3\right) Z_{1,0,0}\left(\sigma _1,\sigma _2,\sigma _3\right)+Z_{0,0,1}\left(\sigma _1,\sigma _2,\sigma _3\right) Z_{1,0,0}\left(\sigma _1,\sigma _2+1,\sigma _3\right)
    	\\ \nonumber
    	+Z_{1,0,0}\left(\sigma _1,\sigma _2,\sigma _3\right)Z_{0,0,1}\left(\sigma _1,\sigma _2,\sigma _3\right)+\frac{\left(2 \sigma _2+1\right){}^2}{4 \sigma _1^2 \sigma _3^2}
    	\\
    	\Rightarrow Z_{1,0,1}(\sigma_1,\sigma_2,\sigma_3)=\frac{\left(m_{1,2}^2+\sigma _1^2-\sigma _2^2\right) \left(m_{2,3}^2-\sigma _2^2+\sigma _3^2\right)}{4 \sigma _1^2 \sigma _3^2}
    	\end{gather}
     Finally, $Z_{1,1,1}$ is found from the $t_1 t_2 t_3$ terms,
    	\begin{gather} \nonumber
    	Z_{1,1,1}\left(\sigma _1,\sigma _2,\sigma _3\right)=\frac{\left(-m_{1,2}^2+\sigma _1^2+\sigma _2^2\right) \left(-m_{2,3}^2+\sigma _2^2+\sigma _3^2\right)}{8 \sigma _1^2 \sigma _2^2 \sigma _3^2}
    	\\ \nonumber
    	= Z_{0,0,1}\left(\sigma _1,\sigma _2,\sigma _3\right) Z_{0,1,0}\left(\sigma _1,\sigma _2,\sigma _3\right) Z_{1,0,0}\left(\sigma _1,\sigma _2,\sigma _3\right)
    	\\ \nonumber
    	+Z_{1,0,0}\left(\sigma _1,\sigma _2,\sigma _3\right)Z_{0,1,1}\left(\sigma _1,\sigma _2,\sigma _3\right) 
    	+Z_{1,1,0}\left(\sigma _1,\sigma _2,\sigma _3\right)Z_{0,0,1}\left(\sigma _1,\sigma _2,\sigma _3\right) 
    	\\ \nonumber
    	-2 Z_{0,1,0}\left(\sigma _1,\sigma _2,\sigma _3\right) Z_{1,0,1}\left(\sigma _1,\sigma _2,\sigma _3\right)
    	\\ \nonumber
    	\Rightarrow Z_{1,1,1}(\sigma_1,\sigma_2,\sigma_3)=
    	\frac{4 \sigma _2^2 \left(\sigma _2^2-\sigma _1^2\right) m_{1,2} m_{2,3} \left(m_{2,3}^2-\sigma _2^2+\sigma _3^2\right)-4 \sigma _2^2 m_{1,2}^3 m_{2,3} \left(m_{2,3}^2-\sigma _2^2+\sigma _3^2\right)}{8 \sigma _1^2 \sigma _2^2 \sigma _3^2}
    	\\ \nonumber
    	+\frac{\left(\sigma _1^4-\left(2 \sigma _2^2+1\right) \sigma _1^2+\sigma _2^4-\sigma _2^2\right) \left(-m_{2,3}^4+m_{2,3}^2+\sigma _2^4-\left(2 \sigma _3^2+1\right) \sigma _2^2+\sigma _3^4-\sigma _3^2\right)}{8 \sigma _1^2 \sigma _2^2 \sigma _3^2}
    	\\
    	-\frac{m_{1,2}^2 \left(m_{1,2}^2-1\right) \left(-m_{2,3}^4+m_{2,3}^2+\sigma _2^4-\left(2 \sigma _3^2+1\right) \sigma _2^2+\sigma _3^4-\sigma _3^2\right)}{8 \sigma _1^2 \sigma _2^2 \sigma _3^2}
    	\end{gather}
 	
 	\color{black}

		\bibliographystyle{unsrtaipauth4-1}
		\bibliography{Biblio}

\begin{thebibliography}{58}%
\makeatletter
\providecommand \@ifxundefined [1]{%
 \@ifx{#1\undefined}
}%
\providecommand \@ifnum [1]{%
 \ifnum #1\expandafter \@firstoftwo
 \else \expandafter \@secondoftwo
 \fi
}%
\providecommand \@ifx [1]{%
 \ifx #1\expandafter \@firstoftwo
 \else \expandafter \@secondoftwo
 \fi
}%
\providecommand \natexlab [1]{#1}%
\providecommand \enquote  [1]{``#1''}%
\providecommand \bibnamefont  [1]{#1}%
\providecommand \bibfnamefont [1]{#1}%
\providecommand \citenamefont [1]{#1}%
\providecommand \href@noop [0]{\@secondoftwo}%
\providecommand \href [0]{\begingroup \@sanitize@url \@href}%
\providecommand \@href[1]{\@@startlink{#1}\@@href}%
\providecommand \@@href[1]{\endgroup#1\@@endlink}%
\providecommand \@sanitize@url [0]{\catcode `\\12\catcode `\$12\catcode
  `\&12\catcode `\#12\catcode `\^12\catcode `\_12\catcode `\%12\relax}%
\providecommand \@@startlink[1]{}%
\providecommand \@@endlink[0]{}%
\providecommand \url  [0]{\begingroup\@sanitize@url \@url }%
\providecommand \@url [1]{\endgroup\@href {#1}{\urlprefix }}%
\providecommand \urlprefix  [0]{URL }%
\providecommand \Eprint [0]{\href }%
\providecommand \doibase [0]{http://dx.doi.org/}%
\providecommand \selectlanguage [0]{\@gobble}%
\providecommand \bibinfo  [0]{\@secondoftwo}%
\providecommand \bibfield  [0]{\@secondoftwo}%
\providecommand \translation [1]{[#1]}%
\providecommand \BibitemOpen [0]{}%
\providecommand \bibitemStop [0]{}%
\providecommand \bibitemNoStop [0]{.\EOS\space}%
\providecommand \EOS [0]{\spacefactor3000\relax}%
\providecommand \BibitemShut  [1]{\csname bibitem#1\endcsname}%
\let\auto@bib@innerbib\@empty
\bibitem [{\citenamefont {Cecotti}\ and\ \citenamefont
  {Vafa}(1991)}]{Cecotti:1991me}%
  \BibitemOpen
  \bibfield  {author} {\bibinfo {author} {\bibnamefont {Cecotti}, \bibfnamefont
  {S.}}\ and\ \bibinfo {author} {\bibnamefont {Vafa}, \bibfnamefont {C.}},\
  }\href {\doibase 10.1016/0550-3213(91)90021-O} {\bibfield  {journal}
  {\bibinfo  {journal} {Nucl. Phys. B}\ }\textbf {\bibinfo {volume} {367}},\
  \bibinfo {pages} {359} (\bibinfo {year} {1991})}\BibitemShut {NoStop}%
\bibitem [{\citenamefont {Gukov}\ and\ \citenamefont
  {Witten}(2006)}]{Gukov:2006jk}%
  \BibitemOpen
  \bibfield  {author} {\bibinfo {author} {\bibnamefont {Gukov}, \bibfnamefont
  {S.}}\ and\ \bibinfo {author} {\bibnamefont {Witten}, \bibfnamefont {E.}},\
  }\href@noop {} {\  (\bibinfo {year} {2006})},\ \Eprint
  {http://arxiv.org/abs/hep-th/0612073} {arXiv:hep-th/0612073} \BibitemShut
  {NoStop}%
\bibitem [{\citenamefont {Seiberg}\ and\ \citenamefont
  {Witten}(1994)}]{Seiberg:1994aj}%
  \BibitemOpen
  \bibfield  {author} {\bibinfo {author} {\bibnamefont {Seiberg}, \bibfnamefont
  {N.}}\ and\ \bibinfo {author} {\bibnamefont {Witten}, \bibfnamefont {E.}},\
  }\href {\doibase 10.1016/0550-3213(94)90214-3} {\bibfield  {journal}
  {\bibinfo  {journal} {Nucl. Phys.}\ }\textbf {\bibinfo {volume} {B431}},\
  \bibinfo {pages} {484} (\bibinfo {year} {1994})},\ \Eprint
  {http://arxiv.org/abs/hep-th/9408099} {arXiv:hep-th/9408099 [hep-th]}
  \BibitemShut {NoStop}%
\bibitem [{\citenamefont {Gorsky}\ \emph {et~al.}(1995)\citenamefont {Gorsky},
  \citenamefont {Krichever}, \citenamefont {Marshakov}, \citenamefont
  {Mironov},\ and\ \citenamefont {Morozov}}]{Gorsky:1995zq}%
  \BibitemOpen
  \bibfield  {author} {\bibinfo {author} {\bibnamefont {Gorsky}, \bibfnamefont
  {A.}}, \bibinfo {author} {\bibnamefont {Krichever}, \bibfnamefont {I.}},
  \bibinfo {author} {\bibnamefont {Marshakov}, \bibfnamefont {A.}}, \bibinfo
  {author} {\bibnamefont {Mironov}, \bibfnamefont {A.}}, \ and\ \bibinfo
  {author} {\bibnamefont {Morozov}, \bibfnamefont {A.}},\ }\href {\doibase
  10.1016/0370-2693(95)00723-X} {\bibfield  {journal} {\bibinfo  {journal}
  {Phys. Lett.}\ }\textbf {\bibinfo {volume} {B355}},\ \bibinfo {pages} {466}
  (\bibinfo {year} {1995})},\ \Eprint {http://arxiv.org/abs/hep-th/9505035}
  {arXiv:hep-th/9505035 [hep-th]} \BibitemShut {NoStop}%
\bibitem [{\citenamefont {Martinec}\ and\ \citenamefont
  {Warner}(1996)}]{Martinec:1995by}%
  \BibitemOpen
  \bibfield  {author} {\bibinfo {author} {\bibnamefont {Martinec},
  \bibfnamefont {E.~J.}}\ and\ \bibinfo {author} {\bibnamefont {Warner},
  \bibfnamefont {N.~P.}},\ }\href {\doibase 10.1016/0550-3213(95)00588-9}
  {\bibfield  {journal} {\bibinfo  {journal} {Nucl. Phys. B}\ }\textbf
  {\bibinfo {volume} {459}},\ \bibinfo {pages} {97} (\bibinfo {year} {1996})},\
  \Eprint {http://arxiv.org/abs/hep-th/9509161} {arXiv:hep-th/9509161}
  \BibitemShut {NoStop}%
\bibitem [{\citenamefont {Jimbo}, \citenamefont {Miwa},\ and\ \citenamefont
  {Ueno}(1981)}]{Jimbo:1981zz}%
  \BibitemOpen
  \bibfield  {author} {\bibinfo {author} {\bibnamefont {Jimbo}, \bibfnamefont
  {M.}}, \bibinfo {author} {\bibnamefont {Miwa}, \bibfnamefont {T.}}, \ and\
  \bibinfo {author} {\bibnamefont {Ueno}, \bibfnamefont {a.~K.}},\ }\href@noop
  {} {\bibfield  {journal} {\bibinfo  {journal} {Physica}\ }\textbf {\bibinfo
  {volume} {D2}},\ \bibinfo {pages} {306} (\bibinfo {year} {1981})}\BibitemShut
  {NoStop}%
\bibitem [{\citenamefont {Jimbo}, \citenamefont {Miwa},\ and\ \citenamefont
  {Ueno}(1982)}]{Jimbo:1982zz}%
  \BibitemOpen
  \bibfield  {author} {\bibinfo {author} {\bibnamefont {Jimbo}, \bibfnamefont
  {M.}}, \bibinfo {author} {\bibnamefont {Miwa}, \bibfnamefont {T.}}, \ and\
  \bibinfo {author} {\bibnamefont {Ueno}, \bibfnamefont {a.~K.}},\ }\href@noop
  {} {\bibfield  {journal} {\bibinfo  {journal} {Physica}\ }\textbf {\bibinfo
  {volume} {D2}},\ \bibinfo {pages} {407} (\bibinfo {year} {1982})}\BibitemShut
  {NoStop}%
\bibitem [{\citenamefont {Nakajima}\ and\ \citenamefont
  {Yoshioka}(2005)}]{Nakajima:2003pg}%
  \BibitemOpen
  \bibfield  {author} {\bibinfo {author} {\bibnamefont {Nakajima},
  \bibfnamefont {H.}}\ and\ \bibinfo {author} {\bibnamefont {Yoshioka},
  \bibfnamefont {K.}},\ }\href {\doibase 10.1007/s00222-005-0444-1} {\bibfield
  {journal} {\bibinfo  {journal} {Invent. Math.}\ }\textbf {\bibinfo {volume}
  {162}},\ \bibinfo {pages} {313} (\bibinfo {year} {2005})},\ \Eprint
  {http://arxiv.org/abs/math/0306198} {arXiv:math/0306198 [math.AG]}
  \BibitemShut {NoStop}%
\bibitem [{\citenamefont {Kim}\ \emph {et~al.}(2019)\citenamefont {Kim},
  \citenamefont {Kim}, \citenamefont {Lee}, \citenamefont {Lee},\ and\
  \citenamefont {Song}}]{Kim:2019uqw}%
  \BibitemOpen
  \bibfield  {author} {\bibinfo {author} {\bibnamefont {Kim}, \bibfnamefont
  {J.}}, \bibinfo {author} {\bibnamefont {Kim}, \bibfnamefont {S.-S.}},
  \bibinfo {author} {\bibnamefont {Lee}, \bibfnamefont {K.-H.}}, \bibinfo
  {author} {\bibnamefont {Lee}, \bibfnamefont {K.}}, \ and\ \bibinfo {author}
  {\bibnamefont {Song}, \bibfnamefont {J.}},\ }\href {\doibase
  10.1007/JHEP11(2019)092} {\bibfield  {journal} {\bibinfo  {journal} {JHEP}\
  }\textbf {\bibinfo {volume} {11}},\ \bibinfo {pages} {092} (\bibinfo {year}
  {2019})},\ \bibinfo {note} {[Erratum: JHEP 06, 124 (2020)]},\ \Eprint
  {http://arxiv.org/abs/1908.11276} {arXiv:1908.11276 [hep-th]} \BibitemShut
  {NoStop}%
\bibitem [{\citenamefont {Jeong}\ and\ \citenamefont
  {Nekrasov}(2020)}]{Jeong:2020uxz}%
  \BibitemOpen
  \bibfield  {author} {\bibinfo {author} {\bibnamefont {Jeong}, \bibfnamefont
  {S.}}\ and\ \bibinfo {author} {\bibnamefont {Nekrasov}, \bibfnamefont {N.}},\
  }\href@noop {} {\  (\bibinfo {year} {2020})},\ \Eprint
  {http://arxiv.org/abs/2007.03660} {arXiv:2007.03660 [hep-th]} \BibitemShut
  {NoStop}%
\bibitem [{\citenamefont {Bonelli}, \citenamefont {Globlek},\ and\
  \citenamefont {Tanzini}(2021)}]{Bonelli:2021rrg}%
  \BibitemOpen
  \bibfield  {author} {\bibinfo {author} {\bibnamefont {Bonelli}, \bibfnamefont
  {G.}}, \bibinfo {author} {\bibnamefont {Globlek}, \bibfnamefont {F.}}, \ and\
  \bibinfo {author} {\bibnamefont {Tanzini}, \bibfnamefont {A.}},\ }\href
  {\doibase 10.1103/PhysRevLett.126.231602} {\bibfield  {journal} {\bibinfo
  {journal} {Phys. Rev. Lett.}\ }\textbf {\bibinfo {volume} {126}},\ \bibinfo
  {pages} {231602} (\bibinfo {year} {2021})},\ \Eprint
  {http://arxiv.org/abs/2102.01627} {arXiv:2102.01627 [hep-th]} \BibitemShut
  {NoStop}%
\bibitem [{\citenamefont {Gaiotto}(2012)}]{Gaiotto:2009we}%
  \BibitemOpen
  \bibfield  {author} {\bibinfo {author} {\bibnamefont {Gaiotto}, \bibfnamefont
  {D.}},\ }\href {\doibase 10.1007/JHEP08(2012)034} {\bibfield  {journal}
  {\bibinfo  {journal} {JHEP}\ }\textbf {\bibinfo {volume} {08}},\ \bibinfo
  {pages} {034} (\bibinfo {year} {2012})},\ \Eprint
  {http://arxiv.org/abs/0904.2715} {arXiv:0904.2715 [hep-th]} \BibitemShut
  {NoStop}%
\bibitem [{\citenamefont {Tsuda}(2003)}]{Tsuda03birationalsymmetries}%
  \BibitemOpen
  \bibfield  {author} {\bibinfo {author} {\bibnamefont {Tsuda}, \bibfnamefont
  {T.}},\ }\href@noop {} {\bibfield  {journal} {\bibinfo  {journal} {J. Math.
  Sci. Univ. Tokyo}\ ,\ \bibinfo {pages} {2341}} (\bibinfo {year}
  {2003})}\BibitemShut {NoStop}%
\bibitem [{\citenamefont {Bonelli}, \citenamefont {Grassi},\ and\ \citenamefont
  {Tanzini}(2017)}]{Bonelli:2016idi}%
  \BibitemOpen
  \bibfield  {author} {\bibinfo {author} {\bibnamefont {Bonelli}, \bibfnamefont
  {G.}}, \bibinfo {author} {\bibnamefont {Grassi}, \bibfnamefont {A.}}, \ and\
  \bibinfo {author} {\bibnamefont {Tanzini}, \bibfnamefont {A.}},\ }\href
  {\doibase 10.1007/s11005-016-0893-z} {\bibfield  {journal} {\bibinfo
  {journal} {Lett. Math. Phys.}\ }\textbf {\bibinfo {volume} {107}},\ \bibinfo
  {pages} {1} (\bibinfo {year} {2017})},\ \Eprint
  {http://arxiv.org/abs/1603.01174} {arXiv:1603.01174 [hep-th]} \BibitemShut
  {NoStop}%
\bibitem [{\citenamefont {Bershtein}, \citenamefont {Gavrylenko},\ and\
  \citenamefont {Grassi}(2021)}]{Bershtein:2021uts}%
  \BibitemOpen
  \bibfield  {author} {\bibinfo {author} {\bibnamefont {Bershtein},
  \bibfnamefont {M.}}, \bibinfo {author} {\bibnamefont {Gavrylenko},
  \bibfnamefont {P.}}, \ and\ \bibinfo {author} {\bibnamefont {Grassi},
  \bibfnamefont {A.}},\ }\href@noop {} {\  (\bibinfo {year} {2021})},\ \Eprint
  {http://arxiv.org/abs/2105.00985} {arXiv:2105.00985 [math-ph]} \BibitemShut
  {NoStop}%
\bibitem [{\citenamefont {Bonelli}\ \emph {et~al.}(2020)\citenamefont
  {Bonelli}, \citenamefont {Del~Monte}, \citenamefont {Gavrylenko},\ and\
  \citenamefont {Tanzini}}]{Bonelli:2019boe}%
  \BibitemOpen
  \bibfield  {author} {\bibinfo {author} {\bibnamefont {Bonelli}, \bibfnamefont
  {G.}}, \bibinfo {author} {\bibnamefont {Del~Monte}, \bibfnamefont {F.}},
  \bibinfo {author} {\bibnamefont {Gavrylenko}, \bibfnamefont {P.}}, \ and\
  \bibinfo {author} {\bibnamefont {Tanzini}, \bibfnamefont {A.}},\ }\href
  {\doibase 10.1007/s00220-020-03743-y} {\bibfield  {journal} {\bibinfo
  {journal} {Commun. Math. Phys.}\ }\textbf {\bibinfo {volume} {377}},\
  \bibinfo {pages} {1381} (\bibinfo {year} {2020})},\ \Eprint
  {http://arxiv.org/abs/1901.10497} {arXiv:1901.10497 [hep-th]} \BibitemShut
  {NoStop}%
\bibitem [{\citenamefont {Bonelli}\ \emph {et~al.}(2019)\citenamefont
  {Bonelli}, \citenamefont {Del~Monte}, \citenamefont {Gavrylenko},\ and\
  \citenamefont {Tanzini}}]{Bonelli:2019yjd}%
  \BibitemOpen
  \bibfield  {author} {\bibinfo {author} {\bibnamefont {Bonelli}, \bibfnamefont
  {G.}}, \bibinfo {author} {\bibnamefont {Del~Monte}, \bibfnamefont {F.}},
  \bibinfo {author} {\bibnamefont {Gavrylenko}, \bibfnamefont {P.}}, \ and\
  \bibinfo {author} {\bibnamefont {Tanzini}, \bibfnamefont {A.}},\ }\href@noop
  {} {\bibfield  {journal} {\bibinfo  {journal} {Lett. Math. Phys.}\ }\textbf
  {\bibinfo {volume} {83}} (\bibinfo {year} {2019})},\ \Eprint
  {http://arxiv.org/abs/1909.07990} {arXiv:1909.07990 [hep-th]} \BibitemShut
  {NoStop}%
\bibitem [{\citenamefont {Bonelli}, \citenamefont {Grassi},\ and\ \citenamefont
  {Tanzini}(2019)}]{Bonelli:2017gdk}%
  \BibitemOpen
  \bibfield  {author} {\bibinfo {author} {\bibnamefont {Bonelli}, \bibfnamefont
  {G.}}, \bibinfo {author} {\bibnamefont {Grassi}, \bibfnamefont {A.}}, \ and\
  \bibinfo {author} {\bibnamefont {Tanzini}, \bibfnamefont {A.}},\ }\href
  {\doibase 10.1007/s11005-019-01174-y} {\bibfield  {journal} {\bibinfo
  {journal} {Lett. Math. Phys.}\ }\textbf {\bibinfo {volume} {109}},\ \bibinfo
  {pages} {1961} (\bibinfo {year} {2019})},\ \Eprint
  {http://arxiv.org/abs/1710.11603} {arXiv:1710.11603 [hep-th]} \BibitemShut
  {NoStop}%
\bibitem [{\citenamefont {Bershtein}, \citenamefont {Gavrylenko},\ and\
  \citenamefont {Marshakov}(2018)}]{Bershtein:2018srt}%
  \BibitemOpen
  \bibfield  {author} {\bibinfo {author} {\bibnamefont {Bershtein},
  \bibfnamefont {M.}}, \bibinfo {author} {\bibnamefont {Gavrylenko},
  \bibfnamefont {P.}}, \ and\ \bibinfo {author} {\bibnamefont {Marshakov},
  \bibfnamefont {A.}},\ }\href@noop {} {\  (\bibinfo {year} {2018})},\ \Eprint
  {http://arxiv.org/abs/1804.10145} {arXiv:1804.10145 [math-ph]} \BibitemShut
  {NoStop}%
\bibitem [{\citenamefont {Bonelli}, \citenamefont {Del~Monte},\ and\
  \citenamefont {Tanzini}(2021)}]{Bonelli:2020dcp}%
  \BibitemOpen
  \bibfield  {author} {\bibinfo {author} {\bibnamefont {Bonelli}, \bibfnamefont
  {G.}}, \bibinfo {author} {\bibnamefont {Del~Monte}, \bibfnamefont {F.}}, \
  and\ \bibinfo {author} {\bibnamefont {Tanzini}, \bibfnamefont {A.}},\ }\href
  {\doibase 10.1007/s00023-021-01034-3} {\bibfield  {journal} {\bibinfo
  {journal} {Annales Henri Poincare}\ }\textbf {\bibinfo {volume} {22}},\
  \bibinfo {pages} {2721} (\bibinfo {year} {2021})},\ \Eprint
  {http://arxiv.org/abs/2007.11596} {arXiv:2007.11596 [hep-th]} \BibitemShut
  {NoStop}%
\bibitem [{\citenamefont {Nawata}\ and\ \citenamefont
  {Zhu}(2021)}]{Nawata:2021dlk}%
  \BibitemOpen
  \bibfield  {author} {\bibinfo {author} {\bibnamefont {Nawata}, \bibfnamefont
  {S.}}\ and\ \bibinfo {author} {\bibnamefont {Zhu}, \bibfnamefont {R.-D.}},\
  }\href {\doibase 10.1007/JHEP09(2021)190} {\bibfield  {journal} {\bibinfo
  {journal} {JHEP}\ }\textbf {\bibinfo {volume} {09}},\ \bibinfo {pages} {190}
  (\bibinfo {year} {2021})},\ \Eprint {http://arxiv.org/abs/2107.03656}
  {arXiv:2107.03656 [hep-th]} \BibitemShut {NoStop}%
\bibitem [{\citenamefont {Brini}\ and\ \citenamefont
  {Osuga}(2022)}]{Brini:2021wrm}%
  \BibitemOpen
  \bibfield  {author} {\bibinfo {author} {\bibnamefont {Brini}, \bibfnamefont
  {A.}}\ and\ \bibinfo {author} {\bibnamefont {Osuga}, \bibfnamefont {K.}},\
  }\href {\doibase 10.1007/s11005-022-01538-x} {\bibfield  {journal} {\bibinfo
  {journal} {Lett. Math. Phys.}\ }\textbf {\bibinfo {volume} {112}},\ \bibinfo
  {pages} {44} (\bibinfo {year} {2022})},\ \Eprint
  {http://arxiv.org/abs/2110.11638} {arXiv:2110.11638 [hep-th]} \BibitemShut
  {NoStop}%
\bibitem [{\citenamefont {Bonelli}\ \emph {et~al.}(2022)\citenamefont
  {Bonelli}, \citenamefont {Globlek}, \citenamefont {Kubo}, \citenamefont
  {Nosaka},\ and\ \citenamefont {Tanzini}}]{Bonelli:2022dse}%
  \BibitemOpen
  \bibfield  {author} {\bibinfo {author} {\bibnamefont {Bonelli}, \bibfnamefont
  {G.}}, \bibinfo {author} {\bibnamefont {Globlek}, \bibfnamefont {F.}},
  \bibinfo {author} {\bibnamefont {Kubo}, \bibfnamefont {N.}}, \bibinfo
  {author} {\bibnamefont {Nosaka}, \bibfnamefont {T.}}, \ and\ \bibinfo
  {author} {\bibnamefont {Tanzini}, \bibfnamefont {A.}},\ }\href@noop {} {\
  (\bibinfo {year} {2022})},\ \Eprint {http://arxiv.org/abs/2202.10654}
  {arXiv:2202.10654 [hep-th]} \BibitemShut {NoStop}%
\bibitem [{\citenamefont {Its}, \citenamefont {Lisovyy},\ and\ \citenamefont
  {Tykhyy}(2014)}]{Its:2014lga}%
  \BibitemOpen
  \bibfield  {author} {\bibinfo {author} {\bibnamefont {Its}, \bibfnamefont
  {A.}}, \bibinfo {author} {\bibnamefont {Lisovyy}, \bibfnamefont {O.}}, \ and\
  \bibinfo {author} {\bibnamefont {Tykhyy}, \bibfnamefont {Y.}},\ }\href
  {\doibase 10.1093/imrn/rnu209} {\  (\bibinfo {year} {2014}),\
  10.1093/imrn/rnu209},\ \Eprint {http://arxiv.org/abs/1403.1235}
  {arXiv:1403.1235 [math-ph]} \BibitemShut {NoStop}%
\bibitem [{\citenamefont {Bonelli}\ \emph {et~al.}(2017)\citenamefont
  {Bonelli}, \citenamefont {Lisovyy}, \citenamefont {Maruyoshi}, \citenamefont
  {Sciarappa},\ and\ \citenamefont {Tanzini}}]{Bonelli:2016qwg}%
  \BibitemOpen
  \bibfield  {author} {\bibinfo {author} {\bibnamefont {Bonelli}, \bibfnamefont
  {G.}}, \bibinfo {author} {\bibnamefont {Lisovyy}, \bibfnamefont {O.}},
  \bibinfo {author} {\bibnamefont {Maruyoshi}, \bibfnamefont {K.}}, \bibinfo
  {author} {\bibnamefont {Sciarappa}, \bibfnamefont {A.}}, \ and\ \bibinfo
  {author} {\bibnamefont {Tanzini}, \bibfnamefont {A.}},\ }\href {\doibase
  10.1007/s11005-017-0983-6} {\bibfield  {journal} {\bibinfo  {journal}
  {Letters in Mathematical Physics}\ }\textbf {\bibinfo {volume} {107}},\
  \bibinfo {pages} {2359} (\bibinfo {year} {2017})},\ \Eprint
  {http://arxiv.org/abs/1612.06235} {arXiv:1612.06235 [hep-th]} \BibitemShut
  {NoStop}%
\bibitem [{\citenamefont {Bonelli}, \citenamefont {Grassi},\ and\ \citenamefont
  {Tanzini}(2018)}]{Bonelli:2017ptp}%
  \BibitemOpen
  \bibfield  {author} {\bibinfo {author} {\bibnamefont {Bonelli}, \bibfnamefont
  {G.}}, \bibinfo {author} {\bibnamefont {Grassi}, \bibfnamefont {A.}}, \ and\
  \bibinfo {author} {\bibnamefont {Tanzini}, \bibfnamefont {A.}},\ }\href
  {\doibase 10.1007/s00023-017-0643-5} {\bibfield  {journal} {\bibinfo
  {journal} {Annales Henri Poincare}\ }\textbf {\bibinfo {volume} {19}},\
  \bibinfo {pages} {743} (\bibinfo {year} {2018})},\ \Eprint
  {http://arxiv.org/abs/1704.01517} {arXiv:1704.01517 [hep-th]} \BibitemShut
  {NoStop}%
\bibitem [{\citenamefont {McCoy}, \citenamefont {Tracy},\ and\ \citenamefont
  {Wu}(1977)}]{McCoy:1976cd}%
  \BibitemOpen
  \bibfield  {author} {\bibinfo {author} {\bibnamefont {McCoy}, \bibfnamefont
  {B.~M.}}, \bibinfo {author} {\bibnamefont {Tracy}, \bibfnamefont {C.~A.}}, \
  and\ \bibinfo {author} {\bibnamefont {Wu}, \bibfnamefont {T.~T.}},\ }\href
  {\doibase 10.1063/1.523367} {\bibfield  {journal} {\bibinfo  {journal} {J.
  Math. Phys.}\ }\textbf {\bibinfo {volume} {18}},\ \bibinfo {pages} {1058}
  (\bibinfo {year} {1977})}\BibitemShut {NoStop}%
\bibitem [{\citenamefont {Bertola}, \citenamefont {Del~Monte},\ and\
  \citenamefont {Harnad}(2021)}]{Bertola:2021ugs}%
  \BibitemOpen
  \bibfield  {author} {\bibinfo {author} {\bibnamefont {Bertola}, \bibfnamefont
  {M.}}, \bibinfo {author} {\bibnamefont {Del~Monte}, \bibfnamefont {F.}}, \
  and\ \bibinfo {author} {\bibnamefont {Harnad}, \bibfnamefont {J.}},\
  }\href@noop {} {\  (\bibinfo {year} {2021})},\ \Eprint
  {http://arxiv.org/abs/2112.12666} {arXiv:2112.12666 [math-ph]} \BibitemShut
  {NoStop}%
\bibitem [{\citenamefont {Kimura}\ and\ \citenamefont
  {Nieri}(2021)}]{Kimura:2021ngu}%
  \BibitemOpen
  \bibfield  {author} {\bibinfo {author} {\bibnamefont {Kimura}, \bibfnamefont
  {T.}}\ and\ \bibinfo {author} {\bibnamefont {Nieri}, \bibfnamefont {F.}},\
  }\href {\doibase 10.1088/1751-8121/ac2716} {\bibfield  {journal} {\bibinfo
  {journal} {J. Phys. A}\ }\textbf {\bibinfo {volume} {54}},\ \bibinfo {pages}
  {435401} (\bibinfo {year} {2021})},\ \Eprint
  {http://arxiv.org/abs/2105.02776} {arXiv:2105.02776 [hep-th]} \BibitemShut
  {NoStop}%
\bibitem [{\citenamefont {Takasaki}()}]{Takasaki1999}%
  \BibitemOpen
  \bibfield  {author} {\bibinfo {author} {\bibnamefont {Takasaki},
  \bibfnamefont {K.}},\ }\href {\doibase 10.1063/1.533056} {\
  10.1063/1.533056},\ \Eprint {http://arxiv.org/abs/math/9905101v5}
  {math/9905101v5} \BibitemShut {NoStop}%
\bibitem [{\citenamefont {Krichever}(1980)}]{Krichever1980}%
  \BibitemOpen
  \bibfield  {author} {\bibinfo {author} {\bibnamefont {Krichever},
  \bibfnamefont {I.~M.}},\ }\href {\doibase 10.1007/BF01078304} {\bibfield
  {journal} {\bibinfo  {journal} {Functional Analysis and Its Applications}\
  }\textbf {\bibinfo {volume} {14}},\ \bibinfo {pages} {282} (\bibinfo {year}
  {1980})}\BibitemShut {NoStop}%
\bibitem [{\citenamefont {Takasaki}(1999)}]{takasaki1999elliptic}%
  \BibitemOpen
  \bibfield  {author} {\bibinfo {author} {\bibnamefont {Takasaki},
  \bibfnamefont {K.}},\ }\href@noop {} {\bibfield  {journal} {\bibinfo
  {journal} {Journal of Mathematical Physics}\ }\textbf {\bibinfo {volume}
  {40}},\ \bibinfo {pages} {5787} (\bibinfo {year} {1999})}\BibitemShut
  {NoStop}%
\bibitem [{\citenamefont {Manin}(1996)}]{manin1996sixth}%
  \BibitemOpen
  \bibfield  {author} {\bibinfo {author} {\bibnamefont {Manin}, \bibfnamefont
  {Y.~I.}},\ }\href@noop {} {\bibfield  {journal} {\bibinfo  {journal} {arXiv
  preprint alg-geom/9605010}\ ,\ \bibinfo {pages} {131}} (\bibinfo {year}
  {1996})}\BibitemShut {NoStop}%
\bibitem [{\citenamefont {Levin}\ and\ \citenamefont
  {Olshanetsky}(2000)}]{Levin2000}%
  \BibitemOpen
  \bibfield  {author} {\bibinfo {author} {\bibnamefont {Levin}, \bibfnamefont
  {A.~M.}}\ and\ \bibinfo {author} {\bibnamefont {Olshanetsky}, \bibfnamefont
  {M.~A.}},\ }\enquote {\bibinfo {title} {Painlev{\'e}---calogero
  correspondence},}\ in\ \href {\doibase 10.1007/978-1-4612-1206-5_20} {\emph
  {\bibinfo {booktitle} {Calogero---Moser--- Sutherland Models}}},\ \bibinfo
  {editor} {edited by\ \bibinfo {editor} {\bibfnamefont {J.~F.}\ \bibnamefont
  {van Diejen}}\ and\ \bibinfo {editor} {\bibfnamefont {L.}~\bibnamefont
  {Vinet}}}\ (\bibinfo  {publisher} {Springer New York},\ \bibinfo {address}
  {New York, NY},\ \bibinfo {year} {2000})\ pp.\ \bibinfo {pages}
  {313--332}\BibitemShut {NoStop}%
\bibitem [{\citenamefont {Marshakov}, \citenamefont {Mironov},\ and\
  \citenamefont {Morozov}(2009)}]{Marshakov:2009gn}%
  \BibitemOpen
  \bibfield  {author} {\bibinfo {author} {\bibnamefont {Marshakov},
  \bibfnamefont {A.}}, \bibinfo {author} {\bibnamefont {Mironov}, \bibfnamefont
  {A.}}, \ and\ \bibinfo {author} {\bibnamefont {Morozov}, \bibfnamefont
  {A.}},\ }\href {\doibase 10.1016/j.physletb.2009.10.077} {\bibfield
  {journal} {\bibinfo  {journal} {Phys. Lett.}\ }\textbf {\bibinfo {volume}
  {B682}},\ \bibinfo {pages} {125} (\bibinfo {year} {2009})},\ \Eprint
  {http://arxiv.org/abs/0909.2052} {arXiv:0909.2052 [hep-th]} \BibitemShut
  {NoStop}%
\bibitem [{\citenamefont {Silverman}(1994)}]{Silverman1994}%
  \BibitemOpen
  \bibfield  {author} {\bibinfo {author} {\bibnamefont {Silverman},
  \bibfnamefont {J.~H.}},\ }\href@noop {} {\emph {\bibinfo {title} {Advanced
  topics in the arithmetic of elliptic curves}}},\ Vol.\ \bibinfo {volume}
  {151}\ (\bibinfo  {publisher} {Springer Science \& Business Media},\ \bibinfo
  {year} {1994})\BibitemShut {NoStop}%
\bibitem [{\citenamefont {Zagier}(2008)}]{Zagier2008}%
  \BibitemOpen
  \bibfield  {author} {\bibinfo {author} {\bibnamefont {Zagier}, \bibfnamefont
  {D.}},\ }in\ \href@noop {} {\emph {\bibinfo {booktitle} {The 1-2-3 of modular
  forms}}}\ (\bibinfo  {publisher} {Springer},\ \bibinfo {year} {2008})\ pp.\
  \bibinfo {pages} {1--103}\BibitemShut {NoStop}%
\bibitem [{\citenamefont {D'Hoker}\ and\ \citenamefont {Phong}()}]{DHoker1999}%
  \BibitemOpen
  \bibfield  {author} {\bibinfo {author} {\bibnamefont {D'Hoker}, \bibfnamefont
  {E.}}\ and\ \bibinfo {author} {\bibnamefont {Phong}, \bibfnamefont {D.~H.}},\
  }\href {\doibase 10.1143/PTPS.135.75} {\ 10.1143/PTPS.135.75},\ \Eprint
  {http://arxiv.org/abs/hep-th/9906027v1} {hep-th/9906027v1} \BibitemShut
  {NoStop}%
\bibitem [{\citenamefont {D'Hoker}\ and\ \citenamefont
  {Phong}(1999)}]{DHoker:1999yni}%
  \BibitemOpen
  \bibfield  {author} {\bibinfo {author} {\bibnamefont {D'Hoker}, \bibfnamefont
  {E.}}\ and\ \bibinfo {author} {\bibnamefont {Phong}, \bibfnamefont {D.~H.}},\
  }in\ \href@noop {} {\emph {\bibinfo {booktitle} {{Theoretical physics at the
  end of the twentieth century. Proceedings, Summer School, Banff, Canada, June
  27-July 10, 1999}}}}\ (\bibinfo {year} {1999})\ pp.\ \bibinfo {pages}
  {1--125},\ \Eprint {http://arxiv.org/abs/hep-th/9912271}
  {arXiv:hep-th/9912271 [hep-th]} \BibitemShut {NoStop}%
\bibitem [{\citenamefont {Bershtein}\ and\ \citenamefont
  {Shchechkin}(2017)}]{Bershtein:2016uov}%
  \BibitemOpen
  \bibfield  {author} {\bibinfo {author} {\bibnamefont {Bershtein},
  \bibfnamefont {M.~A.}}\ and\ \bibinfo {author} {\bibnamefont {Shchechkin},
  \bibfnamefont {A.~I.}},\ }\href {\doibase 10.1088/1751-8121/aa59c9}
  {\bibfield  {journal} {\bibinfo  {journal} {J. Phys.}\ }\textbf {\bibinfo
  {volume} {A50}},\ \bibinfo {pages} {115205} (\bibinfo {year} {2017})},\
  \Eprint {http://arxiv.org/abs/1608.02568} {arXiv:1608.02568 [math-ph]}
  \BibitemShut {NoStop}%
\bibitem [{\citenamefont {Seiberg}\ and\ \citenamefont
  {Witten}(1994)}]{Seiberg:1994rs}%
  \BibitemOpen
  \bibfield  {author} {\bibinfo {author} {\bibnamefont {Seiberg}, \bibfnamefont
  {N.}}\ and\ \bibinfo {author} {\bibnamefont {Witten}, \bibfnamefont {E.}},\
  }\href {\doibase 10.1016/0550-3213(94)90124-4, 10.1016/0550-3213(94)00449-8}
  {\bibfield  {journal} {\bibinfo  {journal} {Nucl. Phys.}\ }\textbf {\bibinfo
  {volume} {B426}},\ \bibinfo {pages} {19} (\bibinfo {year} {1994})},\ \bibinfo
  {note} {[Erratum: Nucl. Phys.B430,485(1994)]},\ \Eprint
  {http://arxiv.org/abs/hep-th/9407087} {arXiv:hep-th/9407087 [hep-th]}
  \BibitemShut {NoStop}%
\bibitem [{\citenamefont {Edelstein}, \citenamefont {Marino},\ and\
  \citenamefont {Mas}(1999)}]{Edelstein:1998sp}%
  \BibitemOpen
  \bibfield  {author} {\bibinfo {author} {\bibnamefont {Edelstein},
  \bibfnamefont {J.~D.}}, \bibinfo {author} {\bibnamefont {Marino},
  \bibfnamefont {M.}}, \ and\ \bibinfo {author} {\bibnamefont {Mas},
  \bibfnamefont {J.}},\ }\href {\doibase 10.1016/S0550-3213(98)00798-6}
  {\bibfield  {journal} {\bibinfo  {journal} {Nucl. Phys. B}\ }\textbf
  {\bibinfo {volume} {541}},\ \bibinfo {pages} {671} (\bibinfo {year}
  {1999})},\ \Eprint {http://arxiv.org/abs/hep-th/9805172}
  {arXiv:hep-th/9805172} \BibitemShut {NoStop}%
\bibitem [{\citenamefont {Gaiotto}\ \emph {et~al.}(2015)\citenamefont
  {Gaiotto}, \citenamefont {Kapustin}, \citenamefont {Seiberg},\ and\
  \citenamefont {Willett}}]{Gaiotto:2014kfa}%
  \BibitemOpen
  \bibfield  {author} {\bibinfo {author} {\bibnamefont {Gaiotto}, \bibfnamefont
  {D.}}, \bibinfo {author} {\bibnamefont {Kapustin}, \bibfnamefont {A.}},
  \bibinfo {author} {\bibnamefont {Seiberg}, \bibfnamefont {N.}}, \ and\
  \bibinfo {author} {\bibnamefont {Willett}, \bibfnamefont {B.}},\ }\href
  {\doibase 10.1007/JHEP02(2015)172} {\bibfield  {journal} {\bibinfo  {journal}
  {JHEP}\ }\textbf {\bibinfo {volume} {02}},\ \bibinfo {pages} {172} (\bibinfo
  {year} {2015})},\ \Eprint {http://arxiv.org/abs/1412.5148} {arXiv:1412.5148
  [hep-th]} \BibitemShut {NoStop}%
\bibitem [{\citenamefont {Mironov}\ and\ \citenamefont
  {Morozov}(2017)}]{Mironov:2017lgl}%
  \BibitemOpen
  \bibfield  {author} {\bibinfo {author} {\bibnamefont {Mironov}, \bibfnamefont
  {A.}}\ and\ \bibinfo {author} {\bibnamefont {Morozov}, \bibfnamefont {A.}},\
  }\href {\doibase 10.1016/j.physletb.2017.08.004} {\bibfield  {journal}
  {\bibinfo  {journal} {Phys. Lett. B}\ }\textbf {\bibinfo {volume} {773}},\
  \bibinfo {pages} {34} (\bibinfo {year} {2017})},\ \Eprint
  {http://arxiv.org/abs/1707.02443} {arXiv:1707.02443 [hep-th]} \BibitemShut
  {NoStop}%
\bibitem [{\citenamefont {Guest}, \citenamefont {Its},\ and\ \citenamefont
  {Lin}(2012)}]{Guest:2012yg}%
  \BibitemOpen
  \bibfield  {author} {\bibinfo {author} {\bibnamefont {Guest}, \bibfnamefont
  {M.~A.}}, \bibinfo {author} {\bibnamefont {Its}, \bibfnamefont {A.~R.}}, \
  and\ \bibinfo {author} {\bibnamefont {Lin}, \bibfnamefont {C.-S.}},\
  }\href@noop {} {\  (\bibinfo {year} {2012})},\ \Eprint
  {http://arxiv.org/abs/1209.2045} {arXiv:1209.2045 [math.DG]} \BibitemShut
  {NoStop}%
\bibitem [{\citenamefont {Nekrasov}(2003)}]{Nekrasov:2003af}%
  \BibitemOpen
  \bibfield  {author} {\bibinfo {author} {\bibnamefont {Nekrasov},
  \bibfnamefont {N.~A.}},\ }in\ \href@noop {} {\emph {\bibinfo {booktitle}
  {{International Congress of Mathematicians (ICM 2002) Beijing, China, August
  20-28, 2002}}}}\ (\bibinfo {year} {2003})\ \Eprint
  {http://arxiv.org/abs/hep-th/0306211} {arXiv:hep-th/0306211 [hep-th]}
  \BibitemShut {NoStop}%
\bibitem [{\citenamefont {Nekrasov}\ and\ \citenamefont
  {Okounkov}(2006)}]{Nekrasov:2003rj}%
  \BibitemOpen
  \bibfield  {author} {\bibinfo {author} {\bibnamefont {Nekrasov},
  \bibfnamefont {N.}}\ and\ \bibinfo {author} {\bibnamefont {Okounkov},
  \bibfnamefont {A.}},\ }\href {\doibase 10.1007/0-8176-4467-9_15} {\bibfield
  {journal} {\bibinfo  {journal} {Prog. Math.}\ }\textbf {\bibinfo {volume}
  {244}},\ \bibinfo {pages} {525} (\bibinfo {year} {2006})},\ \Eprint
  {http://arxiv.org/abs/hep-th/0306238} {arXiv:hep-th/0306238 [hep-th]}
  \BibitemShut {NoStop}%
\bibitem [{\citenamefont {Marino}\ and\ \citenamefont
  {Wyllard}(2004)}]{Marino:2004cn}%
  \BibitemOpen
  \bibfield  {author} {\bibinfo {author} {\bibnamefont {Marino}, \bibfnamefont
  {M.}}\ and\ \bibinfo {author} {\bibnamefont {Wyllard}, \bibfnamefont {N.}},\
  }\href {\doibase 10.1088/1126-6708/2004/05/021} {\bibfield  {journal}
  {\bibinfo  {journal} {JHEP}\ }\textbf {\bibinfo {volume} {05}},\ \bibinfo
  {pages} {021} (\bibinfo {year} {2004})},\ \Eprint
  {http://arxiv.org/abs/hep-th/0404125} {arXiv:hep-th/0404125} \BibitemShut
  {NoStop}%
\bibitem [{\citenamefont {Keller}\ \emph {et~al.}(2012)\citenamefont {Keller},
  \citenamefont {Mekareeya}, \citenamefont {Song},\ and\ \citenamefont
  {Tachikawa}}]{Keller:2011ek}%
  \BibitemOpen
  \bibfield  {author} {\bibinfo {author} {\bibnamefont {Keller}, \bibfnamefont
  {C.~A.}}, \bibinfo {author} {\bibnamefont {Mekareeya}, \bibfnamefont {N.}},
  \bibinfo {author} {\bibnamefont {Song}, \bibfnamefont {J.}}, \ and\ \bibinfo
  {author} {\bibnamefont {Tachikawa}, \bibfnamefont {Y.}},\ }\href {\doibase
  10.1007/JHEP03(2012)045} {\bibfield  {journal} {\bibinfo  {journal} {JHEP}\
  }\textbf {\bibinfo {volume} {03}},\ \bibinfo {pages} {045} (\bibinfo {year}
  {2012})},\ \Eprint {http://arxiv.org/abs/1111.5624} {arXiv:1111.5624
  [hep-th]} \BibitemShut {NoStop}%
\bibitem [{\citenamefont {Gamayun}, \citenamefont {Iorgov},\ and\ \citenamefont
  {Lisovyy}(2013)}]{Gamayun:2013auu}%
  \BibitemOpen
  \bibfield  {author} {\bibinfo {author} {\bibnamefont {Gamayun}, \bibfnamefont
  {O.}}, \bibinfo {author} {\bibnamefont {Iorgov}, \bibfnamefont {N.}}, \ and\
  \bibinfo {author} {\bibnamefont {Lisovyy}, \bibfnamefont {O.}},\ }\href
  {\doibase 10.1088/1751-8113/46/33/335203} {\bibfield  {journal} {\bibinfo
  {journal} {J. Phys.}\ }\textbf {\bibinfo {volume} {A46}},\ \bibinfo {pages}
  {335203} (\bibinfo {year} {2013})},\ \Eprint {http://arxiv.org/abs/1302.1832}
  {arXiv:1302.1832 [hep-th]} \BibitemShut {NoStop}%
\bibitem [{\citenamefont {Flume}\ and\ \citenamefont
  {Poghossian}(2003)}]{Flume:2002az}%
  \BibitemOpen
  \bibfield  {author} {\bibinfo {author} {\bibnamefont {Flume}, \bibfnamefont
  {R.}}\ and\ \bibinfo {author} {\bibnamefont {Poghossian}, \bibfnamefont
  {R.}},\ }\href {\doibase 10.1142/S0217751X03013685} {\bibfield  {journal}
  {\bibinfo  {journal} {Int. J. Mod. Phys.}\ }\textbf {\bibinfo {volume}
  {A18}},\ \bibinfo {pages} {2541} (\bibinfo {year} {2003})},\ \Eprint
  {http://arxiv.org/abs/hep-th/0208176} {arXiv:hep-th/0208176 [hep-th]}
  \BibitemShut {NoStop}%
\bibitem [{\citenamefont {Bruzzo}\ \emph {et~al.}(2003)\citenamefont {Bruzzo},
  \citenamefont {Fucito}, \citenamefont {Morales},\ and\ \citenamefont
  {Tanzini}}]{Bruzzo:2002xf}%
  \BibitemOpen
  \bibfield  {author} {\bibinfo {author} {\bibnamefont {Bruzzo}, \bibfnamefont
  {U.}}, \bibinfo {author} {\bibnamefont {Fucito}, \bibfnamefont {F.}},
  \bibinfo {author} {\bibnamefont {Morales}, \bibfnamefont {J.~F.}}, \ and\
  \bibinfo {author} {\bibnamefont {Tanzini}, \bibfnamefont {A.}},\ }\href
  {\doibase 10.1088/1126-6708/2003/05/054} {\bibfield  {journal} {\bibinfo
  {journal} {JHEP}\ }\textbf {\bibinfo {volume} {05}},\ \bibinfo {pages} {054}
  (\bibinfo {year} {2003})},\ \Eprint {http://arxiv.org/abs/hep-th/0211108}
  {arXiv:hep-th/0211108 [hep-th]} \BibitemShut {NoStop}%
\bibitem [{\citenamefont {Fucito}, \citenamefont {Morales},\ and\ \citenamefont
  {Poghossian}()}]{Fucito2004}%
  \BibitemOpen
  \bibfield  {author} {\bibinfo {author} {\bibnamefont {Fucito}, \bibfnamefont
  {F.}}, \bibinfo {author} {\bibnamefont {Morales}, \bibfnamefont {J.~F.}}, \
  and\ \bibinfo {author} {\bibnamefont {Poghossian}, \bibfnamefont {R.}},\
  }\href {\doibase 10.1088/1126-6708/2004/10/037} {\
  10.1088/1126-6708/2004/10/037},\ \Eprint
  {http://arxiv.org/abs/hep-th/0408090v2} {hep-th/0408090v2} \BibitemShut
  {NoStop}%
\bibitem [{\citenamefont {Nakamura}()}]{Nakamura2015}%
  \BibitemOpen
  \bibfield  {author} {\bibinfo {author} {\bibnamefont {Nakamura},
  \bibfnamefont {S.}},\ }\href {\doibase 10.1093/ptep/ptv085} {\
  10.1093/ptep/ptv085},\ \Eprint {http://arxiv.org/abs/1502.04188v1}
  {1502.04188v1} \BibitemShut {NoStop}%
\bibitem [{\citenamefont {Bernard}\ \emph {et~al.}(1977)\citenamefont
  {Bernard}, \citenamefont {Christ}, \citenamefont {Guth},\ and\ \citenamefont
  {Weinberg}}]{PhysRevD.16.2967}%
  \BibitemOpen
  \bibfield  {author} {\bibinfo {author} {\bibnamefont {Bernard}, \bibfnamefont
  {C.~W.}}, \bibinfo {author} {\bibnamefont {Christ}, \bibfnamefont {N.~H.}},
  \bibinfo {author} {\bibnamefont {Guth}, \bibfnamefont {A.~H.}}, \ and\
  \bibinfo {author} {\bibnamefont {Weinberg}, \bibfnamefont {E.~J.}},\ }\href
  {\doibase 10.1103/PhysRevD.16.2967} {\bibfield  {journal} {\bibinfo
  {journal} {Phys. Rev. D}\ }\textbf {\bibinfo {volume} {16}},\ \bibinfo
  {pages} {2967} (\bibinfo {year} {1977})}\BibitemShut {NoStop}%
\bibitem [{\citenamefont {Ito}\ and\ \citenamefont {Sasakura}()}]{Ito1996}%
  \BibitemOpen
  \bibfield  {author} {\bibinfo {author} {\bibnamefont {Ito}, \bibfnamefont
  {K.}}\ and\ \bibinfo {author} {\bibnamefont {Sasakura}, \bibfnamefont {N.}},\
  }\href {\doibase 10.1016/S0550-3213(96)00598-6} {\
  10.1016/S0550-3213(96)00598-6},\ \Eprint
  {http://arxiv.org/abs/hep-th/9608054v1} {hep-th/9608054v1} \BibitemShut
  {NoStop}%
\bibitem [{\citenamefont {Nekrasov}\ and\ \citenamefont
  {Shadchin}()}]{Nekrasov2004}%
  \BibitemOpen
  \bibfield  {author} {\bibinfo {author} {\bibnamefont {Nekrasov},
  \bibfnamefont {N.}}\ and\ \bibinfo {author} {\bibnamefont {Shadchin},
  \bibfnamefont {S.}},\ }\href {\doibase 10.1007/s00220-004-1189-1} {\
  10.1007/s00220-004-1189-1},\ \Eprint {http://arxiv.org/abs/hep-th/0404225}
  {hep-th/0404225} \BibitemShut {NoStop}%
\bibitem [{\citenamefont {Hollands}, \citenamefont {Keller},\ and\
  \citenamefont {Song}(2011)}]{Hollands:2011zc}%
  \BibitemOpen
  \bibfield  {author} {\bibinfo {author} {\bibnamefont {Hollands},
  \bibfnamefont {L.}}, \bibinfo {author} {\bibnamefont {Keller}, \bibfnamefont
  {C.~A.}}, \ and\ \bibinfo {author} {\bibnamefont {Song}, \bibfnamefont
  {J.}},\ }\href {\doibase 10.1007/JHEP10(2011)100} {\bibfield  {journal}
  {\bibinfo  {journal} {JHEP}\ }\textbf {\bibinfo {volume} {10}},\ \bibinfo
  {pages} {100} (\bibinfo {year} {2011})},\ \Eprint
  {http://arxiv.org/abs/1107.0973} {arXiv:1107.0973 [hep-th]} \BibitemShut
  {NoStop}%
\end{thebibliography}%

	\end{document}